\documentclass[12pt,a4paper,notitlepage]{article}
\usepackage[utf8]{inputenc}
\usepackage{caption}
\usepackage{graphicx}
\usepackage{float}
\usepackage{epstopdf}
\usepackage{longtable}
\usepackage{multirow}
\usepackage{multicol}
\usepackage{booktabs}
\usepackage{comment}
\usepackage{bbm}
\usepackage{array}
\usepackage[a4paper]{geometry}
\usepackage{setspace}
\usepackage[bottom]{footmisc}
\usepackage{enumitem}
\usepackage{amssymb}
\usepackage{amsmath}
\usepackage{mathtools}
\usepackage{siunitx}
\usepackage{subcaption}
\usepackage{lscape}
\usepackage{pdflscape}
\usepackage{subcaption}
\usepackage{rotating}
\usepackage[colorlinks=true,allcolors=blue]{hyperref}

\usepackage[round]{natbib}

\usepackage{istgame}
\usepackage{adjustbox}
\usepackage{url}
\usepackage{listings}
\usepackage{subfiles}
\usepackage{xcolor}
\usepackage{authblk}
\usepackage{threeparttable}
\usepackage{dcolumn}
\newcolumntype{d}[1]{D..{#1}}
\newtheorem{result}{Result}

\title{Strategy Method Effects in Centipede Games: An Optimal Design Approach\thanks{This 
paper was previously circulated under the title 
``An Experiment on the Representation Effect in Centipede Games.'' Support 
from the National Science Foundation (SES-2243268 and SES-2343948) is gratefully acknowledged.  
Joseph Tao-yi Wang thanks the financial support from the National Science and Technology Council of Taiwan (MOST 110-2628-H-002-006, NSTC 114-2410-H-002-235-MY2) and the Taiwan Social Resilience Research Center (NTU-114L9004).  
We thank Colin Camerer, Charles Holt, Kirby Nielsen, Charles Sprenger, and audiences at the 2024 North American Conference of the Economic Science Association 
for their valuable comments.  
We are also grateful to Laila Delgado, Letty Diaz, Mary Martin, and Christopher Crabbe for their administrative support in conducting the experiments at Caltech and UC Irvine, and to Zhenlin Kang, Polina Detkova, Camila Farres Rodriguez, Jiatong Han, Mitchell Linegar, and Po Hyun Sung for their assistance with the pilot experiment.  
This study was approved by the Caltech IRB (protocol \#23-1388) and the NTU IRB (protocol \# 202101HS002).  
The experimental design and analysis plan are pre-registered with the AEA RCT Registry (AEARCTR-0012336).  }}

\author{Shiang-Hung Hu\thanks{%
{California Institute of Technology, Pasadena, 
CA 91125 USA. shu3@caltech.edu}} \;\;\;\; Po-Hsuan Lin\thanks{%
{University of Virginia, Charlottesville, 
VA 22904 USA. plin@virginia.edu}} \;\;\;\;
Thomas R. Palfrey\thanks{%
{California Institute of Technology, Pasadena, 
CA 91125 USA. trp@hss.caltech.edu}} \\ 
Joseph Tao-yi Wang\thanks{%
{National Taiwan University, Taipei, Taiwan 10617. 
josephw@ntu.edu.tw}} \;\;\;\;
Yu-Hsiang Wang\thanks{%
{California Institute of Technology, Pasadena, 
CA 91125 USA. ywang9@caltech.edu}}}

\date{\today}

\begin{document}
\begin{titlepage}
\maketitle
\setcounter{page}{0}
\thispagestyle{empty}

\begin{abstract}

We explore the twin questions of when and why the strategy method creates behavioral distortions in the elicitation of choices in laboratory studies of sequential games. While such distortions have been widely documented, the theoretical forces driving these distortions remain poorly understood. 
In this paper, we compare behavior in six optimally designed centipede games, implemented under three different choice elicitation methods: 
the direct response method, the reduced strategy method and the full strategy method.  
These methods elicit behavioral strategies, reduced strategies, and complete strategies, respectively.  
We find significant behavioral differences across these elicitation methods---differences that cannot be explained by standard game theory, but are consistent with the predictions of the Dynamic Cognitive Hierarchy solution \citep{lin_cognitive_2024}, combined with quantal responses.  
\end{abstract}

\bigskip


\medskip


\thispagestyle{empty}
\end{titlepage} \setcounter{page}{1}
\onehalfspacing

%\newpage

\begin{quote}
    \emph{The centipede game illustrates an important byproduct of experimentation: Choosing a design disciplines theorizing because it forces one to be crystal clear about the conditions under which theory is really expected to apply.} 
\end{quote}
\rightline{---Colin F. Camerer, Behavioral Game Theory 
(\citeyear{camerer2003behavioral}), p.~219}

\section{Introduction}

A wide variety of sequential games---such as bargaining, dynamic contribution games, the trust game, signaling games, sequential voting, and social learning---have been extensively studied in laboratory experiments.  
Most experimental studies use what is known as the \emph{direct response method}, in which subjects play the game \emph{sequentially}, following the exact timing of moves specified in the theoretical model being investigated.  
However, this approach to eliciting players’ strategies has a significant limitation---\emph{incompleteness}.  
That is, the direct response method often fails to elicit the complete strategy of each player. Information about behavior at unreached information sets is missing. 
For example, in the ultimatum game, if the first player offers \$2 and the responder rejects the offer, the experimenter obtains no information about how that responder would have responded to all other possible offers. \emph{The responder’s strategy is only partially elicited.} 

One way to resolve the incompleteness problem in the ultimatum game is to require the responder to specify a conditional response to all possible offers %at the beginning of the game, 
at the same time the proposer is choosing an offer.\footnote{In fact, such an experiment has been conducted by \citet{oxoby2004sequential}, who refer to this approach as the ``strategy vector method.''  
They find no statistically significant difference between this method and the direct response method.} 
This is an example of the \textit{strategy method}, pioneered by \cite{Selten1967}, where each player of the game simultaneously reports a conditional action at each possible information set.  
The strategy method effectively converts any game in extensive form into a simultaneous move game.  
This approach offers a methodological advantage, as it enables the collection of more experimental data, particularly at histories that are only occasionally reached when the game is played sequentially.  
In many applications, the elicitation procedure is simplified into the \emph{reduced} strategy method by collapsing outcome-equivalent strategies, as in the game's reduced normal form.

The traditional justification for this experimental approach to studying behavior in noncooperative games is that the set of Nash equilibrium outcomes is \textit{reduced normal form invariant}; i.e., if two games in extensive form share the same reduced normal form, then the two games have the same set of Nash equilibrium outcomes.  
In fact, this property of strategic invariance\footnote{The concept of strategic equivalence is first proposed by \cite{thompson_equivalence_1952}, who defines two games as strategically equivalent if they share the same reduced normal form.} has been invoked as a desired requirement for any set-valued solution concept for noncooperative games \citep{kohlberg_1986}.  
One could argue that, to the extent that experiments are designed to test Nash equilibrium predictions, this justification appears reasonable.  
However, if the goal is to understand behavioral deviations from equilibrium in a sequential game, then this justification could be problematic because additional behavioral distortions may arise from eliciting strategic decisions using a simultaneous move game, even if the simultaneous move game is strategically equivalent to the sequential game.

In fact, such distortions attributable to these variations in elicitation procedures have been observed, but when and how these behavioral distortions arise remain open questions.  
\cite{brandts_strategy_2011} document some of the effects of the strategy method by surveying 29 experimental studies of a wide range of noncooperative games that used both the strategy method (either full or reduced) and the direct response method.  
The survey offers some descriptive observations about regularities regarding whether differences were observed and the extent of those differences, but it provides no definitive conclusion about the existence of a strategy method effect, as the evidence is mixed.\footnote{Among the 29 comparisons in \cite{brandts_strategy_2011}, 16 find no difference, four find differences, and nine report mixed evidence.}  
The mixed findings highlight the need to organize and explain these diverse effects in a rigorous theoretical framework that will enable experimenters to better understand the ``when'' and ``how'' questions about behavioral distortions that arise from the use of the strategy method to elicit choices in sequential games.

A theory that predicts the behavioral distortions caused by the use of the strategy method must distinguish between two extensive games that share the same reduced normal form; that is, it must allow for violations of invariance under strategic equivalence.  
To this end, the Dynamic Cognitive Hierarchy (DCH) solution developed by \cite{lin_cognitive_2024} offers a promising theoretical framework.  
DCH can predict when and how the strategy method distorts behavior, as it generates both \textit{qualitative} and \emph{quantitative} predictions regarding violations of invariance under strategic equivalence.

The DCH solution belongs to the ``level‑$k$/cognitive hierarchy'' family of models (e.g., \citealt{stahl1994experimental, stahl1995players,
nagel1995unraveling, costa2001cognition,
camerer_cognitive_2004}), which posit a hierarchical structure of strategic sophistication among players.  
Each player is endowed with a level of sophistication and believes that all other players in the game are less sophisticated. Unlike the standard level‑$k$/cognitive hierarchy solution, which is defined only for simultaneous move games, the DCH solution is defined for general games in extensive form.  
Specifically, a DCH level‑$k$ player holds a prior belief about other players' sophistication levels based on the truncated true distribution of levels, conditional on levels ranging from 0 to $k$–1, i.e., players have ``truncated rational expectations.''  
These beliefs are updated as the history of play unfolds.  
In the DCH solution, level-$0$ players are non-strategic and are assumed to randomize uniformly at every information set. Players with level $k \geq 1$ sequentially best respond to their updated beliefs.  

The DCH solution naturally extends the standard cognitive hierarchy solution to general games in extensive form, and the violation of invariance under strategic equivalence is a surprising theoretical property that emerges from this extension.  
That is, the DCH solution may differ across games that share the same reduced normal form.  
Intuitively, strategically equivalent games that share the same reduced normal form may differ in sizes of their action sets across extensive-form representations. Since level-$0$ players in DCH randomize uniformly, their behavior depends on the size of the action set, which in turn influences the behavior of higher-level players.  
In summary, the key insight of DCH is that \emph{behavioral distortions introduced by the strategy method arise from changes in the number of available actions}.\footnote{On the other hand, the DCH solution is invariant between the extensive form and the corresponding (non-reduced) normal form, implying no behavioral distortion from using the \textit{full} strategy method. Our experiment tests this prediction of a non-effect, in addition to testing 
DCH predictions about the direction and magnitudes of distortions from using the reduced strategy method.}

The direction and magnitude of the violation predicted by DCH generally depend on 
the game structure and the prior distribution of levels of sophistication.
However, in a broad class of ``centipede games,''
DCH makes a bold prediction about the direction of the violation of invariance, making the centipede 
game an ideal experimental environment.
An extensive literature in experimental game theory
has examined whether players can reach the unique subgame perfect 
equilibrium, with both the direct response method and the reduced strategy method being commonly adopted.  
The use of the strategy method is justified by strategic equivalence and the uniqueness of the equilibrium.  
However, the DCH solution 
predicts earlier taking under the direct response method 
than under the reduced strategy method (\citealt{lin_cognitive_2024}, Theorem 1), 
a pattern consistent with the strategy method effect 
observed in the centipede game experiment by \cite{garcia-pola_hot_2020}
at the aggregate level.\footnote{\cite{garcia-pola_hot_2020}'s centipede game experiment 
consists of four different centipede games, each with a subgame perfect equilibrium that 
predicts termination at the first stage. To test for behavioral distortions 
caused by the use of the strategy method, \cite{garcia-pola_hot_2020} 
compare behavior under the direct response method and the strategy method using 
a \emph{between-subjects design}. In three of the four games, those for which the DCH predicts a strategy method effect, termination occurs earlier under the direct response method.  
In the fourth game where no such effect is predicted, the strategy method effect is not observed empirically.}

Building on this result, the goal of this paper is to carefully test whether DCH can serve as a theoretical foundation for understanding behavioral deviations from strategic equivalence, using a fine-tuned centipede game experiment.  
To examine how the number of available actions under different elicitation methods influences behavior, our experiment employs a within-subject design consisting of three treatments, each implementing the same centipede game using one of the three elicitation methods: the direct response method, the reduced strategy method, and the full strategy method.  
This design allows us to observe each subject’s behavior across all three different elicitation methods for the same game, enabling us to measure behavioral differences at the individual level.

Although DCH makes a clear qualitative prediction about the strategy method effect in the centipede game, the magnitude of the treatment effect depends on both the payoff parameters and the distribution of levels of sophistication, with the latter unknown to the experimenter prior to data collection. Therefore, the main challenge of our experimental design is to construct payoff structures that maximize the informativeness of the experiment \emph{a priori}.  
To address this, we adopt the ``optimal design approach'' developed by \cite{lin_cognitive_2023} to select game parameters.\footnote{It is worth noting that the optimal design 
approach adopted in this paper does not refer to the one commonly used in the statistical literature, which aims to maximize the determinant of the information matrix.  
See Chapter 14 of \cite{moffatt_experimetrics_2020} for more on optimal design in the statistical context and its applications in risky lottery experiments.}

The optimal design approach is a two-step procedure: first, we calibrate
the distribution of levels; then, treating this as the true distribution, 
we select centipede game parameters that generate various magnitudes of the strategy method
effects predicted by DCH. Specifically, we begin by conducting 
a pilot experiment using four centipede games theoretically
examined by \cite{lin_cognitive_2024} and estimate the distribution of 
levels based on the pilot data. To systematically search for 
the most informative game parameters, we compute the expected strategy method effects
predicted by DCH across three parameterized classes of centipede games: linear, exponential,
and constant centipede games,\footnote{These classes are named according to 
the growth rate of the pie size and are the most commonly 
studied centipede games in the literature.} based on the calibrated distribution.  From each class, 
we select two games, one predicted to exhibit a strong 
strategy method effect and one a weak effect, yielding six centipede games in total.

With this high-powered experimental design, we first compare the distribution of terminal nodes across the three different elicitation methods at the aggregate level and find a significant strategy method effect.  
This effect is primarily driven by later termination under the reduced strategy method, consistent with the predictions of DCH.  
Second, leveraging the six different payoff configurations, we find that although the magnitudes of the strategy method effects vary across games, but the quantitative patterns never violate the prediction of DCH.  
In all six centipede games, the strategy method effects occur in the direction predicted 
by DCH when comparing the reduced strategy method with either the direct 
response or the full strategy method. When comparing the full strategy method
with the direct response method---where DCH predicts no effect---no significant difference 
is detected in five of the six games.

Although DCH successfully explains the qualitative patterns in the data,
the relative magnitudes of the strategy method effects within each class 
of centipede games do not fully align with the predictions of the calibrated DCH. 
We hypothesize that these discrepancies are driven by quantal responses,
which can also lead to violations of invariance under strategic equivalence.
To disentangle the effect of quantal responses, we compare DCH with 
two alternative behavioral models: the Agent Quantal Response 
Equilibrium (AQRE; \citealt{mckelvey_quantal_1998}) and a hybrid model, 
the Quantal Dynamic Cognitive Hierarchy (QDCH; \citealt{lin_cognitive_2023}) solution.
We structurally estimate these models and find that QDCH not only provides
a better fit to the data than the other models but also successfully explains
the relative magnitudes of the strategy method effects.

In short, despite the value of the strategy method in solving the incompleteness problem in experimental data collection, evidence suggests that studying behavior in sequential games by having subjects play a ``strategically equivalent'' simultaneous move game can lead to distorted findings.  
In particular, we use the centipede game to demonstrate that choice behavior elicited through the reduced strategy method asking players to make (outcome-equivalent) decisions simultaneously can indeed be distorted, but \emph{in predictable ways} that are consistent with the predictions of DCH combined with logit quantal response behavior.

The remainder of the paper is organized as follows. 
The next section reviews the related literature.
To clarify our terminology, Section \ref{section:prelim} introduces the three elicitation methods used in centipede games: the direct response method, the (full) strategy method, and the reduced strategy method.  
It also provides an overview of the DCH strategy method effect.
Sections \ref{sec:experimental_design} describes the optimal selection of centipede games and the experimental design.  
Section \ref{sec:exp_results} presents the experimental results, and Section \ref{section:structural_estimation} compares DCH with alternative behavior solution concepts.  
Finally, Section \ref{sec:conclusion} concludes.

\section{Related Literature}
\label{section:literature}

The centipede game was first introduced by \cite{ROSENTHAL198192} to illustrate how counterintuitive backward induction can be in certain environments.\footnote{The name ``centipede'' was coined by \cite{binmore1987modeling}, referring to a 100-node variant.}  
Since then, it has been described by game theorists as a paradox 
of backward induction \citep{megiddo1986remarks, aumann1992irrational,
reny1992rationality, ben1997rationality}, and several theories have been proposed to explain why people deviate from the unique subgame perfect equilibrium.

\cite{mckelvey_experimental_1992} 
conducted the first centipede game experiment, demonstrating that
behavior is grossly inconsistent with the prediction of 
subgame perfect equilibrium. This divergence from equilibrium behavior 
has been extensively replicated across various environments, including games
of different lengths (e.g., \citealt{mckelvey_experimental_1992} and
\citealt{fey1996experimental}), different subject pools (e.g., 
\citealt{palacios2009field, levitt2011checkmate, li2021conducting} and 
\citealt{brocas2025children}), 
different payoff configurations (e.g., \citealt{fey1996experimental, zauner1999payoff, 
kawagoe2012level, healy2017epistemic} and \citealt{garcia2020non}), and different 
numbers of players (e.g., \citealt{rapoport2003equilibrium} and
\citealt{bornstein2004individual}).

Although most centipede game experiments, following \cite{mckelvey_experimental_1992}, use the direct response method, there are a few notable exceptions.  \cite{nagel_experimental_1998} was the first to implement the centipede game as a simultaneous move game employing the reduced strategy method.  
While they observed similar deviation from the immediately taking equilibrium found in \cite{mckelvey_experimental_1992}, their experiment was not designed to test for strategy method effects and did not collect data under the direct response method.  
However, the authors themselves did conjecture the existence of such an effect: ``...There might be substantial differences in behavior in the extensive form game and in the
normal form game...” (\citealt{nagel_experimental_1998}, p. 357).

More recently, to explore the effect of using the strategy method in centipede game experiments, \cite{garcia-pola_hot_2020} conduct a direct comparison between the direct-response method and the strategy method---an experiment most closely related to ours.\footnote{It is worth noting that \cite{kawagoe2012level} also compare behavior under both methods in the working paper version of their study.  
In their experiment, \cite{kawagoe2012level} compare the two methods using the exponential centipede game from \cite{mckelvey_experimental_1992} and the constant centipede game from \cite{fey1996experimental}, employing a within-subject design.  
They detect no behavioral differences between the two methods.  
A plausible reason is that, in their comparison, the same subjects experience one elicitation procedure followed immediately by the other using the same game, potentially eliciting virtually identical behavior across both methods.  
By contrast, in each treatment of our experiment, each subject plays six different games in randomized order, which effectively mitigates potential spillover effects.}  
Both aim to examine whether the strategy method is behaviorally equivalent to the direct response method in the centipede game, though we employ different experimental design approaches.  
\cite{garcia-pola_hot_2020} conduct an exploratory experiment to detect the existence of any strategy method effect.  
To enable a broad search and avoid spillover effects across treatments, they implement four completely different centipede games using a between-subject design.  
Their finding of a strategy method effect consistent with DCH serves as the foundation for our experiment.

In contrast to \cite{garcia-pola_hot_2020}, our goal is to test both the qualitative and quantitative predictions of the strategy method effect as predicted by DCH.  
To achieve this, we compare behavior across three different elicitation methods of the game and systematically vary the payoff parameters.  
These design differences allow us to identify when and how a strategy method effect occurs.  
Furthermore, our within-subject design enables us to analyze the strategy method effect at the individual level.  
In this way, our experiment complements \cite{garcia-pola_hot_2020}, offering a more comprehensive understanding of the behavioral distortions introduced by the use of the strategy method.  

Furthermore, our finding of a violation of invariance under strategic equivalence connects to recent developments in behavioral solution concepts and their applications.  
From the perspective of standard game theory, reduced-normal-form invariance has long been considered a desirable property.  
However, recent work in behavioral game theory has shown that violations of this invariance are a common feature of several behavioral solution concepts, including the agent quantal response equilibrium \citep{mckelvey_quantal_1998}, the dynamic cognitive hierarchy solution \citep{lin_cognitive_2023, lin_cognitive_2024}, and the cursed sequential equilibrium \citep{fong2023cursed}.  
Our findings in this paper provide empirical support for this growing literature.

Lastly, our finding of behavioral distortions across different, yet strategically equivalent, elicitation methods is important not only for behavioral game theory but also for the field of mechanism design.  
Strategic equivalence is widely invoked in many institutional settings where reduced normal form mechanisms are often implemented, rather than extensive form mechanisms, due to simplicity or practicality concerns.  
However, growing evidence from various contexts suggests that different representations of a mechanism can lead to different behavior---for example, sealed-bid auctions versus clock auctions \citep{li2017obviously}, runoff elections versus contingent elections \citep{richie2023case}, and dynamic versus static matching mechanisms \citep{gong2025dynamic}.  
This paper contributes to this literature by demonstrating both the empirical relevance and the predictive power of the DCH violation of invariance under strategic equivalence.  

\section{The Dynamic Cognitive Hierarchy (DCH) Strategy Method Effect in Centipede Games}
\label{section:prelim}

In this section, we first define the three classes of centipede games:
linear, exponential, and constant. These are the most commonly studied 
classes in the literature and, more importantly, can be \emph{parameterized} using a small number of variables. Using these parameterizations substantially facilitate our search for optimally designed centipede games. Second, we explain the differences between
the direct response method, the reduced strategy method, and the full strategy method in the context of centipede games. Third, we provide an overview of the theoretical DCH strategy method effects in these games.

\subsection{Three Classes of Centipede Games}

\begin{figure}[htbp!]
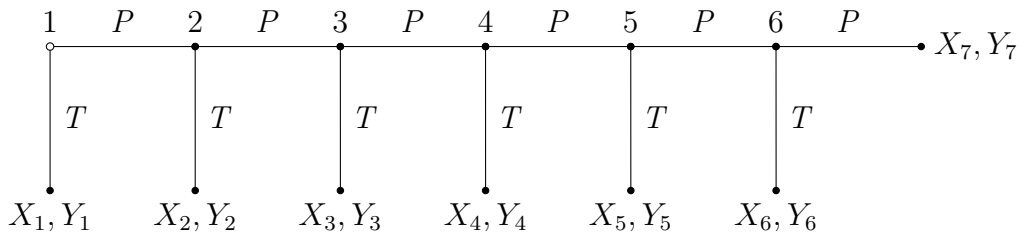

\centering
\begin{istgame}[scale=1.5]
\setistmathTF*001
\xtShowEndPoints
\setistgrowdirection{south east}
\xtdistance{9mm}{18mm}
\istroot(0)[initial node]{1}
\istb{T}[r]{X_1, Y_1}[b] \istb{P}[a] \endist
\istroot(1)(0-2){2}
\istb{T}[r]{X_2, Y_2}[b] \istb{P}[a] \endist
\istroot(2)(1-2){3}
\istb{T}[r]{X_3, Y_3}[b] \istb{P}[a] \endist
\istroot(3)(2-2){4}
\istb{T}[r]{X_4, Y_4}[b] \istb{P}[a] \endist
\istroot(4)(3-2){5}
\istb{T}[r]{X_5, Y_5}[b] \istb{P}[a] \endist
\istroot(5)(4-2){6}
\istb{T}[r]{X_6, Y_6}[b] \istb{P}[a]{X_7, Y_7}[r] \endist
\end{istgame}
\caption{The Generic Game Tree of a 6-Stage Centipede Game}
\label{fig:generic_tree}
\end{figure}

The centipede game is a two-person sequential game in which Player 1, the first mover, and Player 2, the second mover, alternate over a sequence of moves.  
At each turn, the player whose move it is can either end the game by choosing to ``take'' and receive the larger of the two payoffs or allow the game to continue by choosing to ``pass.''  
When a player passes, the larger payoff strictly increases and the difference between the larger and smaller payoffs (weakly) increases.  
The game continues for $2D$ decision nodes (stages) where $D\geq 2$, and we label the decision nodes by $\{1,2, ..., 2D \}$.  
Player 1 moves at odd nodes and Player 2 moves at even nodes.  
If the game is ended by a player at stage $j\leq 2D$, the payoffs are $(X_j, Y_j)$.  
If no player ever takes, the payoffs are $(X_{2D+1}, Y_{2D+1})$.  
The centipede games that are of interest have the property that $X_j$ increases over time, and $X_j > Y_{j+1}$ for $j=1,...,2D$, so that the unique Nash equilibrium outcome is $(X_1, Y_1)$.
The generic game tree of a 6-stage (the length implemented in our experiment) centipede game 
is illustrated in Figure~\ref{fig:generic_tree}.

In this paper, we focus on three commonly adopted classes of centipede games: linear, 
exponential and constant centipede games, which can be parameterized as follows.  
\begin{enumerate}
    \item \textbf{Linear Centipede Games:}
    
    In a linear centipede game, the \emph{difference} between the large and small payoffs is normalized to 1 and remains constant.  
    When a player chooses to pass, both the large and small payoffs are increased by an amount $0<c<1$.  
    Therefore,
    \begin{align*}
        (X_j, Y_j) = \begin{cases}
            \big( 1 + (j - 1)c, \; (j - 1)c \big) \;\;\;\mbox{if}\; j \mbox{ is odd}\phantom{.} \\
            \big( (j - 1)c, \; 1+ (j -1)c \big) \;\;\;\mbox{if}\; j \mbox{ is even.} 
        \end{cases}
    \end{align*}

    \item \textbf{Exponential Centipede Games:}

    In an exponential centipede game, the \emph{ratio} between the large and the small payoffs is equal to $c > 1$ and does not change as the game progresses.  
    As a player passes, both the large and small (positive) payoffs are multiplied by $1<\pi<c$.  
    To prevent the payoffs from becoming explosively large in the later stages, 
    we fix the multiplier at $\pi = 2$ and vary the ratio $c$ in our experiment.
    Thus,
    \begin{align*}
        (X_j, Y_j) = \begin{cases}
            (c \pi^{j-1}, \; \pi^{j-1}) = (c\cdot 2^{j-1}, \; 2^{j-1}) \;\;\;\mbox{if}\; j \mbox{ is odd}\phantom{.} \\
            (\pi^{j-1}, \; c \pi^{j-1}) = (2^{j-1}, \; c\cdot 2^{j-1}) \;\;\;\mbox{if}\; j \mbox{ is even.} 
        \end{cases}
    \end{align*}

    \item \textbf{Constant Centipede Games:}

    In a constant centipede game, the \emph{sum} of the large and the small payoffs remains the same at every stage.  
    When a player passes, the smaller payoff is multiplied by a factor $0<c<1$.  
    As a result,
    \begin{align*}
        (X_j, Y_j) = \begin{cases}
            (2-c^{j-1}, \; c^{j-1}) \;\;\;\mbox{if}\; j \mbox{ is odd}\phantom{.} \\
            (c^{j-1}, \; 2-c^{j-1}) \;\;\;\mbox{if}\; j \mbox{ is even.} 
        \end{cases}
    \end{align*}    
\end{enumerate}

Across all classes of centipede games, the unique
Nash equilibrium outcome from the perspective of standard game theory
is for Player 1 to end the game at Stage 1.
In what follows, we introduce the three elicitation methods in our experiment.

\subsection{Three Elicitation Methods}

\begin{figure}[htbp!]
    \centering
    \includegraphics[width=\linewidth]{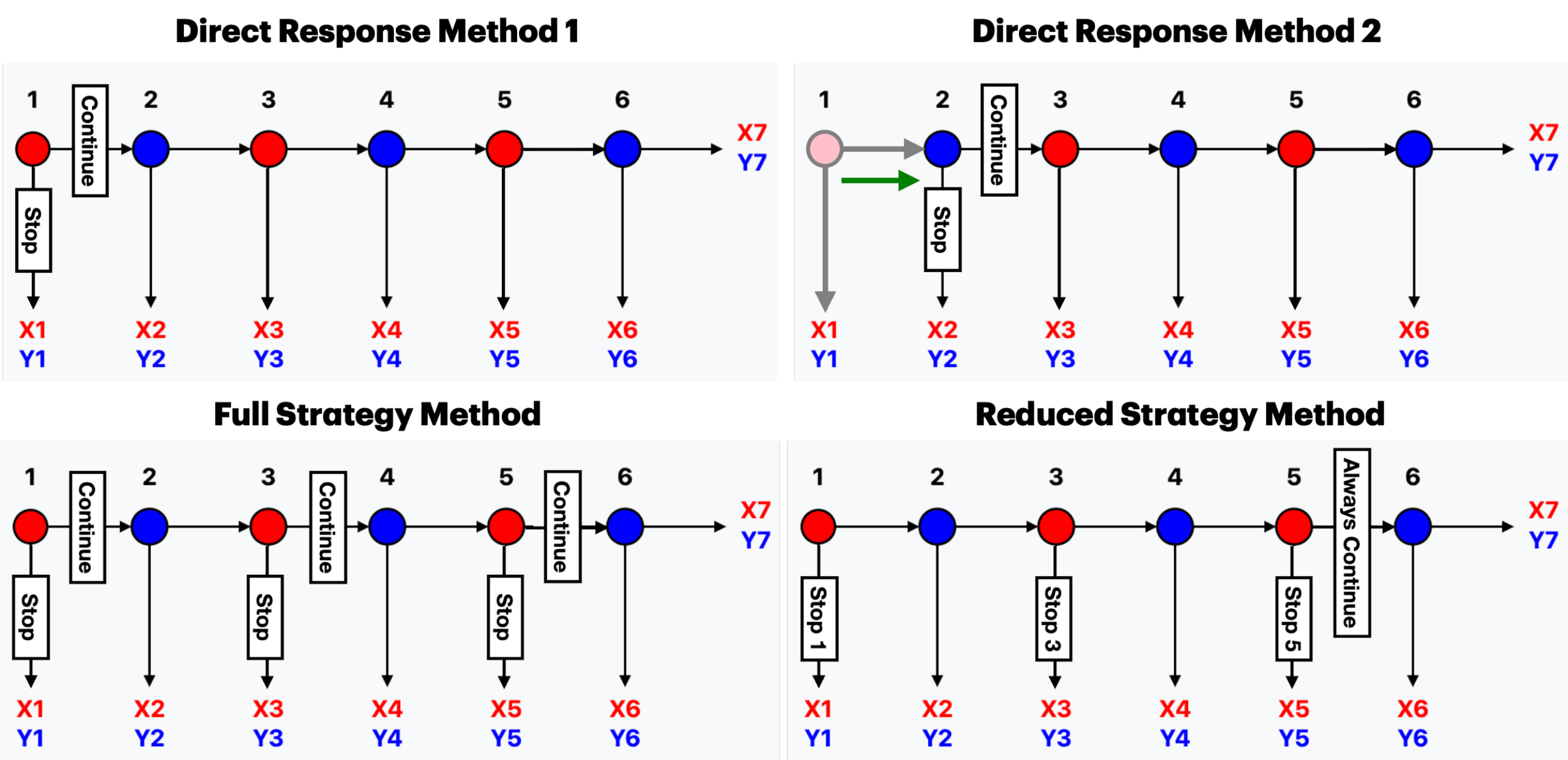}
    \caption{Screenshots of the Direct Response Method (top), Full Strategy Method  
    (bottom left) and Reduced Strategy Method (bottom right)}
    \label{fig:exp_screenshots}
\end{figure}

In this subsection, we describe the three elicitation methods using actual screenshots from the experiment, shown in Figure~\ref{fig:exp_screenshots}.  

The top row of panels in Figure~\ref{fig:exp_screenshots} show the screenshots for the \textbf{direct response method}, which elicits the \emph{behavioral strategy} by implementing the extensive-form game tree.  
Under this method, at the beginning of a 6-stage centipede game, Player 1 decides whether to stop or continue.  
If Player 1 chooses to stop, the game ends immediately, and Player 2 does not have the opportunity to make a decision. Otherwise, the game proceeds to the second stage, where Player 2 decides whether to stop or continue.  
If Player 2 also chooses to continue, it becomes Player 1’s turn again, and so on.\footnote{As the game progresses to the next stage, earlier stages are shaded, and arrows are added to indicate the flow of the game.}  

In contrast, the \textbf{full strategy method} elicits the 
\emph{full strategy} by converting the sequential centipede
game into a simultaneous move game. In a 6-stage centipede game, each 
player has three decision nodes and therefore eight possible strategies, as they can choose either to stop or continue at each node. 
To elicit the full strategy, players are asked to make a 
decision---stop or continue---at each of their decision nodes simultaneously,
as shown in the bottom left panel of Figure~\ref{fig:exp_screenshots}.

From an experimental implementation perspective, the full strategy method 
can be unnecessarily complicated (and is therefore rarely used in the literature), as it requires
players to make decisions at \emph{every} decision node---including those that 
cannot be reached given their own earlier choices.
For example, even if players choose to stop at Stage 1 (as Player 1) and thereby ends the game, a complete strategy still requires players to specify their hypothetical decisions at Stages 3 and 5.\footnote{These strategies are also referred to as ``cloned strategies'' in \cite{camerer_cognitive_2004}.}

To this end, the \textbf{reduced strategy method} is more appealing, as it 
substantially simplifies the elicitation task by asking players to 
simultaneously choose one of their a \emph{reduced strategies}.
In the 6-stage centipede game example, each player has four reduced strategies: taking at their first, second, or third decision node, or always passing.
As shown in the bottom right panel of Figure~\ref{fig:exp_screenshots}, 
the reduced strategy method converts the sequential centipede game into a 
simultaneous move game in which all players simultaneously select one of the 
four reduced strategies.

From a game-theoretic perspective, the centipede game implemented under the
three elicitation methods constitutes three distinct, yet strategically equivalent, games.
Some traditional set-valued solution concepts for noncooperative games other than 
Nash equilibrium fail to satisfy invariance under strategic equivalence.\footnote{For 
example, \cite{kohlberg_1986} observe that many widely used refinements of Nash equilibrium, 
including perfect equilibrium, proper equilibrium, sequential equilibrium, 
and essentially all belief-based refinements such as the intuitive criterion, 
violate invariance under strategic equivalence.}
However, in the centipede games we study, all refinements of Nash equilibrium predict 
the same outcome---termination at the first node---regardless of the elicitation method used.
Therefore, any differences observed between the three elicitation methods in the 
experiment cannot be explained by standard game-theoretic solution concepts.

Departing from standard solution concepts, the DCH solution makes a clear qualitative prediction
about how behavior would differ across the three elicitation methods. In the next
subsection, we provide a brief overview of the DCH strategy method effect.

\subsection{The DCH Framework and Strategy Method Effects}
\label{subsec:Model-DCH}

The DCH framework provides a behavioral solution concept that extends the cognitive hierarchy framework of \cite{camerer_cognitive_2004} from simultaneous-move games in strategic form to general sequential games in extensive form.  
In this model, each player is endowed with a level of sophistication $k\in \{0,1,2,... \}$, assumed to be drawn independently from a Poisson distribution with mean $\tau>0$.\footnote{\cite{lin_cognitive_2024} characterize the DCH solution for a general class of prior level distributions, with the Poisson specification being a commonly assumed special case in the literature for estimation purposes.}  
We denote $f_\tau(k) \equiv e^{-\tau}\tau^k / k!$ as the prior probability of being assigned level $k$ under Poisson($\tau$).  

Each player $i$ with level $k>0$ has a prior belief about all other players' levels and these prior beliefs satisfy \emph{truncated rational expectations}.  
Specifically, for any player $i$ and level $k>0$, a level $k$ player $i$ believes all other players are at most level $k-1$, and their prior belief about any other player $j\neq i$ being level $\kappa$ is 
\begin{align*}
    \mu_{ij}^k(\kappa) \equiv \begin{cases}
        \frac{f_{\tau}(\kappa)}{\sum_{m=0}^{k-1}f_{\tau}(m)} \;\; \mbox{if}\; \kappa< k\\
        0 \;\;\;\;\;\;\;\;\;\;\;\;\;\;\;\; \mbox{if}\; \kappa\geq  k,
    \end{cases}
\end{align*}
which is the lower truncation of the true prior distribution of levels.  
The underlying assumption of this specification is that level-$k$ type players have correct beliefs about the relative proportion of players who are less sophisticated than themselves, while incorrectly believing that other players of level $\kappa \geq k$ do not exist.  

The DCH solution is defined as a level-dependent behavioral strategy profile
such that 
(1) level-$0$ players uniformly randomize at every history,\footnote{Because uniform randomization is non-degenerate, a notable implication is that there is no off-path event in the DCH solution.}  and (2) level $k>0$ players form posterior beliefs about other players' levels via Bayes' rule at every history and sequentially best respond everywhere.  
The DCH solution can be solved recursively, starting with the lowest level and iteratively working up to higher levels.

An important feature of the DCH solution is its violation of invariance under strategic equivalence, since the DCH solution can differ between two games that share the same reduced normal form.  
The intuition behind this effect is that collapsing outcome-equivalent strategies into structurally reduced strategies changes the cardinalities of the action sets.  
Since level-$0$ players are assumed to randomize uniformly over all available actions, their strategies are generally not outcome-equivalent when the number of available actions is reduced. Hence, this property of DCH is referred to as the \emph{strategy reduction effect} \citep{lin_cognitive_2024}.  
The strategy reduction effect on level-0 players then triggers a chain reaction of indirect effects on the behavior of higher-level players.

The DCH solution makes a bold prediction about the \emph{strategy method effect}
in the centipede game: players tend to choose to take earlier under the direct 
response method than under the reduced strategy method, while their behavior 
remains the same between the direct response method and the full strategy method.
To illustrate the rationale behind this prediction, 
consider the 6-stage centipede game.  
When choices are elicited using the direct response method, DCH posits that level-$0$ players will take (or pass) with a 50\% chance at each stage.  
However, under the reduced strategy method, level-$0$ players uniformly randomize across four reduced strategies: taking at their first, second, or third decision node, or always passing,
leading level-$0$ players to choose each reduced strategy with a probability of $1/4$.  
In other words, level-$0$ players’ behavior is not outcome-equivalent between the direct response method and the reduced strategy method---they tend to take later under the reduced strategy method.  
Since higher-level players never rule out the possibility that the other player is level 0, the later taking behavior of level-$0$ players results in later taking behavior for all higher levels, as payoffs increase in the later stages (Theorem 1 of \citealt{lin_cognitive_2024}).

Conversely, under the full strategy method, level-$0$ players uniformly randomize across all pure strategies, which is outcome-equivalent to uniformly randomizing over the set of actions at each history 
under the direct response method. 
Since DCH players with level $k>0$ are forward-looking, as shown by \cite{battigalli_note_2023}, the DCH solution is outcome-equivalent between games that share the same 
(non-reduced) normal form. 
In other words, the DCH solution is \emph{normal form invariant} (Proposition 1 of \citealt{battigalli_note_2023}). This invariance implies that the full strategy method 
does not induce behavioral distortions like those caused by the reduced strategy method.

\bigskip

\noindent\textbf{DCH Strategy Method Effect:} 
\emph{For any linear, exponential, or constant centipede game in which 
the parameters satisfy a mild condition,}\footnote{Specifically, for linear centipede games, we require that the pie-size increment satisfies $1/3 < c < 1$. If $c > 1$, the unique equilibrium is for every player to always pass; if $c < 1/3$, then all players with level $k > 0$ will always take. Similarly, for constant centipede games, we require $0 < c < 1$. Lastly, for exponential centipede games, we require that $\frac{-1 + \sqrt{1 + 8c^2}}{2c} < \pi < c$, which reduces to the condition $c > 2$ when setting $\pi = 2$.}
\emph{DCH predicts that 
\begin{itemize}
    \item[1.] players tend to take earlier under the direct response method than 
    under the reduced strategy method; 
    \item[2.] players will behave the same under the direct response method and the full strategy method.
\end{itemize}
}

As shown by \cite{lin_cognitive_2024}, this prediction is robust to any prior 
distribution of levels. However, the expected magnitude of the difference between
the behavior elicited under the direct response method and the reduced
strategy method depends on both the prior distribution of levels and the 
payoff structure. Therefore, we adopt an optimal design approach to construct 
the payoff structure in order to maximize the informativeness of the experiment.

\section{Design Optimization and Implementation}
\label{sec:experimental_design}

The objective of this paper---and its key methodological innovation---is to use DCH as a theoretical basis for rigorously developing a high-powered experimental design to systematically examine whether it can explain when and how the use of strategy method creates distortions.  
We optimize the design by using the theory to carefully select two centipede games from each of the three different classes of centipede games.  
In particular, we first introduce the optimal design framework in detail in Section \ref{subsec:optimal_design}, and then describe our design implementation and experimental procedures in Section \ref{subsec:experimental_design_implementation}.

\subsection{The Optimal Design Framework}
\label{subsec:optimal_design}

As discussed earlier, in all three classes of these games, the \emph{direction} of the 
DCH strategy method effects are unambiguous: players will take later
if choices are elicited under the reduced strategy method then either 
the direct response method or the full strategy method; no such difference 
is predicted between the direct response method and the full strategy method. 

While these directional effects are unambiguous, the magnitude of these effects varies widely 
depending on both the class of game and the exact payoff details. For this reason, we employ an optimal design 
approach in order to choose payoffs parameters, one for each class of game, that are predicted to give our 
experiment great power by inducing relatively large strategy method effects. 
In addition, as a quasi-placebo test, we use the same optimal design to 
identify payoff parameters that, according to the theory, should not induce measurable 
strategy method effects, thereby enabling a testable comparative static prediction about 
the magnitude of such effects.

Our optimal design approach comprises two steps: (1) model calibration 
and (2) optimal selection of game parameters, with the 
specific procedures described below. Technical details are relegated to 
Appendix \ref{appendix:game_selection}.

\begin{enumerate}
    \item \textbf{Model Calibration:}
    \begin{enumerate}
        \item We first conduct a small-scale centipede experiment to obtain a preliminary dataset across a range of different parameters. 
        \item Using this dataset, we estimate the distribution of levels among the players in the population, assuming the data is generated by a Poisson DCH model.
    \end{enumerate}
    
    \item \textbf{Optimal Game Parameter Selection:}
    \begin{enumerate}
        \item Using the estimated Poisson parameter of the level distribution, we compute the predicted magnitude of the strategy method effect for each of the three classes of games, as a function of the single free parameter of that class.\footnote{Recall that the free parameter is the rate of payoff growth for the linear games, the rate at which the difference in payoffs between the two players increases for the constant games, and the ratio between the large and small payoffs for the exponential games.}
        \item For each class of games, we select two payoff parameters for the experiment: one in which the strategy method effect is predicted to be large, and one in which it is predicted to be small.
    \end{enumerate}
\end{enumerate}

\subsubsection*{Step 1: Model Calibration}

Unlike other optimally design approaches that calibrate models using data from previously published studies by other researchers based on results under different procedures and interfaces (e.g., \citealt{bland_optimizing_2023, 
lin_cognitive_2023}), we purposely obtain our pilot data under the same experimental conditions as the actual experiment. Calibrating the model with data collected in a virtually identical setting provides greater control and hence precision to estimating the theoretical parameters that are used to select the games, as these parameters may be sensitive to the specific experimental procedures.\footnote{For DCH, the model parameter is the population distribution of sophistication levels.}

Two pilot sessions with a total of 16 participants were conducted at the Taiwan Social Sciences Experimental Laboratory (TASSEL) at National Taiwan University using nearly identical procedures to the fully-designed experiment.\footnote{The pilot sessions were conducted during the COVID 
pandemic, when the TASSEL lab was one of the few physical locations in the world where in-person experiments were still feasible, albeit under strict social distancing restrictions.} 
The pilot studies produced data from four 6-stage linear centipede games ($c = 0.4$, $0.6$, $0.75$, and $0.9$), theoretically examined by \cite{lin_cognitive_2024}, using two different elicitation 
methods---the reduced strategy method and the direct response method---in a between-subjects design. 
That is, each participant took part in only one of the two sessions and played four different centipede games using only one of the two elicitation methods.  
%See the Online Supplemental Material for details of the pilot experiment.  

With the pilot data, we estimate the Poisson DCH model parameter $\tau$, which is
the mean of the distribution of levels of sophistication. The estimation is conducted 
using maximum likelihood estimation. Below, we briefly describe the construction of 
the log-likelihood function; a detailed description of the estimation procedure is 
provided in Appendix~\ref{appendix:structural_estimation}.

For any participant $i$, let $H_i$ denote the set of histories that 
participant $i$ encountered during the experiment.\footnote{In the reduced 
strategy method session, $H_i$ is the set of empty histories for 
the four linear centipede games. In contrast, in the direct response method 
session, $H_i$ is the set of decision nodes that participant 
$i$ encountered across all four games.} Assuming the prior distribution of 
levels follows a Poisson($\tau$), let $\sigma_i^k(a_i | h_i, \tau)$ 
denote the probability that a level-$k$ player $i$ chooses action $a_i$ at 
history $h_i \in H_i$. This is uniquely determined, as the DCH solution is unique. 
Furthermore, let $f(k | h_i, \tau)$ denote the posterior distribution of
levels at history $h_i$. 
The choice probability for action $a_i$ at history 
$h_i$ predicted by DCH is then given by the aggregation of choice 
probabilities across all levels, 
weighted by the posterior distribution $f(k | h_i, \tau)$:
\begin{align*}
\mathcal{D}(a_i | h_i, \tau) \equiv \sum_k  f(k | h_i, \tau)\sigma_i^k(a_i|h_i, \tau).
\end{align*}
Accordingly, the log-likelihood function for DCH can be constructed by summing over all participants $i$, all histories $h_i$, and all actions $a_i$:
\begin{align*}
    \ln \mathcal{L}(\tau) = \sum_i \sum_{h_i }\sum_{a_i} \mathbf{1}\{a_i, h_i\}
    \ln[\mathcal{D}(a_i | h_i, \tau)]
\end{align*}
where $\mathbf{1}\{a_i, h_i\}$ is an indicator function equal to 1 if participant $i$ 
chooses action $a_i$ at history $h_i$, and 0 otherwise.

This yields an estimate of \textbf{$\hat{\tau}=1.25$} (Obs. $=120$, S.E. $=0.157$, 
log-likelihood = $-57.05$)\footnote{To estimate the model, we cap 
the level at 10, as the proportion of levels above 
10 is negligible.} as the calibrated parameter for the DCH solution.
Note that an implicit assumption in our optimal design approach is 
that the distribution of levels does not depend on the elicitation method. 
To test this assumption, we separately estimate the parameter 
for each session and perform a likelihood ratio test. 
The null hypothesis is not rejected (Likelihood ratio test: $p$-value 
$= 0.26$), suggesting that 
the assumption is supported in our pilot data.

\subsubsection*{Step 2: Optimal Game Parameter Selection}

With the calibrated DCH, the next step is to 
select one game with a large predicted strategy method
effect and one with a small effect within each class of games. 
To quantify the magnitude of the strategy method effect, we
use the expected \emph{sup-norm} distance, i.e., the maximum difference,
between the CDFs of the terminal nodes
under the reduced strategy 
method and the direct response method, as predicted by the 
calibrated DCH. We adopt the sup-norm distance as our metric because
a theoretical result from \cite{lin_cognitive_2024} implies a specific 
relationship between the CDFs generated by different elicitation methods: 
the distribution of terminal nodes under the reduced strategy 
method \emph{first-order stochastically dominates} the distribution under 
the direct response method.

\begin{figure}[htbp!]
    \centering
    \includegraphics[width=\linewidth]{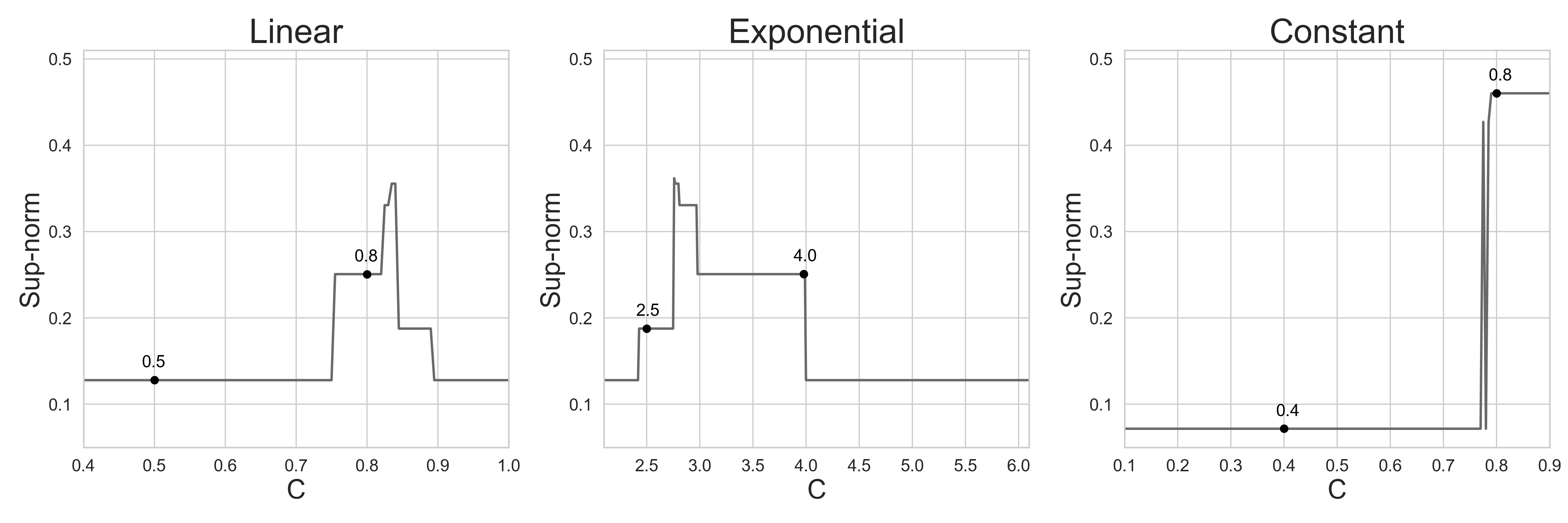}
    \caption{The Sup-norm Function $\mathcal{S}(c)$ for 
    Linear (left) Exponential (middle) and Constant (right) Centipede Games}
    \label{fig:optimal_supnorm}
\end{figure}

For each class of centipede games, we define $F_R(c)$ and $F_D(c)$ 
as the CDFs predicted by the calibrated DCH for the reduced
strategy method and the direct response method, respectively, in the centipede 
game with parameter $c$. The sup-norm distance is thus a function of 
the single free parameter, denoted by
\begin{align*}
\mathcal{S}(c) = \lVert F_{R}(c) - F_{D}(c) \rVert_{\infty}.
\end{align*}
Figure~\ref{fig:optimal_supnorm} plots the sup-norm function 
for each class of centipede games. As shown in the figure, for each class,
there exists a narrow range of \emph{sweet spots} for the parameter 
$c$ that yields a large strategy method effect.

Guided by the calibrated DCH, we select one payoff parameter that is a round number and close to the theoretical maximum sup-norm, and another that yields a negligible strategy method effect, for each class of games.  
For linear games, we select $c = 0.5$ (small effect) and $c = 0.8$ (large effect); for exponential games, $c = 2.5$ (small effect) and $c = 4$ (large effect); and for constant games, $c = 0.4$ (small effect) and $c = 0.8$ (large effect).\footnote{The payoffs used in the experiment are rescaled to avoid decimal values and ensure comparability across all six centipede games.  
See Figure~\ref{fig:six_cg} in the appendix for the actual game trees implemented in the experiment.}  
The predicted sup-norm distances for the selected games are summarized in Table~\ref{tab:RN_E_prediction}. 

\begin{table}[htbp!]
\centering
\caption{The Predicted Sup-norm of the Selected Centipede Games}
\label{tab:RN_E_prediction}
\renewcommand{\arraystretch}{1.35}
\begin{threeparttable}
\begin{tabular}{ccccccccc}
\hline
 & \multicolumn{2}{c}{Linear} &  & \multicolumn{2}{c}{Exponential} &  & \multicolumn{2}{c}{Constant} \\ \cline{2-3} \cline{5-6} \cline{8-9} 
 & $c=0.5$ & $c=0.8$ &  & $c=2.5$ & $c=4$ &  & $c=0.4$ & $c=0.8$ \\
 & (small) & (large) &  & (small) & (large) &  & (small) & (large) \\ \hline
\multirow{2}{*}{Sup-norm} & \multirow{2}{*}{0.128} & \multirow{2}{*}{\textbf{0.251$^\dagger$}} & \multirow{2}{*}{} & \multirow{2}{*}{0.188} & \multirow{2}{*}{\textbf{0.251$^\dagger$}} & \multirow{2}{*}{} & \multirow{2}{*}{0.072} & \multirow{2}{*}{\textbf{0.460$^\dagger$}} \\
 &  &  &  &  &  &  &  &  \\ \hline
\end{tabular}
\begin{tablenotes}
\footnotesize
\item $\dagger$ indicates that the game is expected to yield statistically significant 
strategy method effects under the Kolmogorov–Smirnov test 
at the 1\% significance level, given a sample of 192 participants.
\end{tablenotes}
\end{threeparttable}
\end{table}

In addition, given our sample size of 192 participants, 
the predicted sup-norm distances for all three large-effect 
games are expected to yield statistically significant 
strategy method effects under the Kolmogorov–Smirnov test 
at the 1\% significance level, whereas the three small-effect games
are not. Since the KS test is relatively low-powered, this contrast
between large-effect and small-effect games further justifies 
our selection, as these games are expected to generate statistically
distinguishable differences in the magnitude of the strategy method effect.

Using the optimal design approach, we identify six centipede games: three that are expected to generate large strategy method effects and three that are not. In the following, we describe the implementation of this optimally designed experiment.

\subsection{Design Implementation and Experimental Procedures}
\label{subsec:experimental_design_implementation}

The laboratory experiment consists of three treatments, each implementing the six optimally selected centipede games in one of three elicitation methods: the direct response (DR) method, the full strategy (FS) method and the reduced strategy (RS) method.  
To avoid potential confounding effects from changes in visual display, we maintain a consistent game tree format across all three treatments, varying only the ways for participants to submit their decisions, exactly as shown in Figure~\ref{fig:exp_screenshots}.  

We adopt a \emph{within-subject design} in our laboratory experiment 
in order to stress test the strategy method effect and maximize statistical power. Each participant therefore plays all six optimally selected centipede games under all three elicitation methods. This design enables us to observe how the same participant behaves when the choices are elicited under different elicitation methods.

To mitigate potential spillover effects across treatments, 
participants do not receive any feedback between games or treatments. 
However, under the direct response method, participants inevitably learn 
the outcome of a game, as they make decisions only when the game reaches 
their decision nodes. For this reason, the direct response method is always 
conducted last. 

Additionally, to counterbalance potential order effects between 
the full and the reduced strategy method treatments, 
we alternate the order of these two treatments (RF Order and FR Order) 
across sessions. In the RF Order, participants first play the six games 
under the reduced strategy method, followed by the full strategy method, 
and finally the direct response method. In the FR Order, the sequence of 
the reduced and full strategy methods is reversed.\footnote{Robustness analysis in 
Appendix~\ref{appendix:additional_results} indicates no significant 
differences between the two orders.}

We conducted 16 sessions
at the Social Science Experimental Laboratory (SSEL) at the California Institute of Technology and the Experimental Social Science Laboratory (ESSL) at the University of California, Irvine.  
Eight sessions were held at each laboratory.  
The experiment was programmed with the oTree software \citep{chen2016otree}.  
Each session included 12 participants, for a total of 192 participants in our experimental dataset.  
Half of the sessions were conducted using the RF Order, while the other half under the FR Orders.

At the beginning of each session, participants were randomly assigned to the role of either a first mover or a second mover,\footnote{In our instructions, first movers were referred to as ``red participants'' and second movers as ``blue participants.''} 
and their roles remained unchanged throughout the session.\footnote{Thus, each session consisted of six red participants and six blue participants.}  
In each treatment of every session, all six optimally selected centipede games were played once, and the sequence of games was randomized.  
Additionally, participants were matched with a \emph{different} opponent for each game within a treatment to prevent any potential reputation-building effects.\footnote{To implement this matching protocol, each participant was assigned a unique ID at the beginning of each session.  
Specifically, the six red and six blue participants were randomly assigned IDs ranging from ``Red 1'' to ``Red 6'' and ``Blue 1'' to ``Blue 6,'' respectively.  
In each game of each treatment, participants were informed of their opponent’s ID and reminded that they had not previously been matched with this opponent within the current treatment.}  
Furthermore, participants did not receive any feedback after each game in the full strategy method and reduced strategy method treatments, whereas they automatically learned the outcome of each game in the 
direct response method treatment.  
A summary of all outcomes was provided to participants at the end of the experiment.  
To prevent participants from making inferences about their opponents' actions based on their decision times, a 10-second delay was added at the beginning of each game.  
This adjustment resulted in each session lasting approximately 75 to 90 minutes.

The participants were paid based on the payoffs (in ``points'') they 
received throughout the experiment. Three games (one from each treatment) were randomly selected for payment. Including a show-up fee of USD \$10, subject earnings averaged USD \$26.36.\footnote{The exchange rate was 1 point to USD \$0.02. Final payments were rounded up to the nearest dollar.} See Appendix \ref{appendix:instructions} for the experimental instructions. 
%and the Online Supplemental Material for the complete set of screenshots.

\section{Experimental Results}
\label{sec:exp_results}

\subsection{Data Description and Aggregate-Level Results}

Before presenting the main results, we first summarize the number of 
observations at various levels in Table~\ref{tab:aggregate_level_tests}. 
Recall that in the experiment, all 192 participants 
played six centipede games under 3 elicitation methods, yielding 1,152 observations of strategies 
(players $\times$ games) for each method.  
Under the direct response method, only 63\% (730 out of 1,152) of the strategies elicited are uncensored reduced strategies.\footnote{The term \textit{uncensored reduced strategy} for a player in the direct response treatment means that the sequence of actions in the game unambiguously implies a unique reduced strategy for that player.  
For example, if a game ends at the first node, we can infer the first mover's reduced strategy unambiguously (Stop at Stage 1), but we cannot infer the second mover's reduced strategy.  
If the game ends at the second node, we can infer the second mover's reduced strategy but not the first mover's strategy.  
Only if the game continues all the way to the last node can we infer both players' reduced strategy.}  
Furthermore, under the direct response 
method, we obtain only 2,160 pass/take decision-node choices---28\% fewer than that inferred under the reduced strategy method (2,160 vs. 2,763) and 60\% fewer than under the full strategy method (2,160 vs. 3,456).

\begin{table}[htbp!]
\centering
\caption{Number of Observations at Various Levels and Analysis of Terminal Nodes}
\label{tab:aggregate_level_tests}
\renewcommand{\arraystretch}{1.25}
\begin{adjustbox}{width=0.9\columnwidth,center}
\begin{threeparttable}
\begin{tabular}{rcccccc}
\hline
\multicolumn{1}{c}{} &  & \begin{tabular}[c]{@{}c@{}}Direct Response\\ Method\end{tabular} &  & \begin{tabular}[c]{@{}c@{}}Reduced Strategy\\ Method\end{tabular} &  & \begin{tabular}[c]{@{}c@{}}Full Strategy\\ Method\end{tabular} \\ \hline
\multicolumn{7}{l}{\# of Observations} \\
Players $\times$ Games &  & 1,152 &  & 1,152 &  & 1,152 \\
Uncensored Reduced Strategies$^1$ &  & 730 &  & 1,152 &  & 1,152 \\
(Inferred) Decision-Node Choices$^2$ &  & 2,160 &  & 2,763 &  & 3,456 \\
Terminal Nodes$^3$ &  & 576 &  & 576 &  & 576 \\ \hline
\multicolumn{7}{l}{Matching Pair-Specific Terminal Nodes}  \\
Mean &  & 3.859 &  & 4.002 &  & 3.861 \\
Standard Deviation &  & (2.099) &  & (2.051) &  & (2.050) \\ \hline
\multicolumn{7}{l}{Three-Way Friedman Test} \\
$p$-value &  &  &  & 0.005 &  &  \\
\multicolumn{7}{l}{Pairwise Signed-Rank Tests} \\
vs. Full Strategy Method $p$-value &  & 1.000 &  & 0.025 &  & --- \\
\;\;\;\;vs. Direct Response Method $p$-value &  & --- &  & 0.044 &  & 1.000 \\ \hline
\end{tabular}
\begin{tablenotes}
\footnotesize
\item[1.] The ``Uncensored Reduced Strategies'' row represents the number of strategies that 
allow us to recover the reduced strategies.
\item[2.] The ``(Inferred) Decision-Node Choices'' row represents the number 
of pass/take decisions, either directly observed under the direct response method
or inferred from strategies under the reduced or full strategy method.
\item[3.] The ``terminal node'' refers to the earliest stage at which a player chooses 
to take. If both players pass at all six stages, the terminal node is coded as 7.
\item[4.] $p$-values are adjusted using the Bonferroni correction. 
\end{tablenotes}
\end{threeparttable}
\end{adjustbox}
\end{table}

These findings highlight a key limitation of the direct response method: 
\emph{incompleteness}, especially in our within-subject experiment, where it yields 
significantly less data from the \emph{same} group of players.
To enable a fair comparison across all three
elicitation methods, we therefore focus in this section on the analysis of 
terminal nodes---the only outcome variable for which all three methods
yield the same number of observations: 576 (96 pairs of players $\times$ 6 games).

Nonetheless, unlike in the direct response method, where terminal nodes 
are \emph{observed}, those under the reduced and full strategy methods 
must be \emph{inferred} from strategies. Moreover, the distribution 
of terminal nodes is sensitive to how players (and their strategies) are paired. 
To address this, we fully leverage our within-subject, no-between-game-feedback 
design to 
compute terminal nodes for the reduced and full strategy methods by 
assuming that \emph{players are paired exactly as they were in the direct response method}.
For example, if Ann and Bob are paired in game $X$ under the direct response method, 
we assume that Ann and Bob are 
paired in game $X$ under the reduced and full strategy methods when obtaining matching pair-specific terminal nodes.
This construction enables a clean analysis of strategy method effects by
eliminating confounds arising from different pairing realizations across elicitation methods.
 
Table~\ref{tab:aggregate_level_tests} summarizes the matching pair-specific terminal node data under our 
construction. As a first cut at analyzing strategy method effects, we find
that terminal nodes vary significantly across elicitation methods, as indicated 
by the Friedman test, which tests the null hypothesis that 
the distributions of terminal nodes are identical across treatments 
(Friedman test: $Q = 10.628$, $p$-value $= 0.005$).
This result implies that behavior under at least one elicitation method differs significantly from the others.  
In fact, the average terminal node is 4.002 under the reduced strategy method, but 3.859 under the direct response method and 3.861 under the full strategy method.  

To further identify which pairs of elicitation methods differ significantly, we
conduct Wilcoxon signed-rank tests \citep{wilcoxon1945} to compare
each pair of elicitation methods. As shown in Table~\ref{tab:aggregate_level_tests},
games played using the reduced strategy method terminate significantly later than both the direct response method (signed-rank test: $p$-value $= 0.044$) and the full
strategy method (signed-rank test: $p$-value $= 0.025$). In contrast, there is no significant difference between the
direct response and full strategy methods. 
These findings are consistent with the qualitative predictions of DCH.

\begin{result}
Games played using the reduced strategy method terminate significantly later than both the direct response method (signed-rank test: $p$-value $= 0.044$) and the full
strategy method (signed-rank test: $p$-value $= 0.025$). There is no significant difference between the
direct response and full strategy methods.
\end{result}

The calibrated DCH makes clear predictions about the comparative statics 
of strategy method effects across different game parameters. To test these
predictions, we compare terminal nodes under each elicitation method across games
in the next section.

\subsection{Game-Level Results}

\subsubsection*{Overview of Strategy Method Effects Across Different Games}

\begin{figure}[htbp!]
    \centering
    \includegraphics[width=0.8\linewidth]{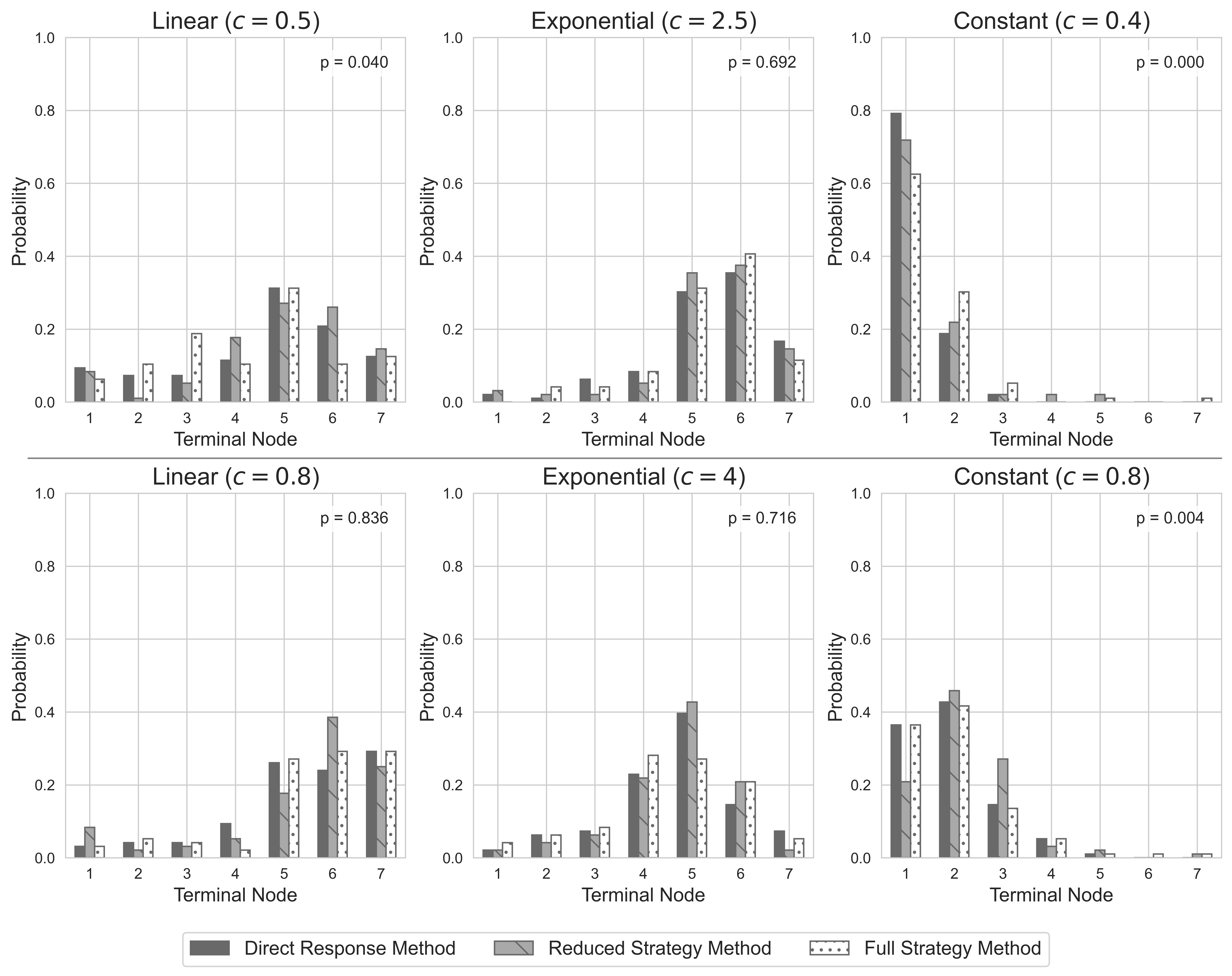}
    \caption{Distribution of Terminal Nodes by Elicitation Method and Game.
    Each panel reports the $p$-value from the Friedman test for the corresponding class of games.
    See Table~\ref{tab:pair_level_terminal_node_pdf} for the corresponding table.}
    \label{fig:pair_level_terminal_node_pdf}
\end{figure}

To provide an overview of strategy method effects across different games, Figure~\ref{fig:pair_level_terminal_node_pdf}, its corresponding Table~\ref{tab:pair_level_terminal_node_pdf}, and the cumulative distribution in Figure~\ref{fig:pair_level_empirical_cdf} illustrate the distributions of (inferred) terminal nodes under the three elicitation methods for all six games. The results reveal substantial heterogeneity in strategy method effects across different classes of games: the distributions of terminal nodes differ significantly across methods in the Small Linear Game and in both the Small and Large Constant Games (Friedman test: $p$-value $= 0.040$ for Small Linear; $p$-value $< 0.001$ for Small Constant; $p$-value $= 0.004$ for Large Constant), but not in the remaining three game classes.

\begin{figure}[htbp!]
\centering
\includegraphics[width=.8\linewidth]{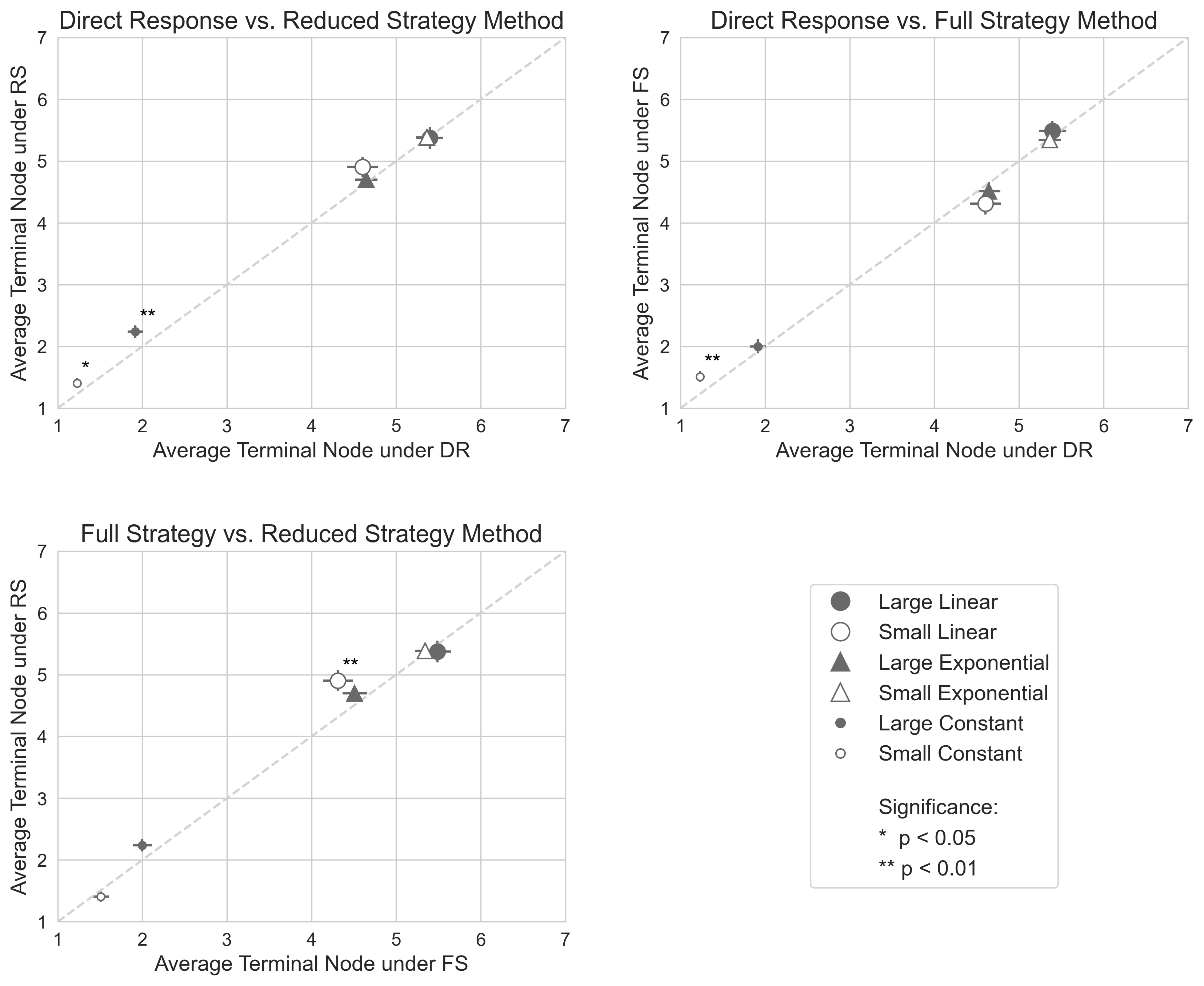}
\caption{Pairwise Comparison of Terminal Nodes Across Elicitation Methods. Each point 
represents the average terminal node for a given game, with standard error 
bars overlaid. Stars indicate statistical significance based on 
two-sided signed-rank tests with Bonferroni correction:  $^{*} \; p < 0.05$, $^{**}\; p < 0.01$.
See Table~\ref{tab:end_node_data_signed_rank} for the corresponding table.}
\label{fig:game_level_pairwise_comparison}
\end{figure}

Building on these observations, we further decompose the observed strategy 
method effects by conducting pairwise comparisons among the three elicitation
methods within each class of games. To summarize this analysis, 
Figure~\ref{fig:game_level_pairwise_comparison} plots the average terminal nodes under each method. 
Each panel presents a specific pairwise comparison, and each point represents 
the average terminal node for a game under the two corresponding methods. This figure
provides a clear visualization of strategy method effects, as any deviation 
from the 45-degree line indicates the presence of such effects.

We first compare the reduced strategy method with the direct response method (top left panel of Figure~\ref{fig:game_level_pairwise_comparison}) and with the full strategy method (bottom left panel).  
For both comparisons, the DCH solution makes clear predictions: later termination should occur under the reduced strategy method.  
Accordingly, DCH predicts that all points in these two panels should lie \emph{above} the 45-degree line, as the average terminal nodes under the reduced strategy method are plotted on the y-axis, although points for the three small games may not deviate significantly from the line.  

From the top left panel, we observe that all points lie on or above the 45-degree line.  
This suggests that the reduced strategy method leads to later termination compared to the direct response method, consistent with the qualitative predictions of DCH.  
However, both exponential games and both linear games do not deviate significantly from the 45-degree line, whereas 
the two constant games lie significantly above it
(signed-rank test: $p$-value $=0.005$ for Large Constant; $p$-value $=0.022$ for Small Constant).

Furthermore, a similar pattern emerges when comparing the reduced strategy method with the full strategy method.  
In the bottom left panel, we again observe that all points lie on or above the 45-degree line.  
This pattern again aligns with the qualitative predictions of DCH.  
Nevertheless, the Small Linear Game is the only one significantly above the line (signed-rank test: $p$-value $= 0.005$), indicating later termination under the reduced strategy method compared to the full strategy method.  

Lastly, when comparing the full strategy method to the direct response method, 
DCH makes a null prediction---there should be no strategy method effect in any
centipede game. In other words, in the top right panel, DCH predicts that all points
should lie on the 45-degree line. The figure shows that five of the six centipede games
indeed do not deviate from this line, consistent with the DCH prediction. The only
exception is the Small Constant Game, where there is no behavioral difference 
between the full and reduced strategy methods (signed-rank test: $p$-value $= 0.374$), 
but there is significant later termination under the full strategy method 
compared to the direct response method (signed-rank test: $p$-value $= 0.001$).

\begin{result} Overall, the distributions of terminal nodes differ significantly 
across elicitation methods in three centipede games (Small Linear, Small Constant, 
and Large Constant). Moreover, pairwise comparisons of the elicitation methods 
reveal the following patterns.
\begin{itemize}
    \item \textbf{Direct Response Method vs. Reduced Strategy Method}: There is significant later termination 
    under the reduced strategy method in the Small and Large Constant Games, with no significant 
    differences in the other four games.
    \item \textbf{Full Strategy Method vs. Reduced Strategy Method}: Significant later termination under the reduced strategy method is observed only in the Small Linear Game, with no significant differences 
    in the other games.
    \item \textbf{Direct Response Method vs. Full Strategy Method}: No significant differences are observed, 
    except in the Small Constant Game---the only case that deviates from the qualitative predictions of DCH.
\end{itemize}
\end{result}

Despite the high consistency between the observed strategy method effects and the 
qualitative predictions of DCH, the relative magnitudes of these effects do not align 
with the calibrated DCH. The calibrated model predicts that strategy method effects should 
appear in all three Large Games but not in the Small Games, yet significant effects are 
observed in two Small Games. These discrepancies are detectable only because we 
first calibrate the model and obtain a \emph{quantitative} benchmark. 
In what follows, we address the puzzle of the relative magnitudes of strategy method effects.

\subsubsection*{Relative Magnitudes of Strategy Method Effects}

The calibrated DCH predicts significantly later termination under the reduced strategy method compared to the other two elicitation methods in all three Large Games. To evaluate this prediction, we group the Large Linear, Large Exponential, and Large Constant Games as those with large effects, and classify the remaining games as having small effects.
Due to the construction of our (inferred) terminal node data, for any terminal node (and pair of players) observed under the direct response method, we also observe the terminal nodes generated by the same pair under the other two elicitation methods.
Using this matched data, we compute the \emph{difference} in terminal nodes for each pair of players across any two elicitation methods. We denote the average difference in the Large group of games by $\Delta_L$, and in the Small group by $\Delta_S$.

According to the predictions of the calibrated DCH, when comparing
the reduced strategy method with the direct response method 
($\mbox{RS} - \mbox{DR}$), we should observe $\Delta_L > 0$ and 
$\Delta_S \approx 0$. This implies that the difference
$\Delta_L - \Delta_S$, which measures the relative magnitude of the effect, 
should be positive. Similarly, because DCH makes the same prediction for the 
comparison between the reduced and full strategy methods ($\mbox{RS} - \mbox{FS}$), it 
predicts $\Delta_L > 0$, $\Delta_S \approx 0$, and hence $\Delta_L - \Delta_S > 0$. In 
contrast, for the comparison between the full strategy method and the direct response 
method ($\mbox{FS} - \mbox{DR}$), DCH predicts $\Delta_L \approx 0$, $\Delta_S \approx 
0$, and therefore $\Delta_L - \Delta_S \approx 0$.

\begin{table}[htbp!]
\centering
\caption{Relative Magnitudes of Strategy Method Effects (Grouped Under DCH)}
\label{tab:diff_method_effect_HvsL}
\renewcommand{\arraystretch}{1.35}
\begin{threeparttable}
\begin{tabular}{cccccc}
\hline
 &  & \multicolumn{2}{c}{$\Delta$ in Terminal Nodes} &  &  \\ \cline{3-4}
 &  & $\Delta_L$ & $\Delta_S$ &  & $\Delta_L - \Delta_S$ \\ \hline
\multirow{2}{*}{$\mbox{RS} - \mbox{DR}$} & \multirow{2}{*}{} & 0.118 & 0.167 &  & -0.049\phantom{-} \\
 &  & $p=0.141$ & $p=0.044$ &  & $p=0.997$ \\
& & & & & \\
\multirow{2}{*}{$\mbox{RS} - \mbox{FS}$} & \multirow{2}{*}{} & 0.104 & 0.177 &  & -0.073\phantom{-} \\
 &  & $p=0.105$ & $p=0.031$ &  & $p=0.700$ \\
& & & & & \\
\multirow{2}{*}{$\mbox{FS} - \mbox{DR}$} & \multirow{2}{*}{} & 0.014 & -0.010\phantom{-} &  & 0.024 \\
 &  & $p=0.862$ & $p=0.948$ &  & $p=0.724$ \\ \hline
\end{tabular}
\begin{tablenotes}
\footnotesize
\item Statistical inferences are based on the signed-rank test, with $p$-values reported below.
\end{tablenotes}
\end{threeparttable}
\end{table}

Table~\ref{tab:diff_method_effect_HvsL} summarizes the relative magnitudes 
across all pairs of elicitation methods. As shown, the empirical patterns 
do not fully align with the predictions of the calibrated DCH. The null 
difference between the full strategy method and the direct response method is 
consistent with the model. However, the comparisons involving the reduced strategy 
method deviate significantly: for both $\mbox{RS} - \mbox{DR}$ and $\mbox{RS} - 
\mbox{FS}$, we observe $\Delta_L \approx 0$ (signed-rank test: $p$-value $=0.141$
for $\mbox{RS} - \mbox{DR}$; $p$-value $=0.105$
for $\mbox{RS} - \mbox{FS}$) and $\Delta_S > 0$ (signed-rank test: 
$p$-value $=0.044$ for $\mbox{RS} - \mbox{DR}$; $p$-value $=0.031$
for $\mbox{RS} - \mbox{FS}$), contrary to the 
prediction of the calibrated DCH. In addition, the 
negative values of $\Delta_L - \Delta_S$ for both 
comparisons---though not statistically significant---further 
diverge from the predictions of the calibrated DCH.

To conclude, while the observed qualitative patterns in 
the data are mostly consistent with DCH, the relative magnitudes of 
the strategy method effects do not align with the calibrated 
DCH---specifically, the effects in the optimally selected large games are not significantly larger than those in the quasi-placebo small games.
This finding suggests that the observed strategy method 
effect is not solely driven by DCH but is also influenced by 
other behavioral forces that contribute to similar effects. Therefore, 
in the next section, we explore two alternative behavioral models 
to assess whether they can better account for the variation in relative magnitudes of elicitation effects across different games.

\section{Alternative Behavioral Models}
\label{section:structural_estimation}

From the analysis in the previous section, 
we find that although the qualitative patterns 
are generally consistent with DCH, the relative magnitudes of the strategy method 
effect do not fully align with the 
predictions of the calibrated DCH. We conjecture that these discrepancies are at least partly driven by stochastic choice, or quantal response effects, which can also contribute to strategy method effects. These models of stochastic choice relax the perfect best response assumption of Nash equilibrium and DCH with quantal responses, whereby individuals choose higher expected payoff actions more frequently than lower expected payoff actions, but do not always choose the highest payoff action.

To decompose the strategy method effects of quantal response behavior and DCH, we compare DCH with two specific behavioral models that incorporate stochastic choice: the agent quantal response equilibrium (AQRE) of \citet{mckelvey_quantal_1998} 
and the quantal dynamic cognitive hierarchy (QDCH) solution, a hybrid model of DCH and logit quantal responses
introduced by \citet{lin_cognitive_2023} using structural estimation. We then revisit our optimal design approach (hypothetically) under these alternative models to examine how incorporating quantal responses could further refine our methodology.

\subsection{Models with Quantal Responses}
\subsubsection*{Agent Quantal Response Equilibrium (AQRE)}

AQRE \citep{mckelvey_quantal_1998} predicts strategy method effects in centipede games in a somewhat different manner than the pure strategy reduction effect of DCH.  
Unlike the DCH solution, AQRE is an equilibrium concept in which players are assumed to make logit quantal responses (rather than best 
responses) to the equilibrium strategies of the other players, at all histories.
Specifically, for any player $i \in N$, any history $h$, and any action $a$, the probability that player $i$ chooses action $a$ in logit specification of AQRE is given by: 
$$\sigma_i(a|h) = \frac{e^{\lambda \bar{u}_a}}{\sum_{a'\in A(h)} e^{\lambda \bar{u}_{a'}}} 
$$
where $A(h)$ is the set of available actions at history $h$, $\bar{u}_{a'}$ is the equilibrium (expected) continuation value of choosing any $a'\in A(h)$, and $\lambda \in [0, \infty)$ is the precision parameter.  
When $\lambda = 0$, players are completely insensitive to payoffs and uniformly randomize, behaving similarly to level-$0$ players.  
As $\lambda$ increases, players' choices are closer to optimal. In the limit as $\lambda \to \infty$, players become fully rational and make best responses that correspond to a sequential equilibrium of the game.

There are two distinct kinds of strategy method effects of AQRE in our 6-stage centipede games. 

The first kind of effect arises because the different elicitation methods imply different systems of equilibrium conditions that characterize the AQRE.  
In the direct response method, AQRE is a profile of behavioral strategies (conditional take probabilities).  
Since each player has three information sets, each with a binary choice this implies that the (logit) AQRE is the solution to a system of 6 nonlinear equations.  
In contrast, under the full strategy method\footnote{Since the centipede game under both the full and reduced strategy methods is a static game, AQRE reduces to the QRE of the corresponding strategic form game \citet{mckelvey_quantal_1995}.}  
each player has 8 possible strategies so AQRE is characterized by a system of 16 nonlinear equations---specifically, the choice probabilities of the eight full strategies for each player.  
Finally, under the \textit{reduced} strategy method, each player has only 4 strategies, so AQRE is computed as the solution to a smaller system of 8 nonlinear equations with only eight choice probabilities of the four reduced strategies for each player.  
This difference affects the implied conditional take probabilities at each stage, leading to different distributions over terminal nodes.  
For instance, when $\lambda=0$, the probability the game is terminated at the first node under both the full strategy method and the direct response method is $1/2$, while under the reduced strategy method, it is $1/4$.  
Thus for very low values of $\lambda=0$ this first kind of elicitation effect is similar\footnote{It is similar, but not identical to the DCH strategy reduction effect when $\lambda>0$, since logit response do not result in uniform randomization.  
Unlike the strategy reduction effect in DCH which predicts no elicitation effect between the full strategy method and the direct response method, AQRE differentiates between the two for all $\lambda>0$, }  
to the strategy reduction effect in DCH.

The second kind of strategy method effect in AQRE is completely different from the strategy reduction effect, and arises because the direct response method elicits action choices sequentially, so the AQRE analysis for the direct response method is carried out separately for each history, exploiting the timing structure inherent in the extensive form. 
Specifically, in AQRE players quantal respond to the \textit{expected continuation values conditional on reaching an information set} rather than \textit{ex ante expected payoffs}, and the latter are deflated by the probability of reaching the information set.  
For example, in our centipede games, the probability of reaching the last node of the game is very low.  
As a result, the last mover may find the \textit{ex ante} expected payoff difference between choosing strategies that take or pass at that node to be very small, and hence, the probability of them choosing an ex ante strategy that (suboptimally) passes at the last node can be quite large.  
In contrast, the expected payoff difference between taking and passing \textit{conditional on reaching the last node} is large, so the AQRE probability of passing at the last node is much lower.  
This feeds back into the expected continuation values of taking and passing at earlier nodes, which generally leads to earlier termination if the game is played sequentially than if it is played simultaneously according to either of the two strategy methods.\footnote{For a more detailed discussion of AQRE strategy method effects in centipede games see Chapter 3 of \citet{goeree2016quantal}.}

For these reasons, AQRE makes different predictions than DCH regarding the magnitude of strategy method effects, 
where the differences depend on the logit response parameter and the payoff function.

\subsubsection*{Quantal Dynamic Cognitive Hierarchy (QDCH) Solution}

The QDCH proposed by \cite{lin_cognitive_2023} is a stochastic extension of DCH that replaces best responses with logit quantal responses at all histories, while retaining the non-equilibrium cognitive hierarchy of beliefs.  
Specifically, QDCH is a two-parameter model in which $\tau > 0$ represents the mean of the Poisson prior distribution of levels, and $\lambda \in [0, \infty)$ is the precision parameter for logit quantal responses.\footnote{We assume that all levels $k > 0$ share the same precision parameter $\lambda$.}

It is worth noting that QDCH includes the DCH solution as a boundary case, as QDCH converges to DCH when 
$\lambda \rightarrow \infty$.  
However, AQRE is not nested within QDCH, even when $\tau \rightarrow \infty$ (i.e., when all players have infinitely high levels of sophistication).  
This distinction arises because DCH and QDCH are non-equilibrium models, where players believe others are strictly less sophisticated than they are, whereas AQRE is an equilibrium model defined as a fixed point.

Although QDCH and AQRE are non-nested models, they share the feature 
that the solution can differ even across games with the same (non-reduced) 
normal form due to the nature of logit quantal responses. In other words, 
logit quantal responses 
can change the original DCH strategy method effect, causing QDCH to yield 
different predictions across all three elicitation methods. The expected magnitude 
of these strategy method effects again depends on both the model
parameters and the payoff structure.

\subsection{Structural Estimation and Model Comparison}

To systematically compare DCH with AQRE and QDCH, we estimate 
these models using maximum likelihood estimation. The construction 
of their likelihood functions is provided in Appendix \ref{appendix:structural_estimation}. 
The estimation results and model comparisons are presented in 
Table~\ref{tab:simplified_model_estimation}.

\begin{table}[htbp!]
\centering
\caption{Estimation Results and Model Comparisons Across Treatments and Pooled Data}
\label{tab:simplified_model_estimation}
\renewcommand{\arraystretch}{1.4}
\begin{adjustbox}{width=1.2\columnwidth,center}
\begin{threeparttable}
\begin{tabular}{cccccccccccccccc}
\hline
 & \multicolumn{3}{c}{Reduced Strategy Method} &  & \multicolumn{3}{c}{Full Strategy Method} &  & \multicolumn{3}{c}{Direct Response Method} &  & \multicolumn{3}{c}{Pooled Data} \\ \cline{2-4} \cline{6-8} \cline{10-12} \cline{14-16} 
 & QDCH & DCH & AQRE &  & QDCH & DCH & AQRE &  & QDCH & DCH & AQRE &  & QDCH & DCH & AQRE \\ \hline
$\tau$ & 3.637 & 1.250 & --- &  & 7.318 & 1.121 & --- &  & 3.528 & 1.236 & --- &  & 4.033 & 1.250 & --- \\
S.E. & (0.243) & (0.036) & --- &  & (1.665) & (0.027) & --- &  & (0.667) & (0.060) & --- &  & (0.350) & (0.013) & --- \\
$\lambda$ & 0.017 & --- & 0.015 &  & 0.021 & --- & 0.019 &  & 0.018 & --- & 0.013 &  & 0.019 & --- & 0.015 \\
S.E. & (0.001) & --- & (0.001) &  & (0.002) & --- & (0.002) &  & (0.007) & --- & (0.001) &  & (0.002) & --- & (0.001) \\ \hline
LL & \textbf{-1166\phantom{-}} & -1311\phantom{-} & -1194\phantom{-} &  & \textbf{-1786\phantom{-}} & -1939\phantom{-} & -1789\phantom{-} &  & \textbf{-920\phantom{-}} & -1009\phantom{-} & -975\phantom{-} &  & \textbf{-3927\phantom{-}} & -4270\phantom{-} & -3986\phantom{-} \\
$\Delta\%$ &  & 12.4\% & 2.4\% &  &  & 8.6\% & 0.2\% &  &  & 9.7\% & 6.0\% &  &  & 8.7\% & 1.5\% \\ \hline
LRT &  & 290.9 & --- &  &  & 306.1 & --- &  &  & 178.1 & --- &  &  & 686.1 & --- \\
Vuong &  & --- & 5.688 &  &  & --- & 1.271 &  &  & --- & 5.538 &  &  & --- & 4.347 \\
$p$-value &  & $<0.001$\phantom{$<$} & $<0.001$\phantom{$<$} &  &  & $<0.001$\phantom{$<$} & 0.204 &  &  & $<0.001$\phantom{$<$} & $<0.001$\phantom{$<$} &  &  & $<0.001$\phantom{$<$} & $<0.001$\phantom{$<$} \\ \hline
\end{tabular}
\begin{tablenotes}
\small
\item[1.] The estimations for the reduced and full strategy 
method treatments are based on 1,152 choices of reduced strategies and 
complete strategies, respectively. The estimation for the direct response 
method treatment is based on 2,160 (binary) decision-node choices.
\item[2.] $\Delta\%$ indicates the percentage improvement in likelihood relative to QDCH.
\item[3.] For the Vuong test of the direct response method treatment,
we account for the interdependence of decision-node choices by 
conducting the test on the 576 observations of terminal nodes.
\end{tablenotes}
\end{threeparttable}
\end{adjustbox}
\end{table}

Intuitively speaking, QDCH relaxes two key requirements of the
standard subgame perfect equilibrium: (1) perfect best response 
and (2) mutual consistency of beliefs. In comparison, DCH retains theif former
but relaxes the latter, replacing it with  dynamic cognitive hierarchy beliefs, whereas AQRE maintains 
mutual consistency and replaces best response 
with quantal responses. Therefore, the comparison between QDCH and DCH identifies the effect of stochastic choice, while the difference between QDCH and AQRE captures the effect of relaxing mutual consistency.

From Table~\ref{tab:simplified_model_estimation}, we first observe that incorporating quantal responses into the DCH solution significantly improves model fit: QDCH fits the data better than DCH in all elicitation methods and in the pooled data.  
Since QDCH and DCH are nested, we conduct likelihood ratio tests (LRTs) and find that QDCH fits the data significantly better in the pooled data (LRT: $p$-value $< 0.001$).  
In fact, LRTs yield $p$-values $< 0.001$ in each elicitation method as well.  
These results indicate that stochastic choice plays a significant role in the behavior of subjects in centipede games.

Given this observation, a natural next question is whether quantal response is the driving mechanism behind the observed violation of invariance under strategic equivalence.  
To investigate this, we compare QDCH and AQRE using the Vuong test \citep{vuong_likelihood_1989}, as they are non-nested models, and 
find that QDCH fits the pooled data significantly better than AQRE ($V = 4.347$, $p$-value $< 0.001$).  
Combined with the earlier finding that QDCH fits the data significantly better than DCH, we conclude that both the dynamic cognitive hierarchy model of beliefs and quantal response jointly contribute to the observed strategy method effect.

While both DCH and quantal responses partially account for
the observed strategy method effect in the pooled data, we find that the
relative strength of these two effects varies across elicitation methods.
To quantify this, we compute the improvement in likelihood scores of DCH and
AQRE relative to QDCH. As shown in Table~\ref{tab:simplified_model_estimation}, 
incorporating quantal responses into DCH results in a substantial improvement 
in log-likelihood, ranging from 8.6\% to 12.4\% across treatments.

Compared to the effect of quantal response, the DCH mechanism exhibits greater variation across treatments.  
In both the reduced strategy method and direct response method treatments, the log-likelihood of QDCH is significantly higher than that of AQRE---by 2.4\% ($V = 5.688$, $p$-value $< 0.001$) and 6.0\% ($V = 5.540$, $p$-value $< 0.001$), respectively.  
However, in the full strategy method treatment, the likelihood scores of AQRE and QDCH are nearly identical ($V = 1.271$, $p$-value $= 0.204$), suggesting that the effect of quantal response dominates the DCH mechanism under this elicitation method.\footnote{We report estimates for
each class of centipede games separately in Tables~\ref{tab:simplified_model_estimation_Linear_appendix}, \ref{tab:simplified_model_estimation_Exponential_appendix} and \ref{tab:simplified_model_estimation_Constant_appendix}
and find that Result 3 is robust across all classes of centipede games.}

\begin{result}
    QDCH fits the data better than both DCH and AQRE across all treatments, highlighting 
    the significance of quantal response effects, while the relative strength of 
    quantal response effects varies across treatments.
\end{result}

It is worth noting that in the pooled data, the estimated 
value of $\tau$, the average level of sophistication in the population, 
under DCH is 1.25 (S.E. $=0.013$), which coincides with the estimate of $\tau$ 
from the pilot data. While our estimate falls within the range of \emph{regular}
$\tau$ values (between 1 and 2) estimated by \citet{camerer_cognitive_2004}, 
we hypothesize that the close estimate between the pilot and main sessions 
arises from the use of a virtually identical setting in the pilot, which enables
a precise estimate of the true distribution of levels.  

Finally, it is instructive to revisit the pilot data using QDCH and to evaluate whether a calibration of the QDCH model from that data would predict the observed relative magnitudes of the strategy method effect within each class of games.\footnote{Section \ref{subsec:appendix_sensitivity} of the Appendix provides additional details and discussion of this QDCH calibration exercise.}
The Appendix \ref{subsec:appendix_sensitivity} presents the calibration of QDCH on the pilot data, in exactly the same way as described for DCH in Section \ref{subsec:optimal_design}. The QDCH calibrated model parameters are: $\hat{\tau} = 2.60$ and $\hat{\lambda} = 0.05$ (Obs. = 120, S.E. of $\hat{\tau} = 0.76$, S.E. of $\hat{\lambda} = 0.01$, log-likelihood $= -45.29$).  
We then compute the CDFs predicted by the calibrated QDCH for each class of games under the three elicitation methods, and calculate the sup-norm distances between these distributions. These are also reported in the Appendix (Table~\ref{tab:RN_E_prediction_QDCH}).  
The calibrated QDCH predicts: (1) no significant difference between the direct response method and the full strategy method across all six games; and (2) the magnitudes of the strategy method effects are larger in the Small Linear game and the Large Constant game than in the other four.

To evaluate this prediction of the QDCH calibration, we reproduce the analysis of the relative magnitudes of the effects in Table~\ref{tab:diff_method_effect_HvsL} but instead group the Small Linear game ($c = 0.5$) and the Large Constant game ($c = 0.8$) as games with large effects, and classifying the remaining four games as games with small effects. Table~\ref{tab:diff_method_effect_HvsL_QDCH} shows that, for the comparison between the direct response method/full strategy method and the reduced strategy method, the strategy method effects are significantly stronger in the small linear game and the large constant game than in the other four games, consistent with the qualitative prediction of the calibrated QDCH.

\section{Concluding Remarks}
\label{sec:conclusion}

We conclude by highlighting the key motivation of this paper: to understand when and how 
the use of the strategy method introduces behavioral distortions to choice behavior in sequential games. Due to the incompleteness problem, the question of how to reliably collect 
behavioral data at rarely observed information sets has long been an important question in experimental methodology.
The strategy method pioneered by \citet{Selten1967} offers a novel and promising approach to solve the incompleteness problem: 
to elicit choices in a sequential game, the experimenter can instead implement a strategically equivalent simultaneous move game, allowing experimenters to observe the action choices of subjects even at off-path information sets.
However, the questions of when and why choices elicited via the strategy method can create behavioral distortions compared with standard (direct response) elicitation methods is not well understood and remains an important open questions. In a partial answer to these questions, this study demonstrates that the strategy method induces \emph{predictable} behavioral distortions---ones that align with the predictions of the Dynamic Cognitive Hierarchy (DCH) solution proposed by \citet{lin_cognitive_2024}.

The centipede game is an ideal environment for our experiment because DCH 
makes a clear qualitative prediction: the reduced strategy method should induce 
later taking compared to both the direct response and full strategy methods. This is 
an unusually strong prediction, as it does not rely on any free parameters in the 
model or on the specific payoff structure of the centipede game. More importantly, 
it is scientifically important: any significant deviation in the opposite direction 
would \emph{falsify} the theory. Thus, our finding that none of the observed 
strategy method effects produce significant differences that run counter to this prediction lends support for the rationale of DCH.

Not only does DCH make unambiguous \textit{qualitative} predictions about the effects of the strategy method, but it further implies that the expected magnitudes of such effects will depend on the specific payoff details of the game and the distribution of levels of sophistication. 
Hence, designing informative payoff parameters is crucial, as one might otherwise fail to detect statistically significant strategy method effects if they are not properly chosen. This is particularly challenging 
in our experiment because the distribution of levels is unknown 
prior to the experiment, and the most informative payoff parameters
depend on that distribution. 

To address this challenge, we adopt an \textit{optimal design} approach 
developed by \citet{lin_cognitive_2023} to select the payoff parameters for the games used in the experiment. 
This approach is a two-step procedure: first, we calibrate the distribution of 
levels using small-scale pilot data; treating this calibrated distribution as the 
true one, we then select a mix of games expected to generate large and small effects---serving as our diagnostic and quasi-placebo games, 
respectively.

We find that although the directions of the 
observed strategy method effects are consistent with DCH, their relative
magnitudes deviate from the calibrated DCH predictions. Specifically, we observe 
significant strategy method effects in two of the three quasi-placebo games (linear 
and constant games), and null effects in two of the three diagnostic
games (linear and exponential games).
These results suggest that the observed strategy method effects cannot be fully 
explained by DCH alone.

Motivated by this finding, we further examine 
whether stochastic choice, in the form of logit quantal responses, can account for the relative magnitudes
of the observed strategy method effects, as the (Agent) Quantal
Response Equilibrium also predicts differences between the direct response, full strategy, and reduced strategy methods, but in different ways than DCH. The QDCH model is also useful since it combines the essential elements of AQRE and DCH. We find that QDCH significantly
outperforms the other two models in terms of model fit, and accounts for the observed relative magnitudes of the elicitation effects. 
Taken together, these findings suggest that the use of the 
(reduced) strategy method can lead to significant behavioral 
distortions---distortions that are predictable under DCH with quantal responses.

As a final remark, although this paper presents the optimal 
design approach as tailored to DCH, its two-step procedure is essentially a general recipe for designing theory-testing experiments---particularly when the theory involves free parameters that must be
estimated \emph{ex post} from the data. This approach enables
experimenters to design highly informative experiments without shooting 
in the dark, allowing for the testing of theoretical predictions that
depend on unknown parameters. The detection of strategy method effects 
in this paper is just one of many potential applications of the optimal 
design approach.

\bibliographystyle{ecta}
\bibliography{reference}

\newpage
\appendix

\section{Appendix}

\setcounter{figure}{0}
\renewcommand{\thefigure}{A.\arabic{figure}}
\setcounter{table}{0}
\renewcommand{\thetable}{A.\arabic{table}}
\setcounter{result}{0}
\renewcommand{\theresult}{A.\arabic{result}}

\subsection[Instructions for RF Order]{Experimental Instructions 
(RF Order)\footnote{
This appendix provides the experimental instructions for the RF Order;
the instructions for the FR Order and the pilot sessions are available 
upon request.
The only difference between the RF and FR Orders lies in the sequence in which the instructions are presented. 
In the RF Order, the instructions are given in the sequence as shown in 
this appendix. In contrast, in the FR order, participants first read the instructions for Main Task---Part Two, which is renamed as ``Main Task---Part One,'' followed by the instructions for Main Task---Part One, which is renamed as ``Main Task---Part Two.''
Additionally, in the FR Order, the last sentence of the current 
Main Task---Part One is changed to ``Now, please click `Next' to proceed to the third part of the experiment.'' Similarly, the last sentence of the current Main Task---Part Two is changed to ``Now, please click `Next' to proceed to the second part of the experiment.''}}
\label{appendix:instructions}
\noindent\textbf{General Instructions}

\bigskip

\noindent Thank you for participating in the experiment. You are about to take part in a
decision-making experiment, in which your earnings will depend partly on your
decisions, partly on the decision of others, and partly on chance.

\bigskip

\noindent The entire session will take place through computer terminals,
and all interactions
between participants will be conducted through the computers. Please turn off your
mobile devices and do not talk or in any way try to communicate with other participants
during the session.

\bigskip

\noindent The main task of the experiment consists of three parts, 
each containing six rounds.
Before each part of the main task, you will be asked to complete some comprehension
questions. If you have any questions, please raise your hand and the question will be
answered so that everyone can hear.

\bigskip

\noindent In this experiment, you will earn ``points'' in each round, 
and your earnings will be
determined by the number of points you earn throughout the three parts of the main
task. Each point has a value of \$0.02. That is, every 100 points generates \$2 in
earnings for you. In addition to your earnings from decisions, 
you will receive a show-up fee of \$10. At the end of the experiment, 
your earnings will be rounded up to the nearest
dollar amount. All your earnings will be paid in cash privately at the end of the
experiment.

\bigskip
\bigskip

\noindent \textbf{Main Task --- General Description}

\bigskip

\noindent \underline{Roles and Groups}

\bigskip

\noindent All participants will be randomly divided into two groups, 
the {\color{red}RED} group and the {\color{blue}BLUE}
group, with equal probability. Before you start making decisions, you will be informed
whether you are in the {\color{red}RED} group or in the {\color{blue}BLUE} group. 
If you are in the {\color{red}RED} group,
you will be called a {\color{red}RED} participant; 
if you are in the {\color{blue}BLUE} group, you will be called a
{\color{blue}BLUE} participant. You will remain in the same group 
for the entire experiment.

\bigskip

\noindent The main task of the experiment consists of three parts, 
each containing six rounds. In
each round, you will be randomly matched with an opponent from the \textbf{other group}.
Furthermore, in each round, you will only be matched with a participant who
has \textbf{not}
been matched with you previously within the same part. In other words, you will be
matched with each participant \textbf{at most once} in each part.

\bigskip

\noindent \underline{Decisions}

\bigskip

\noindent Every round in the experiment has the same format, as represented in Figure 1.

\begin{figure}[H]
    \centering
    \includegraphics[width=0.7\linewidth]{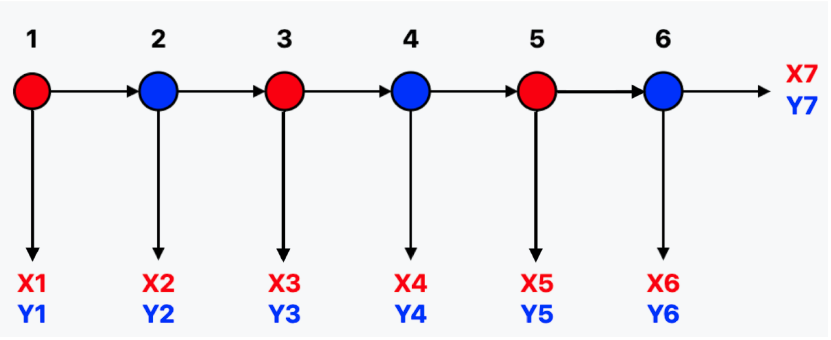}
    \caption*{Figure 1}
\end{figure}

\noindent In each round, there are six circles labeled as ``decision node 1'' to 
``decision node 6'' from left to right. The figure should be 
read from left to right, as indicated by the
directions of the arrows. The ``points'' you and the other participant can earn are
represented by {\color{red}X} and {\color{blue}Y}, 
and colored according to their group colors.

\bigskip

\noindent Throughout these instructions, {\color{red}X} and {\color{blue}Y} are just 
placeholders. In the actual rounds, all
payoffs will be positive integer values.

\bigskip

\noindent At each node, the participant with the same color as the node
can decide to either
``Continue'' or ``Stop.'' In summary, the outcome of a 
round is determined by the following rules:

\begin{itemize}
    \item The participant who chooses ``Stop'' in an earlier node ends the round, 
    and the round ends at the node where this participant chooses ``Stop.'' 
    Participants receive the corresponding payoffs at this node.
    \item If both participants choose ``Continue'' throughout the entire round, 
    the round will end at the last node (decision node 6). 
    The corresponding payoffs are ({\color{red}X7}, {\color{blue}Y7}).
\end{itemize}

\noindent Below, we provide a detailed description.

\bigskip

\noindent As shown in Figure 2, if the {\color{red}RED} participant 
\textbf{chooses ``Stop'' at decision node 1}, the
{\color{red}RED} participant receives {\color{red}X1}, 
the {\color{blue}BLUE} participant receives {\color{blue}Y1}, and this round ends.

\begin{figure}[H]
    \centering
    \includegraphics[width=0.7\linewidth]{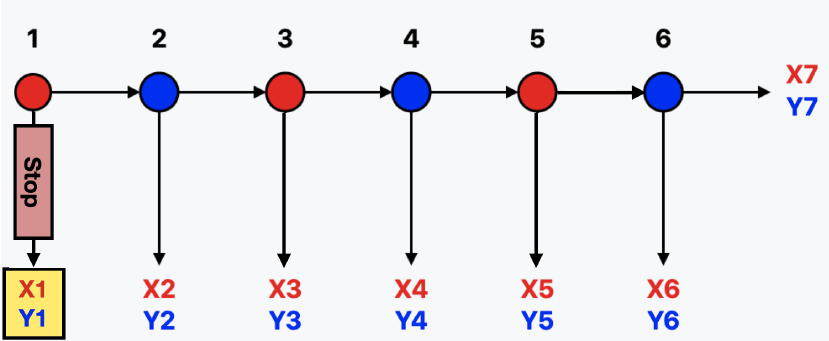}
    \caption*{Figure 2}
\end{figure}

\noindent If the {\color{red}RED} participant chooses ``Continue,'' 
this round proceeds to decision node 2.

\bigskip

\noindent As shown in Figure 3, if the {\color{blue}BLUE} participant 
\textbf{chooses ``Stop'' at decision node 2}, the
{\color{red}RED} participant receives {\color{red}X2}, the {\color{blue}BLUE} 
participant receives {\color{blue}Y2}, and this round ends.

\begin{figure}[H]
    \centering
    \includegraphics[width=0.7\linewidth]{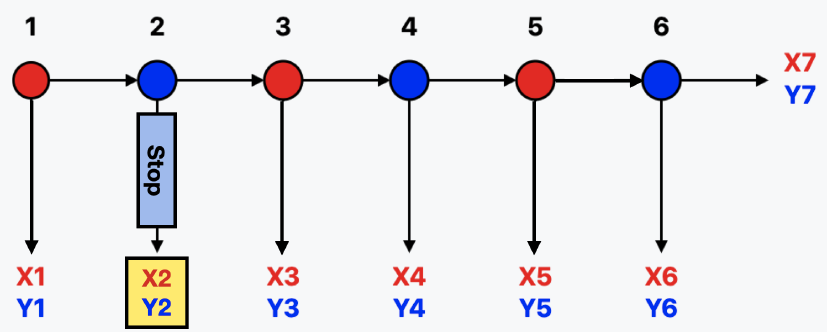}
    \caption*{Figure 3}
\end{figure}

\noindent If the {\color{blue}BLUE} participant chooses ``Continue,'' 
this round proceeds to decision node 3.

\bigskip

\noindent As shown in Figure 4, if the {\color{red}RED} participant 
\textbf{chooses ``Stop'' at decision node 3}, the
{\color{red}RED} participant receives {\color{red}X3}, the {\color{blue}BLUE} 
participant receives {\color{blue}Y3}, and this round ends.

\begin{figure}[H]
    \centering
    \includegraphics[width=0.7\linewidth]{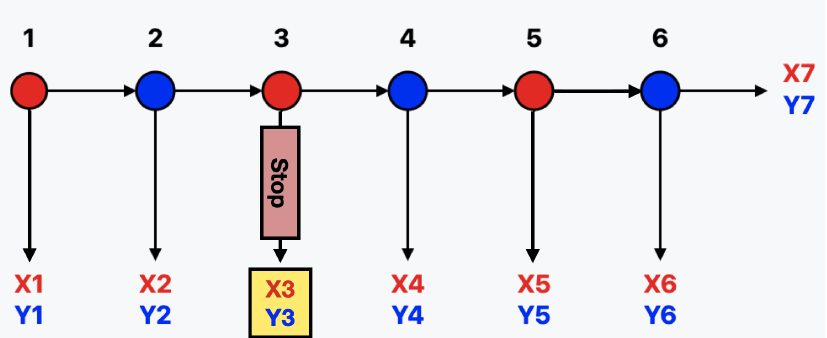}
    \caption*{Figure 4}
\end{figure}

\noindent If the {\color{red}RED} participant chooses ``Continue,'' 
this round proceeds to decision node 4.

\bigskip

\noindent As shown in Figure 5, if the {\color{blue}BLUE} participant 
\textbf{chooses ``Stop'' at decision node 4}, the
{\color{red}RED} participant receives {\color{red}X4}, the {\color{blue}BLUE} 
participant receives {\color{blue}Y4}, and this round ends.

\begin{figure}[H]
    \centering
    \includegraphics[width=0.7\linewidth]{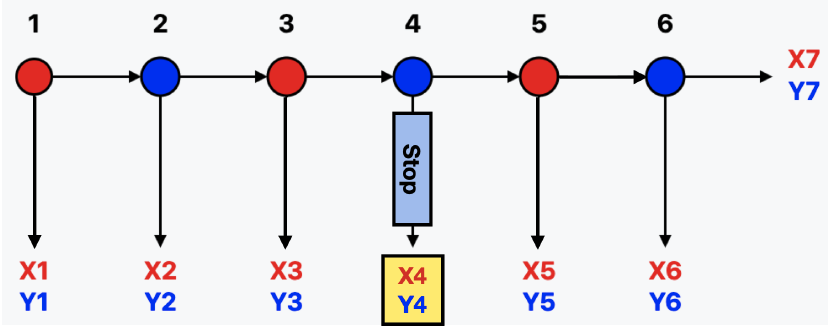}
    \caption*{Figure 5}
\end{figure}

\noindent If the {\color{blue}BLUE} participant chooses ``Continue,'' 
this round proceeds to decision node 5.

\bigskip

\noindent As shown in Figure 6, if the {\color{red}RED} participant 
\textbf{chooses ``Stop'' at decision node 5}, the
{\color{red}RED} participant receives {\color{red}X5}, the {\color{blue}BLUE} 
participant receives {\color{blue}Y5}, and this round ends.

\begin{figure}[H]
    \centering
    \includegraphics[width=0.7\linewidth]{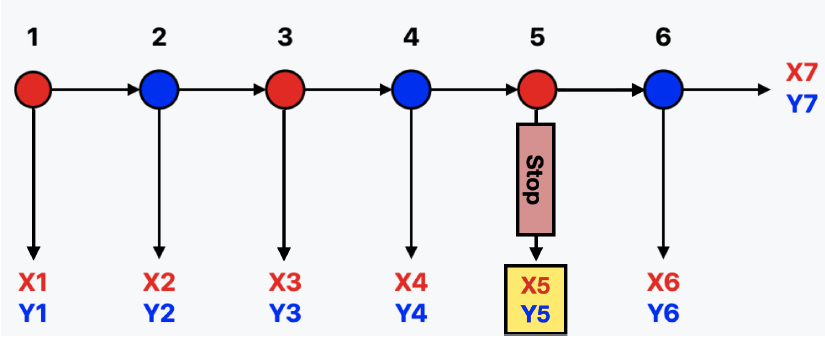}
    \caption*{Figure 6}
\end{figure}

\noindent If the {\color{red}RED} participant chooses ``Continue,'' 
this round proceeds to decision node 6.

\bigskip

\noindent As shown in Figure 7, if the {\color{blue}BLUE} participant 
\textbf{chooses ``Stop'' at decision node 6}, the
{\color{red}RED} participant receives {\color{red}X6}, the {\color{blue}BLUE} 
participant receives {\color{blue}Y6}, and this round ends.

\begin{figure}[H]
    \centering
    \includegraphics[width=0.7\linewidth]{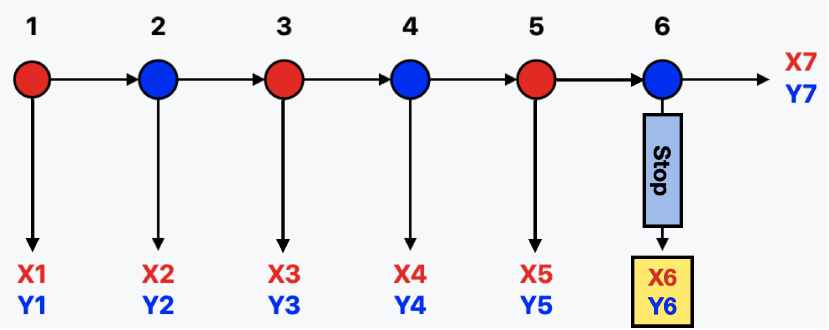}
    \caption*{Figure 7}
\end{figure}

\noindent As shown in Figure 8, if the {\color{blue}BLUE} participant 
\textbf{chooses ``Continue'' at decision node 6}, the
{\color{red}RED} participant receives {\color{red}X7}, the {\color{blue}BLUE} 
participant receives {\color{blue}Y7}, and this round ends.

\begin{figure}[H]
    \centering
    \includegraphics[width=0.7\linewidth]{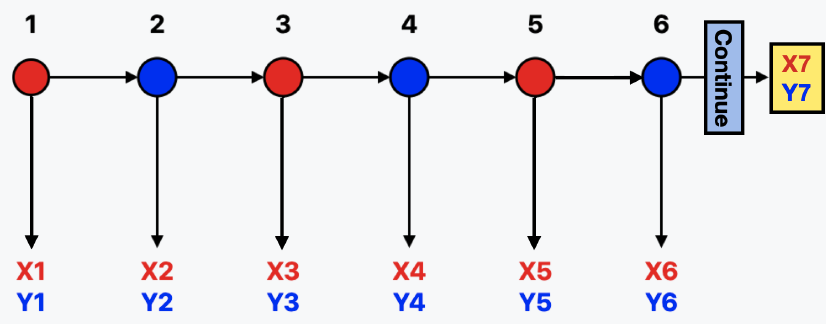}
    \caption*{Figure 8}
\end{figure}

\noindent\underline{Payment}

\begin{itemize}
    \item At the end of the experiment, the computer will 
    randomly select one round from each part, and all participants
    will be paid according to the selected rounds.
    
    \item You will be paid based on your choices and the 
    choices made by the participant matched with you in the selected rounds.
\end{itemize}

\bigskip

\noindent\underline{Reminder}

\bigskip

\noindent The experiment has 3 parts and each part has
6 rounds. In each part, you will be
matched with each participant \textbf{at most once}.

\bigskip
\bigskip

\noindent\textbf{Main Task --- Part One}

\bigskip

\noindent In each round of this part, you and the other
participant you are matched with will make
decisions \textbf{simultaneously}, and both of your 
earnings are determined by your choices.

\bigskip

\noindent If you are in the {\color{red}RED} group, 
you will make your decisions on the following screen,
where you choose between four rectangles corresponding to four choices: 
\textbf{``Stop 1,''}
\textbf{``Stop 3,''} \textbf{``Stop 5''} and \textbf{``Always Continue,''} 
as shown in Figure 9.

\begin{figure}[H]
    \centering
    \includegraphics[width=0.7\linewidth]{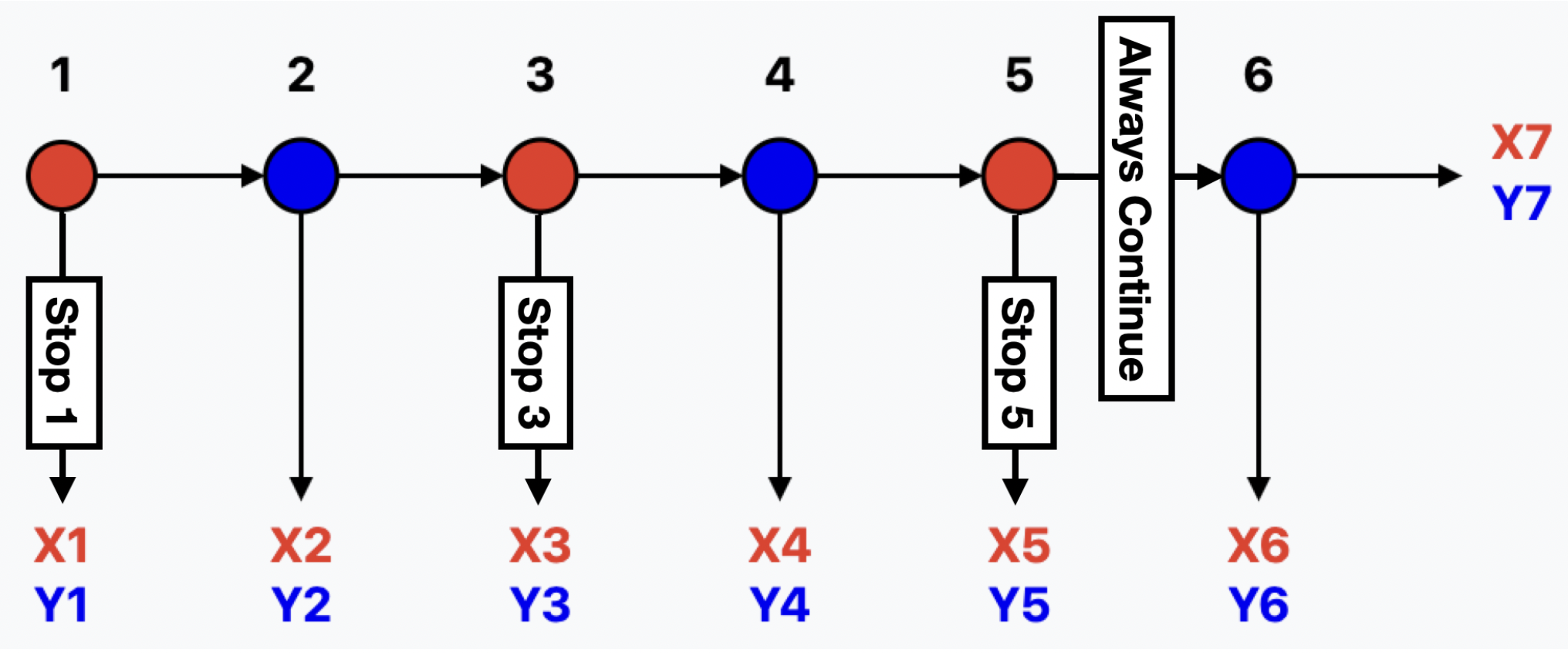}
    \caption*{Figure 9}
\end{figure}

\noindent If you are in the {\color{blue}BLUE} group, 
you will make your decisions on the following screen,
where you choose between four rectangles corresponding to four choices: 
\textbf{``Stop 2,''}
\textbf{``Stop 4,''} \textbf{``Stop 6''} and \textbf{``Always Continue,''} 
as shown in Figure 10.

\begin{figure}[H]
    \centering
    \includegraphics[width=0.7\linewidth]{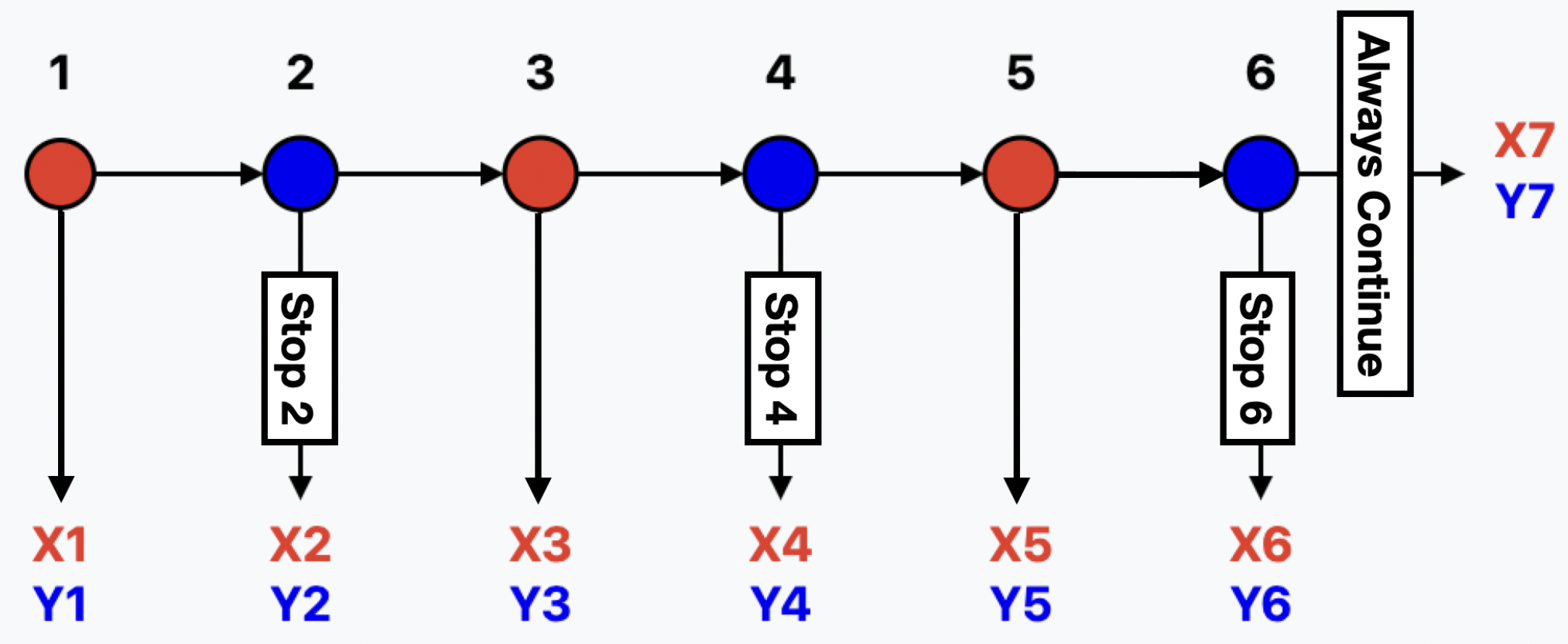}
    \caption*{Figure 10}
\end{figure}

\noindent In summary, the outcome of a round is determined by the following rules:
\begin{itemize}
    \item The participant who chooses ``Stop'' in an earlier node ends 
    the round, and the round ends at the node where this participant
    chooses ``Stop.'' Participants receive the corresponding payoffs at this node.

    \item If both participants choose ``Always Continue,'' 
    the corresponding payoffs are ({\color{red}X7}, {\color{blue}Y7}).
\end{itemize}

\bigskip

\noindent Note that at the beginning of each round,
there is a 10-second delay before you can
submit your decisions.

\bigskip

\noindent\textbf{Now, please click ``Next'' to proceed to 
the first part of the experiment.}

\bigskip
\bigskip

\noindent\textbf{Main Task --- Part Two}

\bigskip

\noindent In each round of this part, you 
and the other participant you are matched with will make
decisions \textbf{simultaneously}, and both of your 
earnings are determined by your choices.

\bigskip

\noindent If you are in the {\color{red}RED} group, you will 
make your decisions on the following screen. At
each node of your color, you choose between two rectangles corresponding to either
\textbf{``Stop''} or \textbf{``Continue''} at the given node, as shown in Figure 11.

\begin{figure}[H]
    \centering
    \includegraphics[width=0.7\linewidth]{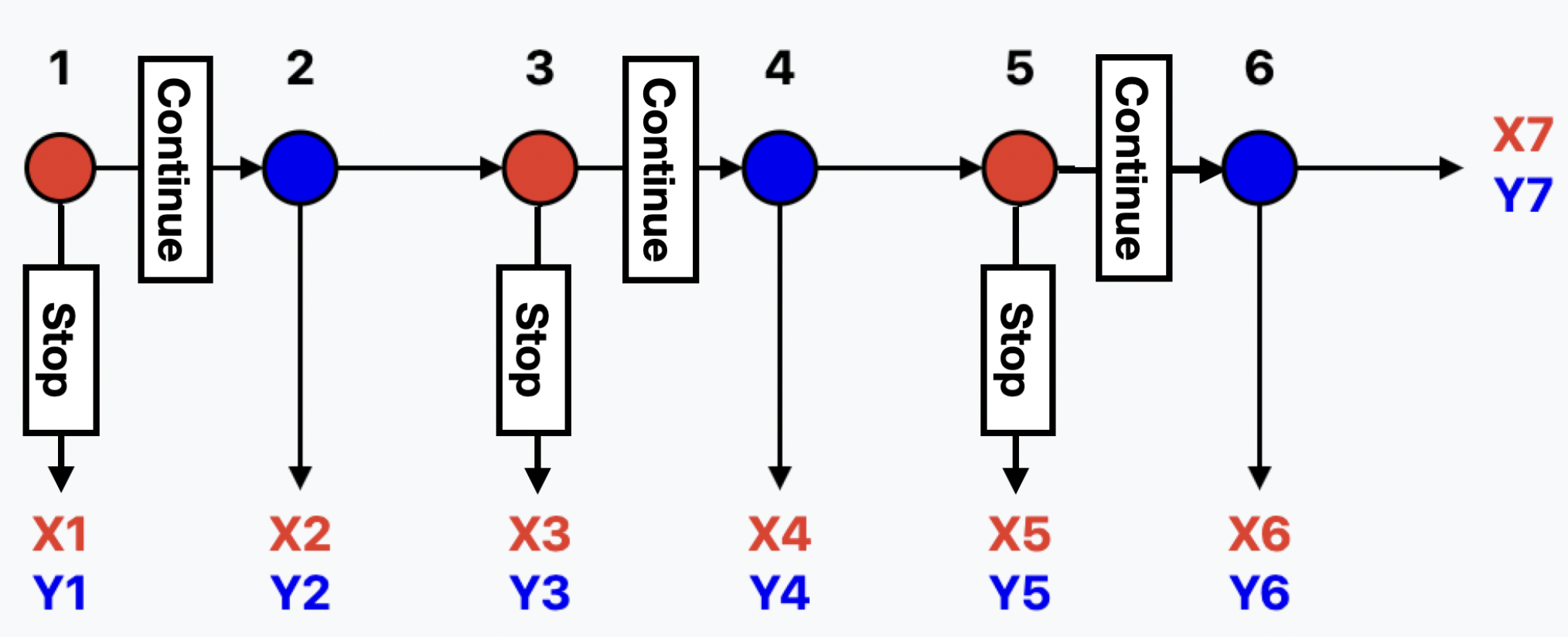}
    \caption*{Figure 11}
\end{figure}

\noindent If you are in the {\color{blue}BLUE} group, 
you will make your decisions on the following screen. At
each node of your color, you choose between two rectangles corresponding to either
\textbf{``Stop''} or \textbf{``Continue''} at the given node, as shown in Figure 12.

\begin{figure}[H]
    \centering
    \includegraphics[width=0.7\linewidth]{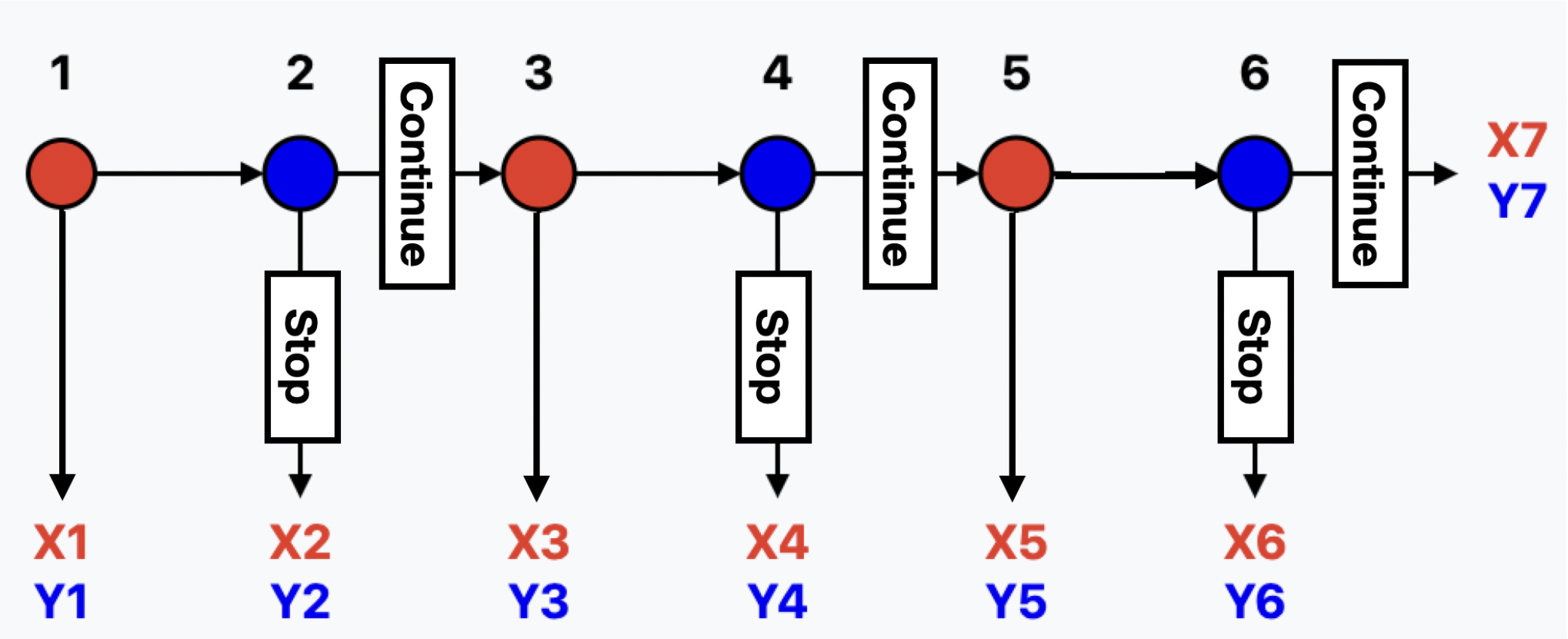}
    \caption*{Figure 12}
\end{figure}

\noindent In summary, the outcome of a round is determined by the following rules:

\begin{itemize}
    \item The participant who chooses ``Stop'' in an earlier node ends the round, 
    and the round ends at the node where this participant chooses ``Stop.''
    Participants receive the corresponding payoffs at this node.

    \item If both participants choose ``Continue'' at every decision node, 
    the corresponding payoffs are ({\color{red}X7}, {\color{blue}Y7}).
    
\end{itemize}

\bigskip

\noindent Note that at the beginning of each round, 
there is a 10-second delay before you can
submit your decisions.

\bigskip

\noindent\textbf{Now, please click ``Next'' to proceed to the 
second part of the experiment.}

\bigskip
\bigskip

\noindent\textbf{Main Task --- Part Three}

\bigskip

\noindent In each round of this part, you and the other 
participant you are matched with will make
decisions \textbf{sequentially} starting from decision node 1,
and both of your earnings are
determined by your choices.

\bigskip

\noindent As shown in Figure 13, if the {\color{red}RED} 
participant \textbf{chooses ``Stop'' at decision node 1}, the
{\color{red}RED} participant receives {\color{red}X1}, 
the {\color{blue}BLUE} participant receives {\color{blue}Y1}, and this round ends.

\begin{figure}[H]
    \centering
    \includegraphics[width=0.7\linewidth]{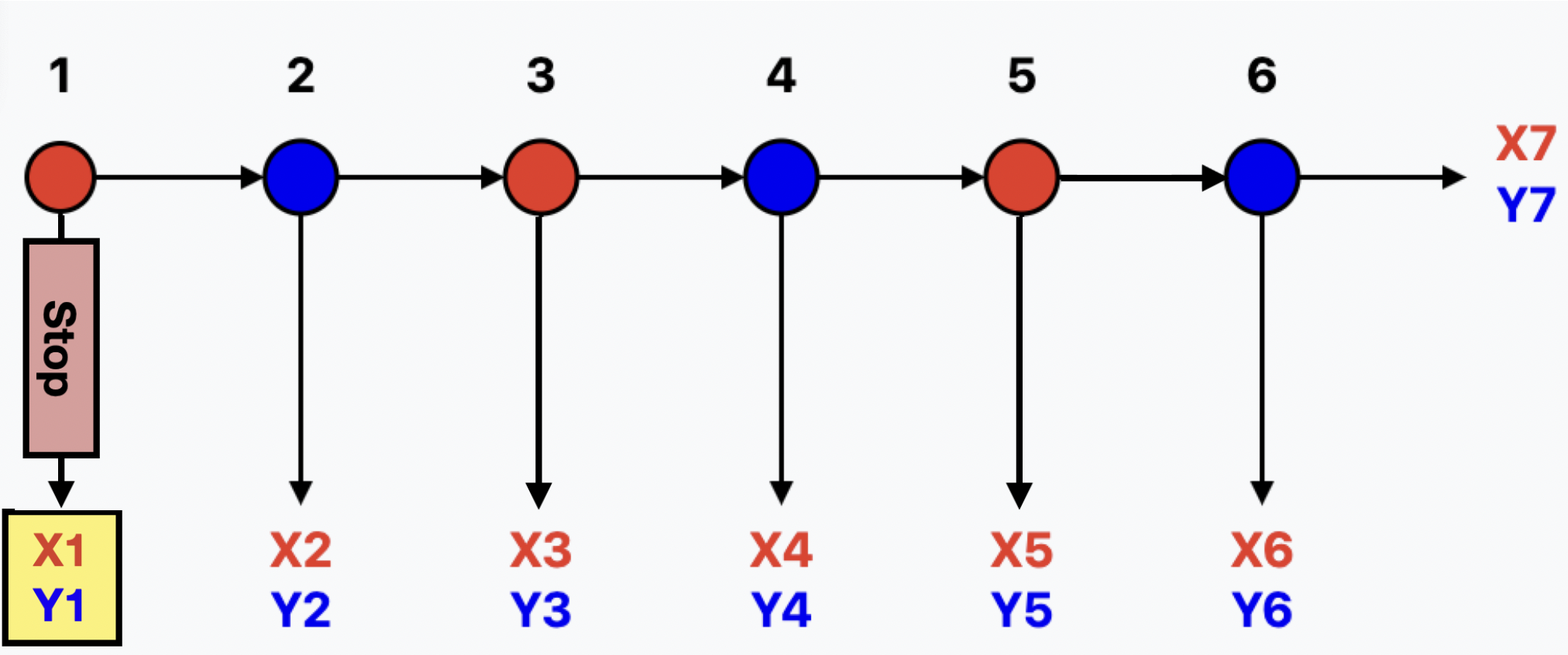}
    \caption*{Figure 13}
\end{figure}

\noindent If the {\color{red}RED} participant chooses ``Continue,''
this round proceeds to decision node 2.

\bigskip

\noindent As shown in Figure 14, if the {\color{blue}BLUE}
participant \textbf{chooses ``Stop'' at decision node 2},
the {\color{red}RED} participant receives {\color{red}X2}, 
the {\color{blue}BLUE} participant receives {\color{blue}Y2}, and this round ends.

\begin{figure}[H]
    \centering
    \includegraphics[width=0.7\linewidth]{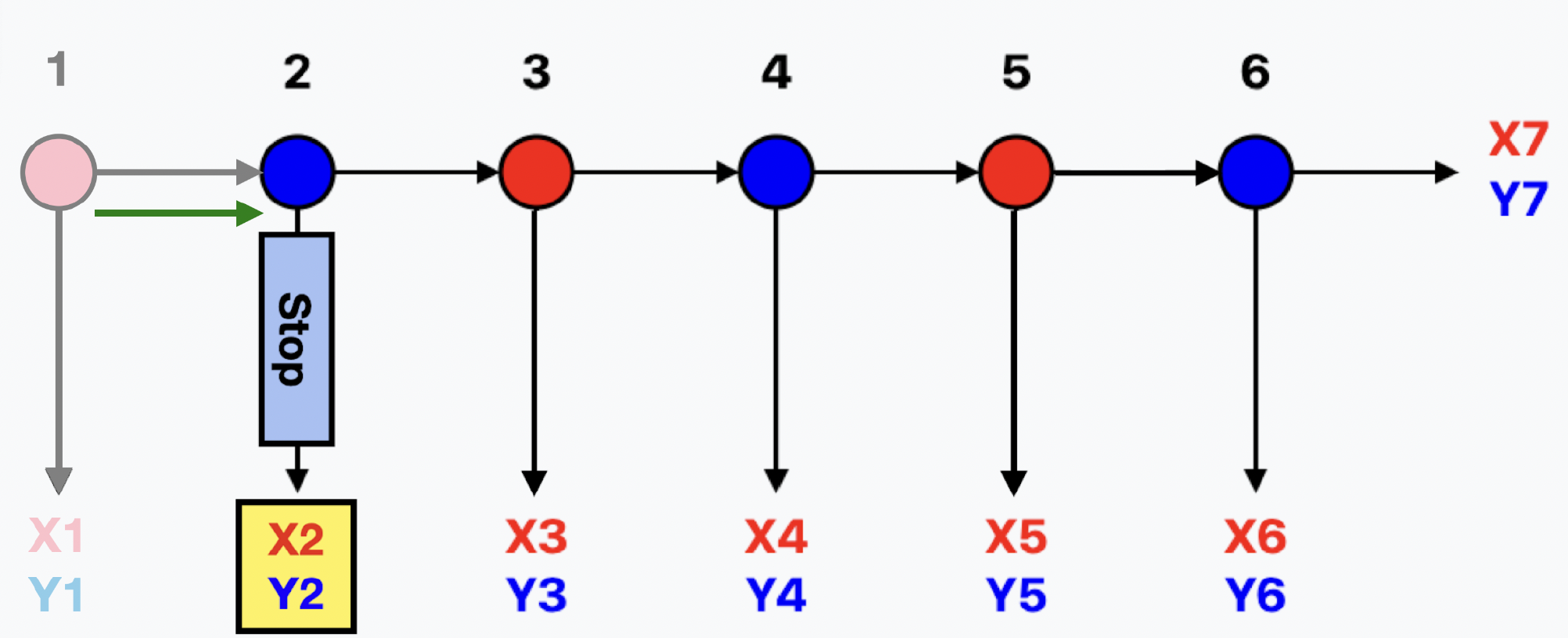}
    \caption*{Figure 14}
\end{figure}

\noindent If the {\color{blue}BLUE} participant chooses 
``Continue,'' this round proceeds to decision node 3.

\bigskip

\noindent As shown in Figure 15, if the {\color{red}RED} 
participant \textbf{chooses ``Stop'' at decision node 3}, the
{\color{red}RED} participant receives {\color{red}X3}, 
the {\color{blue}BLUE} participant receives {\color{blue}Y3}, and this round ends.

\begin{figure}[H]
    \centering
    \includegraphics[width=0.7\linewidth]{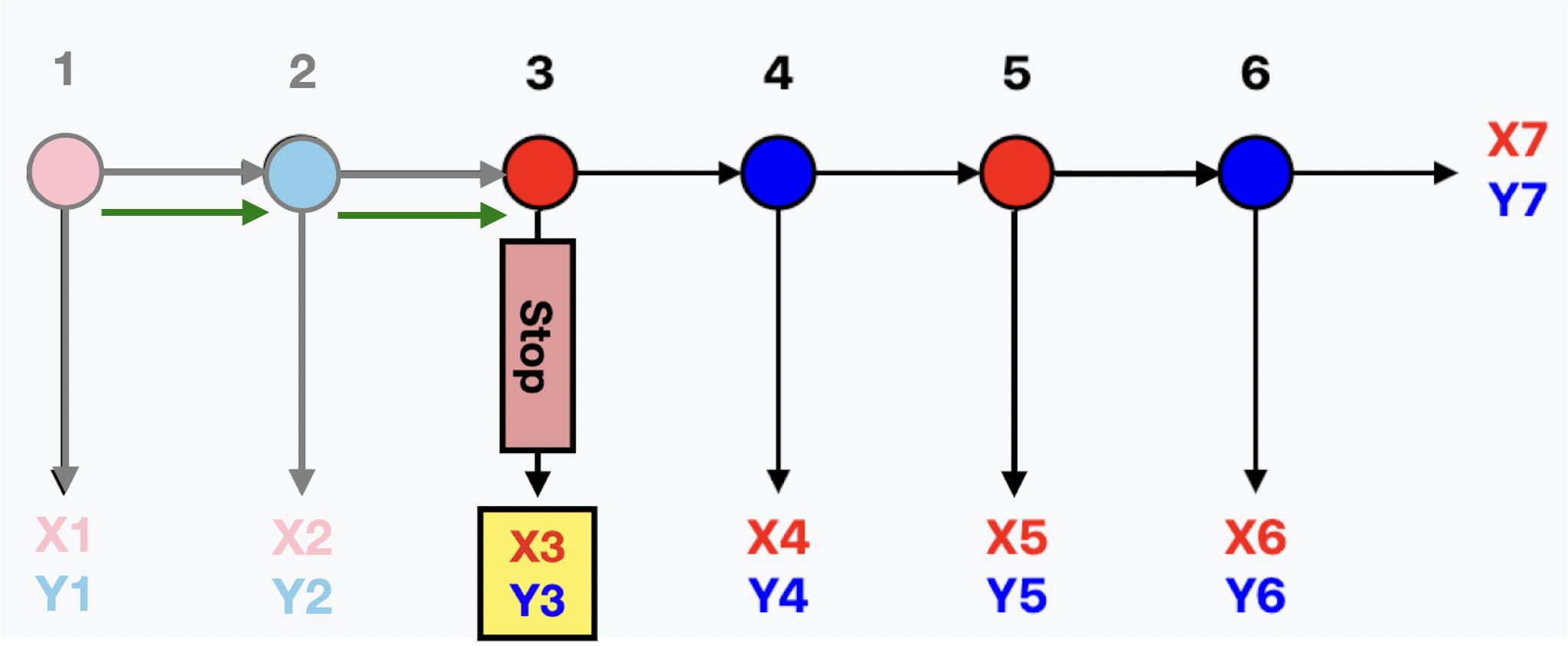}
    \caption*{Figure 15}
\end{figure}

\noindent If the {\color{red}RED} participant chooses ``Continue,''
this round proceeds to decision node 4.

\bigskip

\noindent As shown in Figure 16, if the {\color{blue}BLUE}
participant \textbf{chooses ``Stop'' at decision node 4},
the {\color{red}RED} participant receives {\color{red}X4}, 
the {\color{blue}BLUE} participant receives {\color{blue}Y4}, and this round ends.

\begin{figure}[H]
    \centering
    \includegraphics[width=0.7\linewidth]{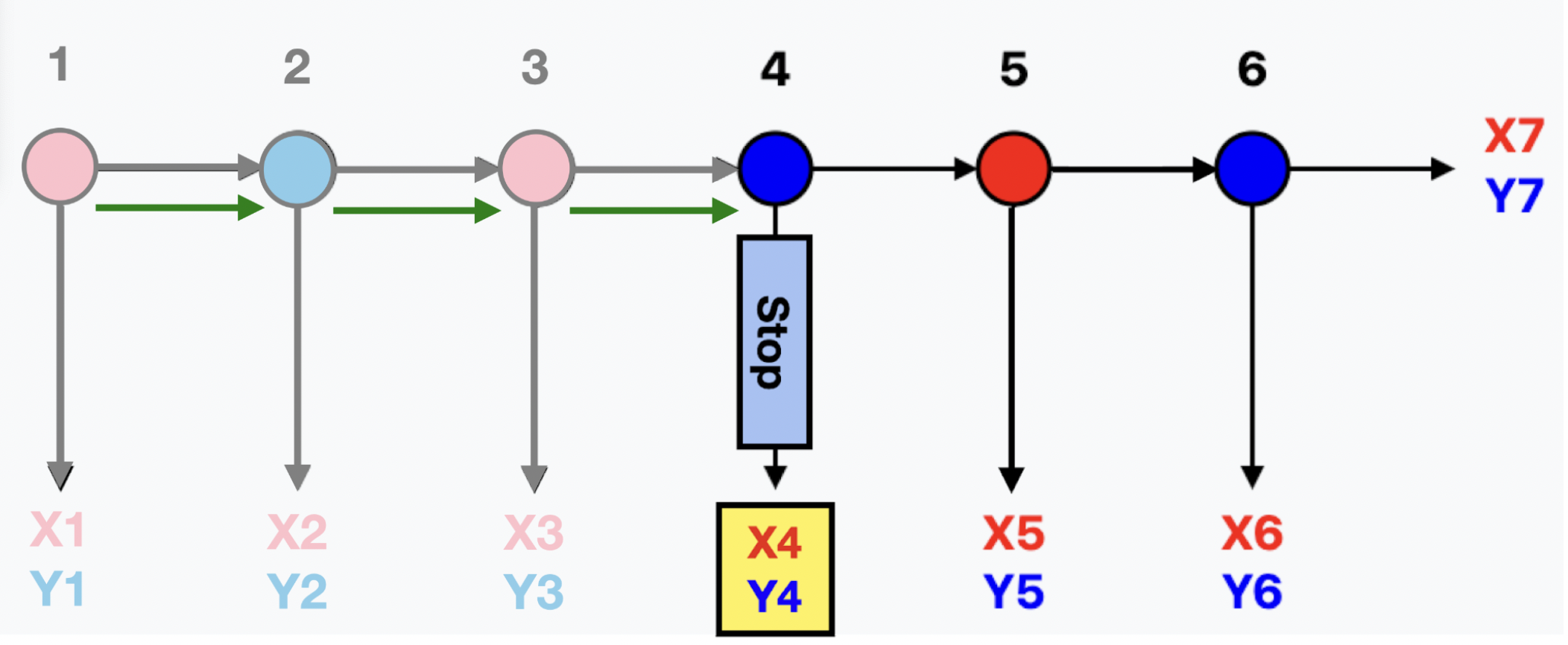}
    \caption*{Figure 16}
\end{figure}

\noindent If the {\color{blue}BLUE} participant chooses 
``Continue,'' this round proceeds to decision node 5.

\bigskip

\noindent As shown in Figure 17, if the {\color{red}RED} 
participant \textbf{chooses ``Stop'' at decision node 5}, the
{\color{red}RED} participant receives {\color{red}X5}, 
the {\color{blue}BLUE} participant receives {\color{blue}Y5}, and this round ends.

\begin{figure}[H]
    \centering
    \includegraphics[width=0.7\linewidth]{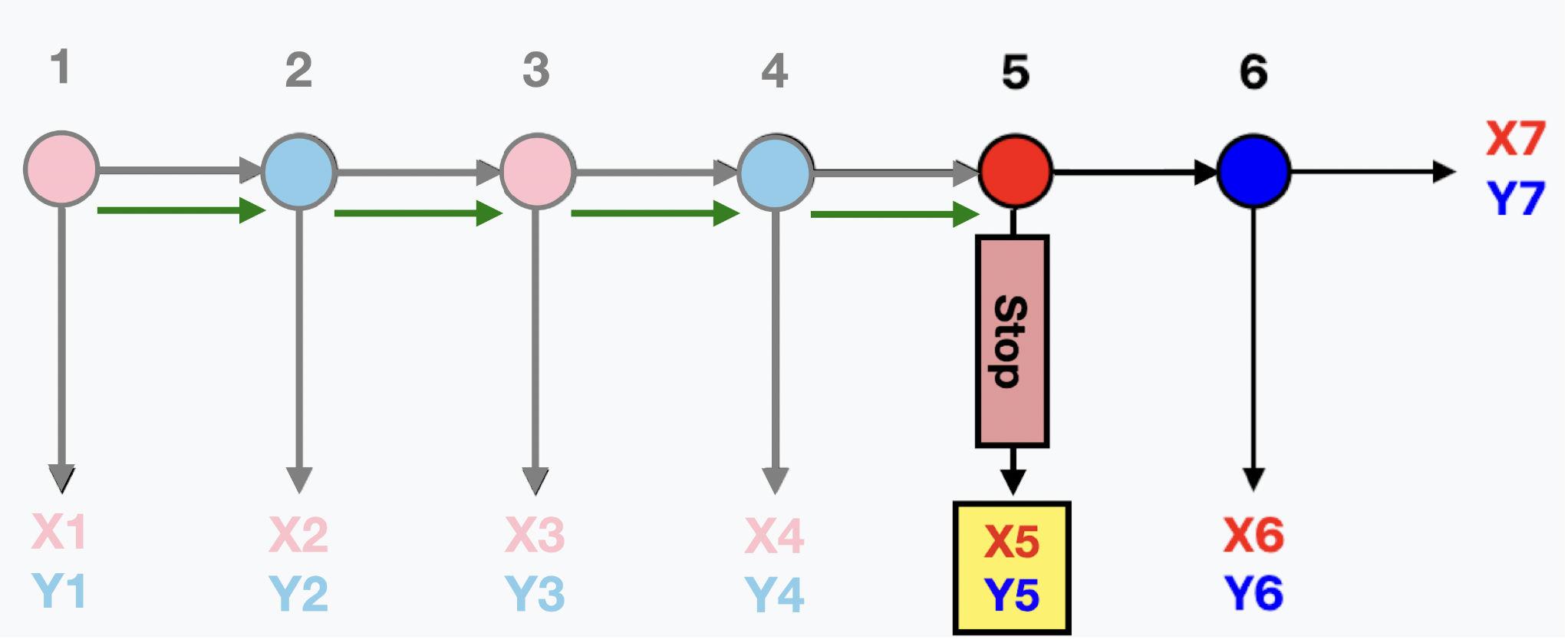}
    \caption*{Figure 17}
\end{figure}

\noindent If the {\color{red}RED} participant chooses ``Continue,''
this round proceeds to decision node 6.

\bigskip

\noindent As shown in Figure 18, if the {\color{blue}BLUE}
participant \textbf{chooses ``Stop'' at decision node 6},
the {\color{red}RED} participant receives {\color{red}X6}, 
the {\color{blue}BLUE} participant receives {\color{blue}Y6}, and this round ends.

\begin{figure}[H]
    \centering
    \includegraphics[width=0.7\linewidth]{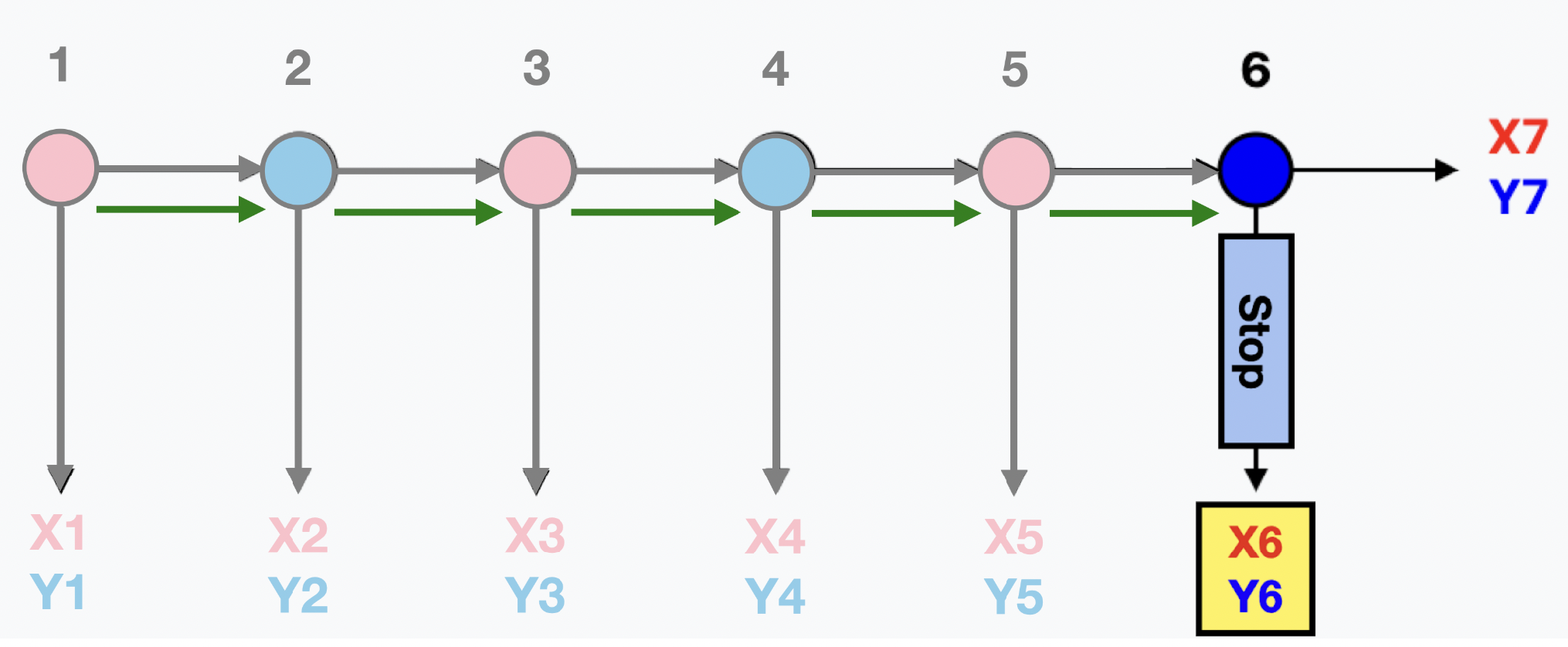}
    \caption*{Figure 18}
\end{figure}

\noindent As shown in Figure 19, if the {\color{blue}BLUE}
participant \textbf{chooses ``Continue'' at decision node 6},
the {\color{red}RED} participant receives {\color{red}X7}, 
the {\color{blue}BLUE} participant receives {\color{blue}Y7}, and this round ends.

\begin{figure}[H]
    \centering
    \includegraphics[width=0.7\linewidth]{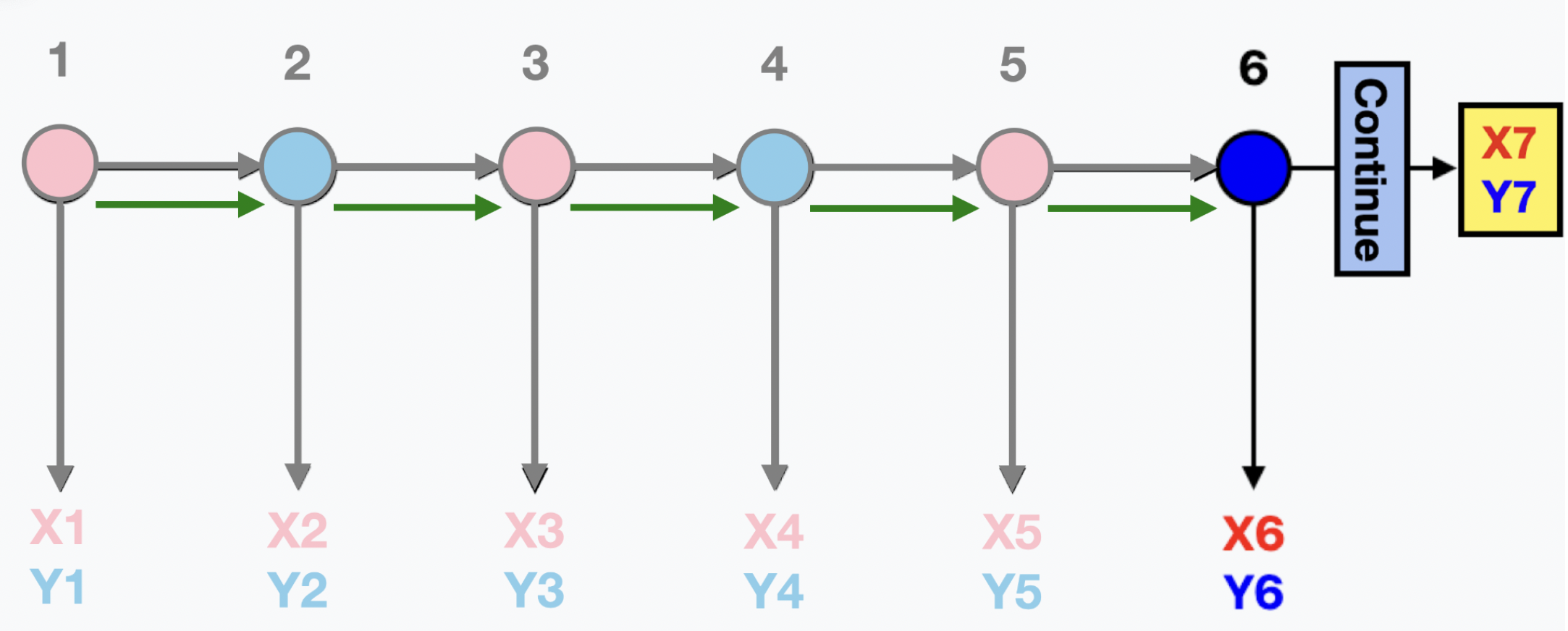}
    \caption*{Figure 19}
\end{figure}

\noindent Note that at the beginning of each round, there is 
a 10-second delay before you can
submit your decisions.

\bigskip

\noindent\textbf{Now, please click 
``Next'' to proceed to the third part of the experiment.}

\subsection{Additional Tables and Figures}

This appendix includes two additional tables and two additional figures that 
supplement the analysis in the main text. Figure~\ref{fig:six_cg} displays the 
six game trees implemented in the experiment. 
Figure~\ref{fig:pair_level_empirical_cdf} presents the CDFs of 
terminal nodes across the three elicitation methods. Lastly, 
Tables~\ref{tab:end_node_data_signed_rank} and 
\ref{tab:pair_level_terminal_node_pdf} provide summary and test statistics
for these distributions of terminal nodes, forming the foundation of Figure~\ref{fig:game_level_pairwise_comparison} and \ref{fig:pair_level_terminal_node_pdf}, respectively.  

\begin{figure}[H]
\centering
\begin{adjustbox}{width=0.9\columnwidth,left}
\begin{minipage}{\textwidth}
\footnotesize % Apply footnotesize to all tree content
% Row 1
\begin{subfigure}[t]{0.45\textwidth}
\centering
\begin{istgame}[scale=1]
\setistmathTF*001
\xtShowEndPoints
\setistgrowdirection{south east}
\xtdistance{8mm}{16mm}
\istroot(0)[initial node]{1}
\istb{T}[r]{\begin{tabular}{c}150 \\ 50\end{tabular}}[b] \istb{P}[a] \endist
\istroot(1)(0-2){2}
\istb{T}[r]{\begin{tabular}{c}100 \\ 200\end{tabular}}[b] \istb{P}[a] \endist
\istroot(2)(1-2){3}
\istb{T}[r]{\begin{tabular}{c}250 \\ 150\end{tabular}}[b] \istb{P}[a] \endist
\istroot(3)(2-2){4}
\istb{T}[r]{\begin{tabular}{c}200 \\ 300\end{tabular}}[b] \istb{P}[a] \endist
\istroot(4)(3-2){5}
\istb{T}[r]{\begin{tabular}{c}350 \\ 250\end{tabular}}[b] \istb{P}[a] \endist
\istroot(5)(4-2){6}
\istb{T}[r]{\begin{tabular}{c}300 \\ 400\end{tabular}}[b] \istb{P}[a]{\begin{tabular}{c}450 \\ 350\end{tabular}}[r] \endist
\end{istgame}
\caption{Small Linear Game ($c=0.5$)}
\end{subfigure}
\hfill
\begin{subfigure}[t]{0.45\textwidth}
\centering
\begin{istgame}[scale=1]
\setistmathTF*001
\xtShowEndPoints
\setistgrowdirection{south east}
\xtdistance{8mm}{16mm}
\istroot(0)[initial node]{1}
\istb{T}[r]{\begin{tabular}{c}150 \\ 50\end{tabular}}[b] \istb{P}[a] \endist
\istroot(1)(0-2){2}
\istb{T}[r]{\begin{tabular}{c}130 \\ 230\end{tabular}}[b] \istb{P}[a] \endist
\istroot(2)(1-2){3}
\istb{T}[r]{\begin{tabular}{c}310 \\ 210\end{tabular}}[b] \istb{P}[a] \endist
\istroot(3)(2-2){4}
\istb{T}[r]{\begin{tabular}{c}290 \\ 390\end{tabular}}[b] \istb{P}[a] \endist
\istroot(4)(3-2){5}
\istb{T}[r]{\begin{tabular}{c}470 \\ 370\end{tabular}}[b] \istb{P}[a] \endist
\istroot(5)(4-2){6}
\istb{T}[r]{\begin{tabular}{c}450 \\ 550\end{tabular}}[b] \istb{P}[a]{\begin{tabular}{c}630 \\ 530\end{tabular}}[r] \endist
\end{istgame}
\caption{Large Linear Game ($c=0.8$)}
\end{subfigure}

\vspace{1cm}

% Row 2
\begin{subfigure}[t]{0.45\textwidth}
\centering
\begin{istgame}[scale=1]
\setistmathTF*001
\xtShowEndPoints
\setistgrowdirection{south east}
\xtdistance{8mm}{16mm}
\istroot(0)[initial node]{1}
\istb{T}[r]{\begin{tabular}{c}10 \\ 4\end{tabular}}[b] \istb{P}[a] \endist
\istroot(1)(0-2){2}
\istb{T}[r]{\begin{tabular}{c}8 \\ 20\end{tabular}}[b] \istb{P}[a] \endist
\istroot(2)(1-2){3}
\istb{T}[r]{\begin{tabular}{c}40 \\ 16\end{tabular}}[b] \istb{P}[a] \endist
\istroot(3)(2-2){4}
\istb{T}[r]{\begin{tabular}{c}32 \\ 80\end{tabular}}[b] \istb{P}[a] \endist
\istroot(4)(3-2){5}
\istb{T}[r]{\begin{tabular}{c}160 \\ 64\end{tabular}}[b] \istb{P}[a] \endist
\istroot(5)(4-2){6}
\istb{T}[r]{\begin{tabular}{c}128 \\ 320\end{tabular}}[b] \istb{P}[a]{\begin{tabular}{c}640 \\ 256\end{tabular}}[r] \endist
\end{istgame}
\caption{Small Exponential Game ($c=2.5$)}
\end{subfigure}
\hfill
\begin{subfigure}[t]{0.45\textwidth}
\centering
\begin{istgame}[scale=1]
\setistmathTF*001
\xtShowEndPoints
\setistgrowdirection{south east}
\xtdistance{8mm}{16mm}
\istroot(0)[initial node]{1}
\istb{T}[r]{\begin{tabular}{c}16 \\ 4\end{tabular}}[b] \istb{P}[a] \endist
\istroot(1)(0-2){2}
\istb{T}[r]{\begin{tabular}{c}8 \\ 32\end{tabular}}[b] \istb{P}[a] \endist
\istroot(2)(1-2){3}
\istb{T}[r]{\begin{tabular}{c}64 \\ 16\end{tabular}}[b] \istb{P}[a] \endist
\istroot(3)(2-2){4}
\istb{T}[r]{\begin{tabular}{c}32 \\ 128\end{tabular}}[b] \istb{P}[a] \endist
\istroot(4)(3-2){5}
\istb{T}[r]{\begin{tabular}{c}256 \\ 64\end{tabular}}[b] \istb{P}[a] \endist
\istroot(5)(4-2){6}
\istb{T}[r]{\begin{tabular}{c}128 \\ 512\end{tabular}}[b] \istb{P}[a]{\begin{tabular}{c}1024 \\ 256\end{tabular}}[r] \endist
\end{istgame}
\caption{Large Exponential Game ($c=4$)}
\end{subfigure}

\vspace{1cm}

% Row 3
\begin{subfigure}[t]{0.45\textwidth}
\centering
\begin{istgame}[scale=1]
\setistmathTF*001
\xtShowEndPoints
\setistgrowdirection{south east}
\xtdistance{8mm}{16mm}
\istroot(0)[initial node]{1}
\istb{T}[r]{\begin{tabular}{c}250 \\ 250\end{tabular}}[b] \istb{P}[a] \endist
\istroot(1)(0-2){2}
\istb{T}[r]{\begin{tabular}{c}100 \\ 400\end{tabular}}[b] \istb{P}[a] \endist
\istroot(2)(1-2){3}
\istb{T}[r]{\begin{tabular}{c}460 \\ 40\end{tabular}}[b] \istb{P}[a] \endist
\istroot(3)(2-2){4}
\istb{T}[r]{\begin{tabular}{c}16 \\ 484\end{tabular}}[b] \istb{P}[a] \endist
\istroot(4)(3-2){5}
\istb{T}[r]{\begin{tabular}{c}494 \\ 6\end{tabular}}[b] \istb{P}[a] \endist
\istroot(5)(4-2){6}
\istb{T}[r]{\begin{tabular}{c}3 \\ 497\end{tabular}}[b] \istb{P}[a]{\begin{tabular}{c}499 \\ 1\end{tabular}}[r] \endist
\end{istgame}
\caption{Small Constant Game ($c=0.4$)}
\end{subfigure}
\hfill
\begin{subfigure}[t]{0.45\textwidth}
\centering
\begin{istgame}[scale=1]
\setistmathTF*001
\xtShowEndPoints
\setistgrowdirection{south east}
\xtdistance{8mm}{16mm}
\istroot(0)[initial node]{1}
\istb{T}[r]{\begin{tabular}{c}250 \\ 250\end{tabular}}[b] \istb{P}[a] \endist
\istroot(1)(0-2){2}
\istb{T}[r]{\begin{tabular}{c}200 \\ 300\end{tabular}}[b] \istb{P}[a] \endist
\istroot(2)(1-2){3}
\istb{T}[r]{\begin{tabular}{c}340 \\ 160\end{tabular}}[b] \istb{P}[a] \endist
\istroot(3)(2-2){4}
\istb{T}[r]{\begin{tabular}{c}128 \\ 372\end{tabular}}[b] \istb{P}[a] \endist
\istroot(4)(3-2){5}
\istb{T}[r]{\begin{tabular}{c}398 \\ 102\end{tabular}}[b] \istb{P}[a] \endist
\istroot(5)(4-2){6}
\istb{T}[r]{\begin{tabular}{c}82 \\ 418\end{tabular}}[b] \istb{P}[a]{\begin{tabular}{c}434 \\ 66\end{tabular}}[r] \endist
\end{istgame}
\caption{Large Constant Game ($c=0.8$)}
\end{subfigure}
\end{minipage}
\end{adjustbox}
\vspace{0.5em}
\caption{The Six Optimally Selected Centipede Games Used in the Experiment. Payoffs are expressed in ``points,'' which were converted to monetary rewards at an exchange rate of 1 point = USD \$0.02.}
\label{fig:six_cg}
\end{figure}

\begin{figure}[H]
    \centering
    \includegraphics[width=0.9\linewidth]{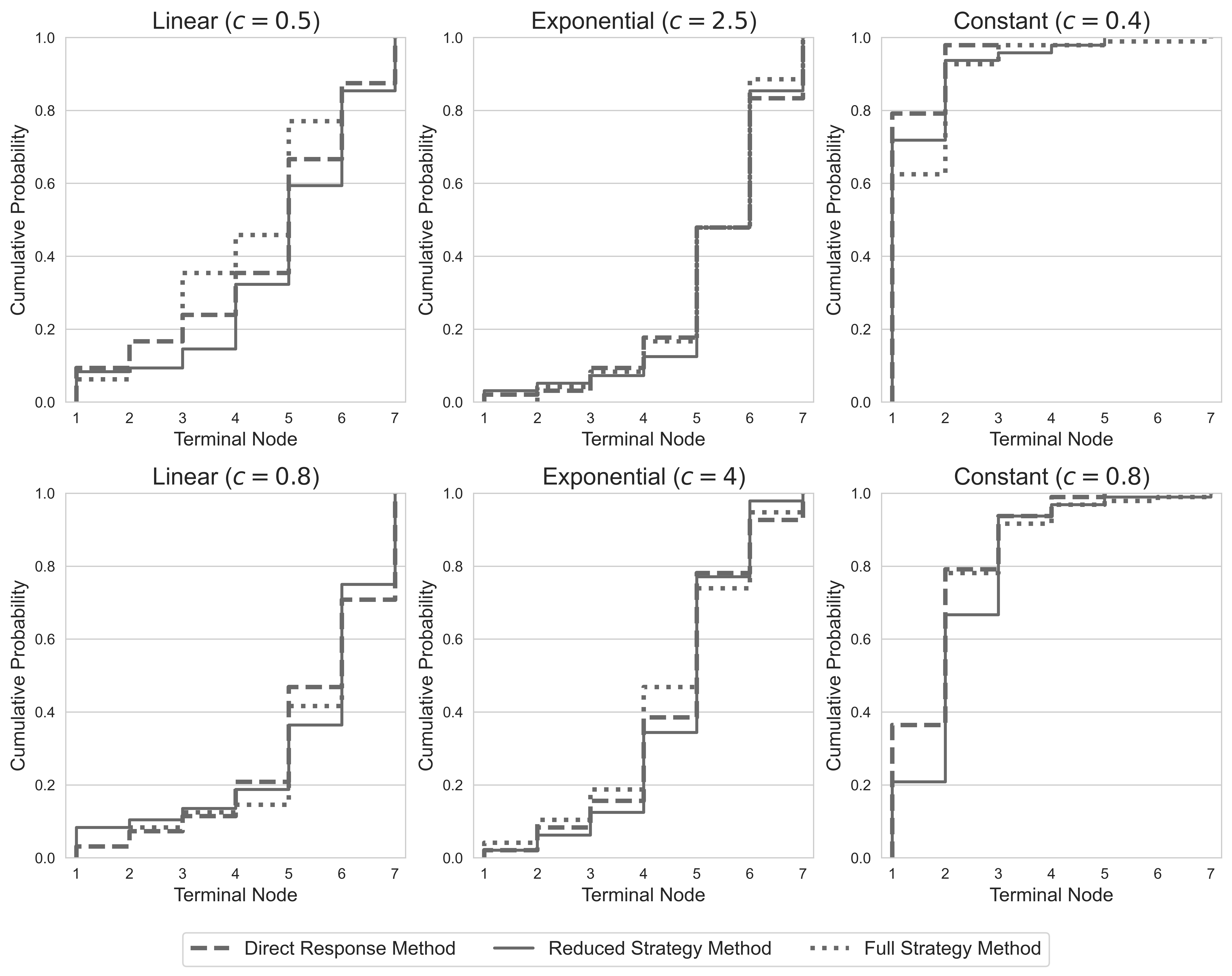}
    \caption{CDFs of Terminal Nodes for All Games and Elicitation Methods}
    \label{fig:pair_level_empirical_cdf}
\end{figure}

\begin{table}[H]\centering 
\caption{Mean Terminal Nodes and Signed-Rank Tests in Figure~\ref{fig:game_level_pairwise_comparison}}
\label{tab:end_node_data_signed_rank}
\renewcommand{\arraystretch}{1.4}
\begin{adjustbox}{width=\columnwidth,center}
\begin{threeparttable}
\begin{tabular}{lcccccc}
\hline
\multirow{2}{*}{} & \multicolumn{2}{c}{Linear} & \multicolumn{2}{c}{Exponential} & \multicolumn{2}{c}{Constant} \\
 & ($c=0.5$) & ($c=0.8$) & ($c=2.5$) & ($c=4$) & ($c=0.4$) & ($c=0.8$) \\ \hline
\multicolumn{7}{l}{\emph{Panel A. Mean Terminal Nodes Across Elicitation Methods}} \\
Direct Response Method & 4.604 & 5.396 & 5.365 & 4.646 & 1.229 & 1.917 \\
 & (1.765) & (1.544) & (1.284) & (1.315) & (0.467) & (0.898) \\
Reduced Strategy Method & 4.906 & 5.375 & 5.385 & 4.698 & 1.406 & 2.240 \\
 & (1.627) & (1.715) & (1.278) & (1.174) & (0.811) & (0.997) \\
Full Strategy Method & 4.312 & 5.490 & 5.344 & 4.510 & 1.510 & 2.000 \\
 & (1.710) & (1.541) & (1.171) & (1.422) & (0.890) & (1.109) \\ \hline
\multicolumn{7}{l}{\emph{Panel B. Pairwise Signed-Rank Tests $p$-Values}} \\
DR vs.~RS & 0.365 & 1.000 & 1.000 & 1.000 & 0.022 & 0.005 \\
DR vs.~FS & 0.486 & 1.000 & 1.000 & 1.000 & 0.001 & 1.000 \\
FS vs.~RS & 0.005 & 1.000 & 1.000 & 0.502 & 0.374 & 0.119 \\ \hline
\end{tabular} 
\begin{tablenotes}
\footnotesize
\item[1.] The standard deviations are reported in parentheses.
\item[2.] The bottom three rows report two-sided signed-rank test $p$-values, adjusted using the Bonferroni correction.
\end{tablenotes}
\end{threeparttable}
\end{adjustbox}
\end{table}

\begin{table}[H]\centering
\caption{Empirical Terminal Node Frequencies and Friedman Tests  
in Figure \ref{fig:pair_level_terminal_node_pdf}}
\label{tab:pair_level_terminal_node_pdf}
\renewcommand{\arraystretch}{1.25}
\begin{adjustbox}{width=\columnwidth,center}
\begin{threeparttable}
\begin{tabular}{cccccccccc}
\hline
 & \multirow{2}{*}{\begin{tabular}[c]{@{}c@{}}Friedman\\ $p$-values\end{tabular}} & \multirow{2}{*}{\begin{tabular}[c]{@{}c@{}}Eliciting\\ Methods\end{tabular}} & \multicolumn{7}{c}{Terminal Node Frequencies} \\ \cline{4-10} 
 &  &  & 1 & 2 & 3 & 4 & 5 & 6 & 7 \\ \hline
\multirow{3}{*}{\begin{tabular}[c]{@{}c@{}}Linear \\ ($c=0.5$)\end{tabular}} & \multirow{3}{*}{0.040} & DR & 0.094 & 0.073 & 0.073 & 0.115 & 0.312 & 0.208 & 0.125 \\
 &  & RS & 0.083 & 0.010 & 0.052 & 0.177 & 0.271 & 0.260 & 0.146 \\
 &  & FS & 0.062 & 0.104 & 0.188 & 0.104 & 0.312 & 0.104 & 0.125 \\ \hline
\multirow{3}{*}{\begin{tabular}[c]{@{}c@{}}Linear \\ ($c=0.8$)\end{tabular}} & \multirow{3}{*}{0.836} & DR & 0.031 & 0.042 & 0.042 & 0.094 & 0.260 & 0.240 & 0.292 \\
 &  & RS & 0.083 & 0.021 & 0.031 & 0.052 & 0.177 & 0.385 & 0.250 \\
 &  & FS & 0.031 & 0.052 & 0.042 & 0.021 & 0.271 & 0.292 & 0.292 \\ \hline
\multirow{3}{*}{\begin{tabular}[c]{@{}c@{}}Exponential \\ ($c=2.5$)\end{tabular}} & \multirow{3}{*}{0.692} & DR & 0.021 & 0.010 & 0.062 & 0.083 & 0.302 & 0.354 & 0.167 \\
 &  & RS & 0.031 & 0.021 & 0.021 & 0.052 & 0.354 & 0.375 & 0.146 \\
 &  & FS & 0.000 & 0.042 & 0.042 & 0.083 & 0.312 & 0.406 & 0.115 \\ \hline
\multirow{3}{*}{\begin{tabular}[c]{@{}c@{}}Exponential \\ ($c=4$)\end{tabular}} & \multirow{3}{*}{0.716} & DR & 0.021 & 0.062 & 0.073 & 0.229 & 0.396 & 0.146 & 0.073 \\
 &  & RS & 0.021 & 0.042 & 0.062 & 0.219 & 0.427 & 0.208 & 0.021 \\
 &  & FS & 0.042 & 0.062 & 0.083 & 0.281 & 0.271 & 0.208 & 0.052 \\ \hline
\multirow{3}{*}{\begin{tabular}[c]{@{}c@{}}Constant \\ ($c=0.4$)\end{tabular}} & \multirow{3}{*}{0.000} & DR & 0.792 & 0.188 & 0.021 & 0.000 & 0.000 & 0.000 & 0.000 \\
 &  & RS & 0.719 & 0.219 & 0.021 & 0.021 & 0.021 & 0.000 & 0.000 \\
 &  & FS & 0.625 & 0.302 & 0.052 & 0.000 & 0.010 & 0.000 & 0.010 \\ \hline
\multirow{3}{*}{\begin{tabular}[c]{@{}c@{}}Constant \\ ($c=0.8$)\end{tabular}} & \multirow{3}{*}{0.004} & DR & 0.365 & 0.427 & 0.146 & 0.052 & 0.010 & 0.000 & 0.000 \\
 &  & RS & 0.208 & 0.458 & 0.271 & 0.031 & 0.021 & 0.000 & 0.010 \\
 &  & FS & 0.365 & 0.417 & 0.135 & 0.052 & 0.010 & 0.010 & 0.010 \\ \hline
\end{tabular}
\begin{tablenotes}
\footnotesize
\item The column ``Friedman $p$-values'' reports the $p$-values from Friedman tests, which test the null hypothesis that the distributions of terminal nodes are identical across the three elicitation methods.
\end{tablenotes}
\end{threeparttable}
\end{adjustbox}
\end{table}

\subsection{Estimation Details for AQRE and QDCH}
\label{appendix:structural_estimation}

This appendix describes the estimation procedures for AQRE and QDCH.  
Since DCH is nested within QDCH, the estimation procedure for DCH is a special case of estimating QDCH, which we discuss at the end.

\bigskip
\noindent\textbf{Data} \; Before we dive into the estimation procedure, we first describe the 
observed data under each elicitation method. Under the full strategy method, each player
in each game is asked to submit one of eight full strategies, 
corresponding to choosing either take ($T$) or pass ($P$) at each of their 
own decision nodes. Thus, each player’s action set under the full strategy method is:
\begin{align*}
\left\{TTT, TTP, TPT, TPP, PTT, PTP, PPT, PPP\right\}.
\end{align*}
In contrast to the full strategy method, under the reduced strategy method, each player in each game 
is asked to submit one of four (structurally) reduced strategies. Since each player has 
three decision nodes and the game ends as soon as someone chooses to take, each player’s
reduced strategy corresponds to taking at their first, second, or third decision node, or never taking. 
The action set is thus given by:
\begin{align*}
\left\{T, PT, PPT, PPP\right\}.
\end{align*}

Lastly, under the direct response method, the centipede game is played sequentially 
in its extensive form. As a result, we observe only the pass/take choices at the decision
nodes along the realized histories.

\bigskip
\noindent\textbf{Estimation for Logit-AQRE} \; The logit-AQRE is defined as the solution 
to a system of nonlinear equations where all players make logit quantal responses. Specifically, for any history $h$ and any action $a$ available at 
$h$, the probability of choosing $a$ in logit-AQRE 
is given by: 
$$\sigma_i(a|h) = \frac{e^{\lambda \bar{u}_a}}{\sum_{a'\in A(h)} e^{\lambda \bar{u}_{a'}}} $$
where $A(h)$ is the set of available actions at history $h$, $\bar{u}_{a'}$ is the equilibrium (expected) continuation value of choosing any $a'\in A(h)$, and $\lambda \in [0, \infty)$ is the precision parameter.

For each participant $i$, let $H_i$ denote the set of histories (or centipede games) that participant
$i$ encountered during the experiment. 
When the centipede game is played under the full or reduced strategy method, the only 
history is the empty history, and the logit-AQRE reduces to the logit-QRE, which specifies
the choice probabilities over all eight full strategies or all four reduced
strategies for both Player 1 and Player 2, respectively. Yet when the game is played under the 
direct response method, there are six histories, corresponding to six terminal nodes.

We use $\mathcal{A}(a_i |h_i, \lambda)$ to denote the choice probability
of action $a_i$ at history $h_i$. This represents the probability of choosing a
full or reduced strategy $a_i$ when the game is played under the full or
reduced strategy method, respectively; and the probability of choosing $a_i \in \{P, T\}$ under 
the direct response method.
Accordingly, the log-likelihood function for logit-AQRE can be constructed by summing over all participants $i$, all histories $h_i$, and all actions $a_i$:
\begin{align*}
    \ln \mathcal{L}^{\mbox{AQRE}}(\lambda) = \sum_i \sum_{h_i }\sum_{a_i} \mathbf{1}\{a_i, h_i\}
    \ln[\mathcal{A}(a_i | h_i, \lambda)]
\end{align*}
where $\mathbf{1}\{a_i, h_i\}$ is an indicator function equal to 1 if participant $i$ 
chooses action $a_i$ at history $h_i$, and 0 otherwise.

\bigskip
\noindent\textbf{Estimation for QDCH} \; The Quantal Dynamic Cognitive Hierarchy (QDCH) Solution
is a natural extension of DCH which assumes all strategic levels of players make quantal responses 
instead of best responses. Therefore, QDCH is a two-parameter model where 
$\lambda \in [0, \infty)$ is the precision parameter and the distribution of levels follows a 
Poisson($\tau$).

We use $\tilde{\sigma}_i^k(a_i | h_i, \tau, \lambda)$ to 
denote the probability that a level-$k$ player $i$ chooses action $a_i$ at 
history $h_i \in H_i$. This is uniquely determined, as the QDCH solution is unique. 
Furthermore, let $f(k | h_i, \tau, \lambda)$ denote the posterior distribution of
levels at history $h_i$. 
The choice probability for action $a_i$ at history 
$h_i$ predicted by QDCH is then given by the aggregation of choice 
probabilities across all levels, 
weighted by the posterior distribution $f(k | h_i, \tau, \lambda)$:
\begin{align*}
\mathcal{Q}(a_i | h_i, \tau, \lambda) \equiv \sum_k  f(k | h_i, \tau, \lambda)\tilde{\sigma}_i^k(a_i|h_i, \tau, \lambda).
\end{align*}
Consequently, the log-likelihood function for QDCH can be constructed by summing over all participants $i$, all histories $h_i$, and all actions $a_i$:
\begin{align*}
    \ln \mathcal{L}^{\mbox{QDCH}}(\tau, 
    \lambda) = \sum_i \sum_{h_i }\sum_{a_i} \mathbf{1}\{a_i, h_i\}
    \ln[\mathcal{Q}(a_i | h_i, \tau, \lambda)]
\end{align*}
where $\mathbf{1}\{a_i, h_i\}$ is an indicator function equal to 1 if participant $i$ 
chooses action $a_i$ at history $h_i$, and 0 otherwise.

Note that DCH is nested in QDCH, as QDCH converges to DCH as $\lambda \rightarrow \infty$.  
Moreover, when estimating QDCH using the main data we cap the level at 50 to ensure more precise estimation.  
When estimating the pilot data, we cap the level at 10 since the proportion of levels above 10 is negligible.

\bigskip
\noindent\textbf{Computing Standard Errors} \; We compute standard errors for the estimates using 1,000 
bootstrap replications. For the data from the full and reduced strategy methods, each replication resamples 
the observed full and reduced strategies with replacement from the original data, respectively. 
For the direct response method, due to the incomplete elicitation of strategies, each
replication resamples the observed terminal nodes with replacement. We then recover the decision 
node choices from the terminal nodes to perform maximum likelihood estimation.

To avoid convergence to local maxima, we estimate the original model parameters using 
a global search procedure. However, due to computational constraints, performing 
a global search in each of the 1,000 bootstrap replications is infeasible. Instead, 
for each replication, we apply a standard optimization procedure using the globally
estimated original parameters as initial values. The bootstrap standard errors are 
then computed as the standard deviations of the resulting estimates across replications.
As a result, the reported standard errors are conservative estimations, as they may be overestimated.

\subsection{Optimal Design Details}
\label{appendix:game_selection}

This appendix provides details of our optimal design procedure. Section~\ref{subsec:appendix_pilot} presents the analysis of our pilot data. Section~\ref{subsec:appendix_optimal_selection} then offers a detailed description of the optimal game selection procedure, along with a robustness analysis of our selection. Finally, Section~\ref{subsec:appendix_sensitivity} revisits our optimal design approach using the Quantal Dynamic Cognitive Hierarchy (QDCH) solution.

\subsubsection{Pilot Data Analysis}
\label{subsec:appendix_pilot}

\begin{table}[htbp!]
\centering
\caption{Mean Terminal Nodes and DCH Estimation Results from the Pilot Data}
\label{tab:pilot_data}
\renewcommand{\arraystretch}{1.25}
\begin{adjustbox}{width=0.9\columnwidth,center}
\begin{threeparttable}
\begin{tabular}{rccc}
\hline
\multicolumn{1}{c}{} & \begin{tabular}[c]{@{}c@{}}Direct Response\\ Method\end{tabular} & \begin{tabular}[c]{@{}c@{}}Reduced Strategy\\ Method\end{tabular} & Pooled Data \\ \hline
\multicolumn{4}{l}{\emph{Panel A. Mean Terminal Nodes$^1$}} \\
Linear ($c=0.40$) & 4.00 & 5.25 & 4.63 \\
Linear ($c=0.60$) & 6.50 & 5.25 & 5.88 \\
Linear ($c=0.75$) & 6.25 & 5.75 & 6.00 \\
Linear ($c=0.90$) & 6.50 & 5.75 & 6.13 \\ \hline
\multicolumn{4}{l}{\emph{Panel B. Structural Estimation of DCH}} \\
$\tau$ & 1.14 & 1.75 & 1.25 \\
S.E. & (0.19) & (0.37) & (0.16) \\
 &  &  &  \\
Observations$^2$ & 88 & 32 & 120 \\
Log-Likelihood & -32.71 & -23.72 & -57.05 \\
Likelihood Ratio Test$^3$ & --- & --- & 1.24 \\
$p$-value & --- & --- & 0.26 \\ \hline
\end{tabular}
\begin{tablenotes}
\footnotesize
\item[1.] Each treatment included eight participants, and each participant played each game once. Therefore, there are only four observations of (inferred) terminal nodes per game in each treatment.
\item[2.] The estimation for the direct response method treatment 
is based on 88 decision-node choices, while the estimation for the reduced strategy method
treatment is based on 32 choices of reduced strategies. 
\item[2.] We perform a likelihood ratio test to evaluate the null hypothesis that the 
estimated $\tau$ values from both treatments are equal.
\end{tablenotes}
\end{threeparttable}
\end{adjustbox}
\end{table}

Table~\ref{tab:pilot_data} summarizes the analysis of our pilot experiment. Panel A reports the mean 
(inferred) terminal nodes for each treatment and for the pooled data. Since each treatment included only eight 
participants, and each participant played each game once, there are just four observations of terminal nodes 
per game in each treatment. As a result, no statistical tests are conducted due to the limited sample size.

Furthermore, we calibrate DCH using maximum likelihood estimation, following the procedure 
described in Section~\ref{subsec:optimal_design} and Appendix~\ref{appendix:structural_estimation}. 
Panel B of Table~\ref{tab:pilot_data} reports the results, including estimates
based on the pooled data and those from each treatment.

Our optimal design approach implicitly assumes that the distribution of levels is 
independent of the elicitation method. To test this assumption, we estimate the parameter 
separately for each treatment and conduct a likelihood ratio test of the null hypothesis
that the estimated $\tau$ values are equal across treatments. As shown in Panel B of 
Table~\ref{tab:pilot_data}, the null is not rejected (Likelihood ratio test: $\chi^2_{(1)} = 1.24$,
$p$-value $= 0.26$), suggesting that this assumption is supported by our pilot data.

\subsubsection{Robustness of Optimal Game Selection}
\label{subsec:appendix_optimal_selection}

The goal of our optimal design approach is to select two games from each class: one expected 
to exhibit a large strategy method effect, as predicted by the calibrated DCH, and another 
with a small effect to serve as a quasi-placebo. In this appendix, we demonstrate the robustness 
of our game selection by showing that (1) the predicted strategy method effects for the 
selected games remain stable across a range of $\tau$ values around the 
estimated $\hat{\tau} = 1.25$, and (2) the games selected for large effects 
are expected to yield statistically significant differences between elicitation methods.
Finally, we describe how we rescale the payoff parameters when implementing these games 
in the experiment to ensure stable expected earnings across all games.

\subsubsection*{A.4.2.1\;\;\;\;Sensitivity Analysis with Respect to $\hat{\tau}$}

Since we calibrate DCH using small-scale pilot data, our estimates of
the sweet spots that generate large strategy method effects may lack precision. 
To assess the robustness of our game selection, given our estimate 
of $\hat{\tau} = 1.25$ with a standard error of 0.16, we plot the 
sup-norm functions for $\tau = 1$ and $\tau = 1.5$ (for each class of games),
which are approximately 1.68 standard errors from $\hat{\tau}$. As we show 
below, the games selected for large effects indeed yield large strategy method
effects at these alternative values of $\tau$.

\begin{figure}[H]
    \centering
    \includegraphics[width=\linewidth]{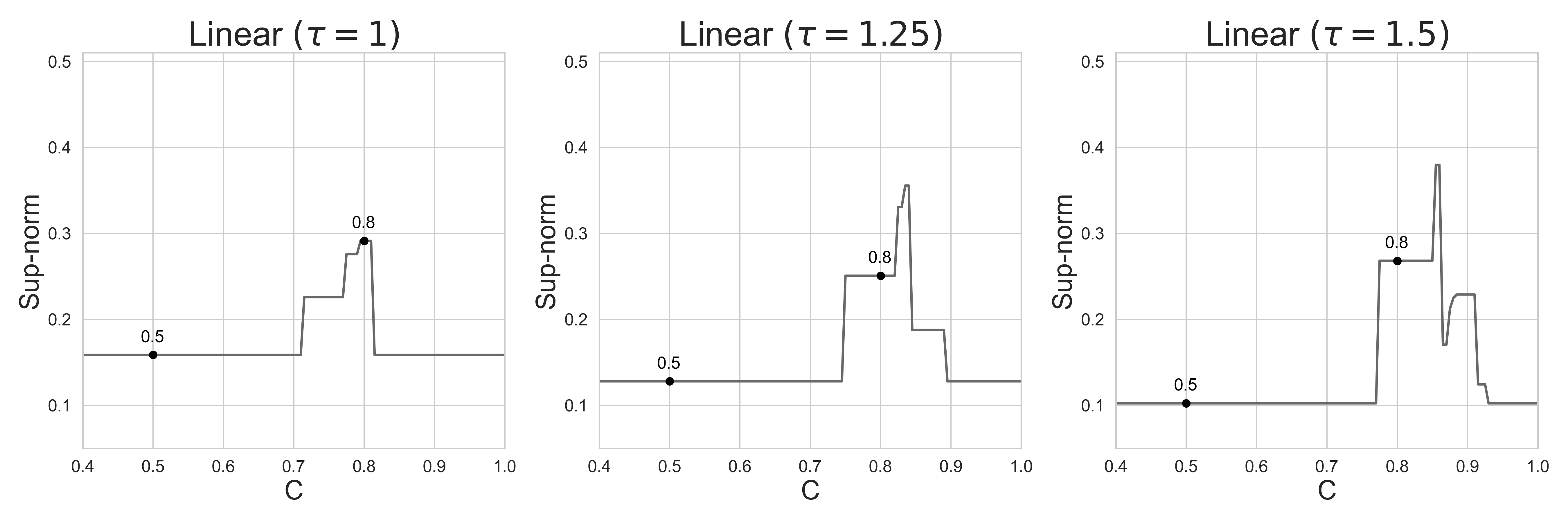}
    \caption{The sup-norm function $\mathcal{S}(c)$ for linear games at $\tau = 1$ 
    (left), $\tau = 1.25$ (middle), and $\tau = 1.5$ (right).}
    \label{fig:Linear_supnorm}
\end{figure}

\bigskip
\noindent\textbf{Linear Centipede Games} \; Figure~\ref{fig:Linear_supnorm} plots 
the sup-norm function $\mathcal{S}(c)$ for linear games under three values
of $\tau$: 1, 1.25, and 1.5. For the linear game with a smaller predicted strategy
method effect, we select $c = 0.5$ to yield more rounded payoffs, thereby enhancing 
comprehensibility. As shown in the figure, $c = 0.5$ consistently produces small effects 
across all three values of $\tau$. In contrast, for the linear game with a larger 
predicted effect, we choose $c = 0.8$, which offers reasonably rounded payoffs 
and stable predicted magnitudes of the strategy method effect across these values of $\tau$.

\begin{figure}[H]
    \centering
    \includegraphics[width=\linewidth]{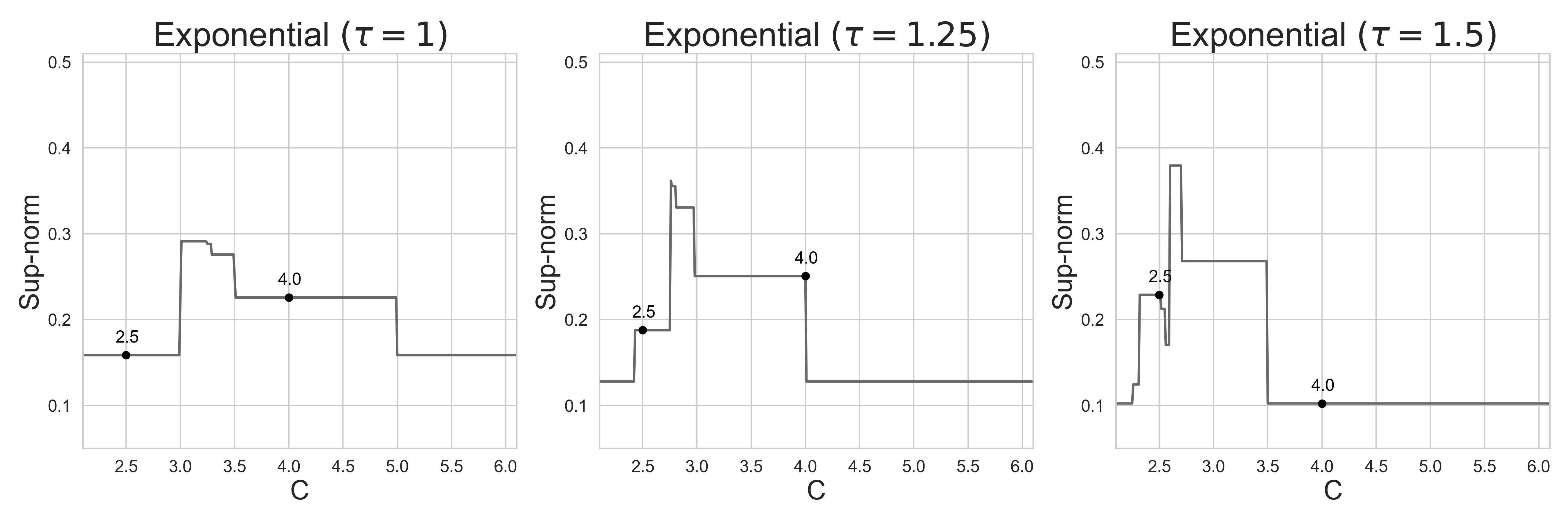}
    \caption{The sup-norm function $\mathcal{S}(c)$ for exponential games at $\tau = 1$ 
    (left), $\tau = 1.25$ (middle), and $\tau = 1.5$ (right).}
    \label{fig:Exponential_supnorm}
\end{figure}

\noindent\textbf{Exponential Centipede Games} \;
For exponential games, recall that we set the multiplier $\pi = 2$ to prevent 
payoffs from becoming excessively large in the later stages.
Figure~\ref{fig:Exponential_supnorm} plots the sup-norm function $\mathcal{S}(c)$ for 
exponential games under three values of $\tau$: $1$, $1.25$, and $1.5$.
For the exponential game with a smaller predicted strategy method effect, we select $c = 2.5$ 
due to budget constraints---even though a larger value, such as $c = 5.5$, yields a slightly 
more stable magnitude of the effect.
For the game with a larger predicted strategy method effect, we select $c = 4$. Although $c = 3.5$ 
is slightly more stable across values of $\tau$, we choose $c = 4$ for its comparability 
with prior literature (e.g., \citealp{mckelvey_experimental_1992, garcia-pola_hot_2020}).

\begin{figure}[H]
    \centering
    \includegraphics[width=\linewidth]{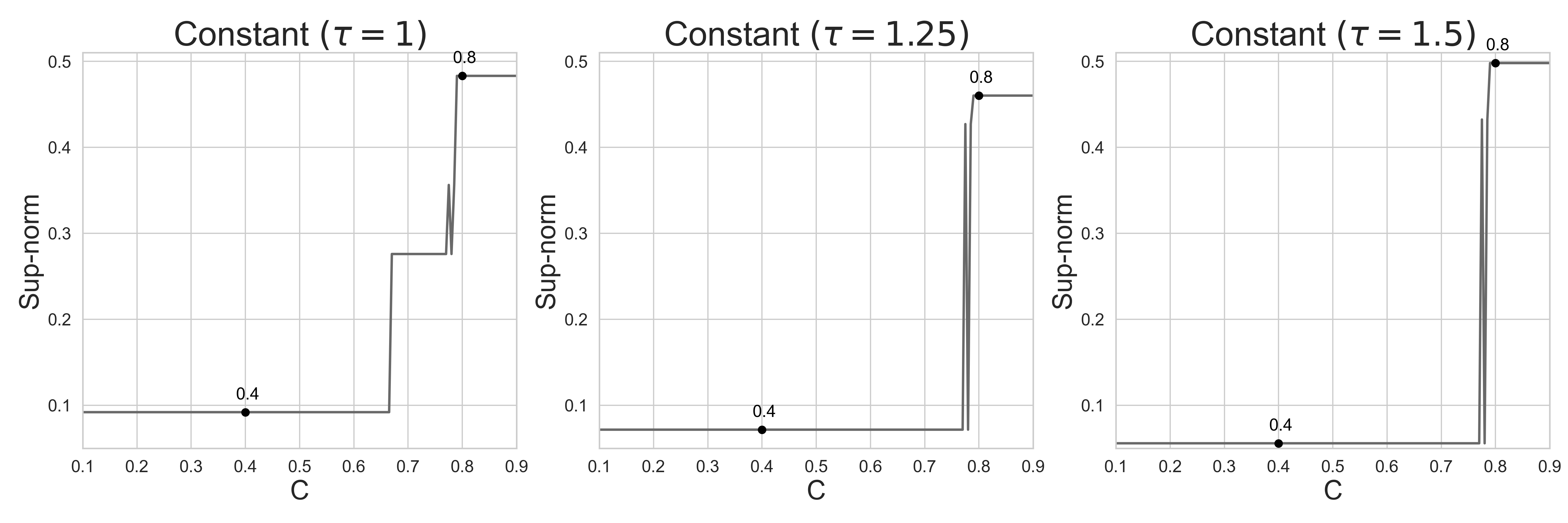}
    \caption{The sup-norm function $\mathcal{S}(c)$ for constant games at $\tau = 1$ 
    (left), $\tau = 1.25$ (middle), and $\tau = 1.5$ (right).}
    \label{fig:Constant_supnorm}
\end{figure}

\noindent\textbf{Constant Centipede Games} \; 
Figure~\ref{fig:Constant_supnorm} displays the sup-norm function $\mathcal{S}(c)$ for 
constant games under each of the three values of $\tau$: $1$, $1.25$, and $1.5$. 
As shown in the figure, the sup-norm is (almost) monotonically increasing in $c$ across 
these values of $\tau$. This implies that DCH predicts larger effects in games with 
higher values of $c$. To avoid selecting a large game too close to the boundary, 
we choose $c = 0.8$. Similarly, for the small game, we select $c = 0.4$, which yields 
reasonably rounded payoffs while remaining well within the interior of the parameter space.

\subsubsection*{A.4.2.2\;\;\;\;Statistical Significance of Selected Games}

In this appendix, we assess whether our selected large and small games are 
sufficiently different such that the large games are expected to yield statistically 
significant effects, while the small games are not. We use the two-sample Kolmogorov–Smirnov (KS)
test as our benchmark because its test statistic is based on the sup-norm and, more importantly,
it serves as a \emph{conservative} benchmark due to its low power.

The expected sup-norm predicted by the calibrated DCH for each game is reported in 
Table~\ref{tab:RN_E_prediction-full}.
We then compute the expected $p$-values of the two-sample KS test using these predicted sup-norms 
and a planned sample size of 192 participants, yielding 96 terminal nodes per
game under each elicitation method.
Specifically, the $p$-values are calculated as follows:
\begin{align*}
    p = 2\sum_{j = 1}^{\infty}(-1)^{j-1}\exp(-2j^2\lambda^2) \quad\mbox{where}\quad
    \lambda = \sqrt{48}\mathcal{S},
\end{align*}
$\mathcal{S}$ denotes the expected sup-norm, and 48 is the effective sample size\footnote{Let $n$ and $m$ be the sample sizes of the two distributions. The effective sample size for the two-sample KS test is $\frac{nm}{n+m}$. Given $n = m = 96$ in our experiment, the effective sample size is $\frac{96 \times 96}{96 + 96} = 48$.} used in the two-sample KS test. Based on this calculation, all large games are expected to yield significant elicitation method effects at the 1\% level, while the small games are not significant, 
as reported in Table~\ref{tab:RN_E_prediction-full}.

\begin{table}[htbp!]
\centering
\caption{Predicted vs. Empirical Sup-norms and KS Test $p$-values (DR vs. RS)}
\label{tab:RN_E_prediction-full}
\renewcommand{\arraystretch}{1.35}
\begin{threeparttable}
\begin{tabular}{rcccccccc}
\hline
\multicolumn{1}{c}{} & \multicolumn{2}{c}{Linear} &  & \multicolumn{2}{c}{Exponential} &  & \multicolumn{2}{c}{Constant} \\ \cline{2-3} \cline{5-6} \cline{8-9} 
\multicolumn{1}{c}{} & $c=0.5$ & $c=0.8$ &  & $c=2.5$ & $c=4$ &  & $c=0.4$ & $c=0.8$ \\
\multicolumn{1}{c}{} & (small) & (large) &  & (small) & (large) &  & (small) & (large) \\ \hline
\multicolumn{9}{l}{\textit{Sup-norm} (\emph{Direct Response vs. Reduced Strategy Methods})} \\
Predicted & 0.128 & 0.251 &  & 0.188 & 0.251 &  & 0.072 & 0.460 \\
Empirical & 0.094 & 0.104 &  & 0.052 & 0.052 &  & 0.073 & 0.156 \\ \hline
\multicolumn{9}{l}{\textit{Kolmogorov-Smirnov Test $p$-value}} \\
\;\;\;\;Predicted & 0.413 & \phantom{$^\dagger$}0.005$^\dagger$ &  & 0.068 & \phantom{$^\dagger$}0.005$^\dagger$ &  & 0.966 & \phantom{$^\dagger$}0.000$^\dagger$\\
Empirical & 0.796 & 0.678 &  & 1.000 & 1.000 &  & 0.962 & 0.192 \\ \hline
\end{tabular}
\begin{tablenotes}
\footnotesize
\item $\dagger$ indicates that the game is expected to yield statistically significant 
strategy method effects under the Kolmogorov–Smirnov test 
at the 1\% significance level, given a sample of 192 participants.
\end{tablenotes}
\end{threeparttable}
\end{table}

To compare these predictions with the experimental data from the main sessions, we compute the empirical 
sup-norm between the terminal node CDFs under the direct response and reduced strategy
methods. The (inferred) terminal nodes for the reduced strategy method are 
calculated by assuming that players are paired exactly as they were in the direct response 
method. The sup-norm for each game, along with the corresponding $p$-value from the two-sample
KS test, is reported in Table~\ref{tab:RN_E_prediction-full}.
From the table, we can find that despite the model predictions, the realized differences are smaller in magnitude and statistically indistinguishable, given the limited power of the two-sample KS test.

\subsubsection*{A.4.2.3\;\;\;\;Rescaling the Payoffs}

When implementing these optimally designed games, we rescale the payoffs to eliminate 
decimal numbers and reduce potential cognitive burdens for participants. Specifically, 
for constant games, we multiply all payoffs by 250 to ensure that each payoff is at
least one point. Under this transformation, the total payoff for both players in each 
constant game is 500 points. Our goal is then to rescale the payoffs in the other 
four games so that expected earnings are comparable across all games.

For linear games, we rescale the payoffs using the transformation 
$f(x) = 100x+50$, so that in the small linear game, the total payoff at the fourth
terminal node is exactly 500 points. Lastly, for exponential games, to prevent payoffs
from exploding in the later stages, we apply a more conservative rescaling 
by multiplying the payoffs by 4, resulting in a total payoff of approximately 
500 points at the sixth terminal node of the small exponential game.
The final game trees after rescaling are shown in Figure~\ref{fig:six_cg}.

In short, we rescale the payoffs to ensure that expected earnings are comparable across 
all games. Since only one game from each treatment is randomly selected for payment, this 
rescaling effectively reduces the variance in realized payments. Moreover, because 
the logit quantal response precision parameter $\lambda$
is sensitive to the scale of payoffs, rescaling also ensures that the estimated 
$\lambda$ comparable across different families of games.

\subsubsection{QDCH Calibration}
\label{subsec:appendix_sensitivity}

In this section, we assess how much incorporating quantal responses into the DCH solution
improves our optimal design approach and predictions about the 
relative magnitudes of the strategy method effect.   
Specifically, we calibrate QDCH on the pilot data to generate qualitative predictions 
for the relative magnitudes of the strategy method effect across our six 
optimally selected centipede games. This exercise is to evaluate the out-of-sample
predictability of QDCH using the pilot data as training data.

\begin{figure}[H]
    \centering
    \includegraphics[width=\linewidth]{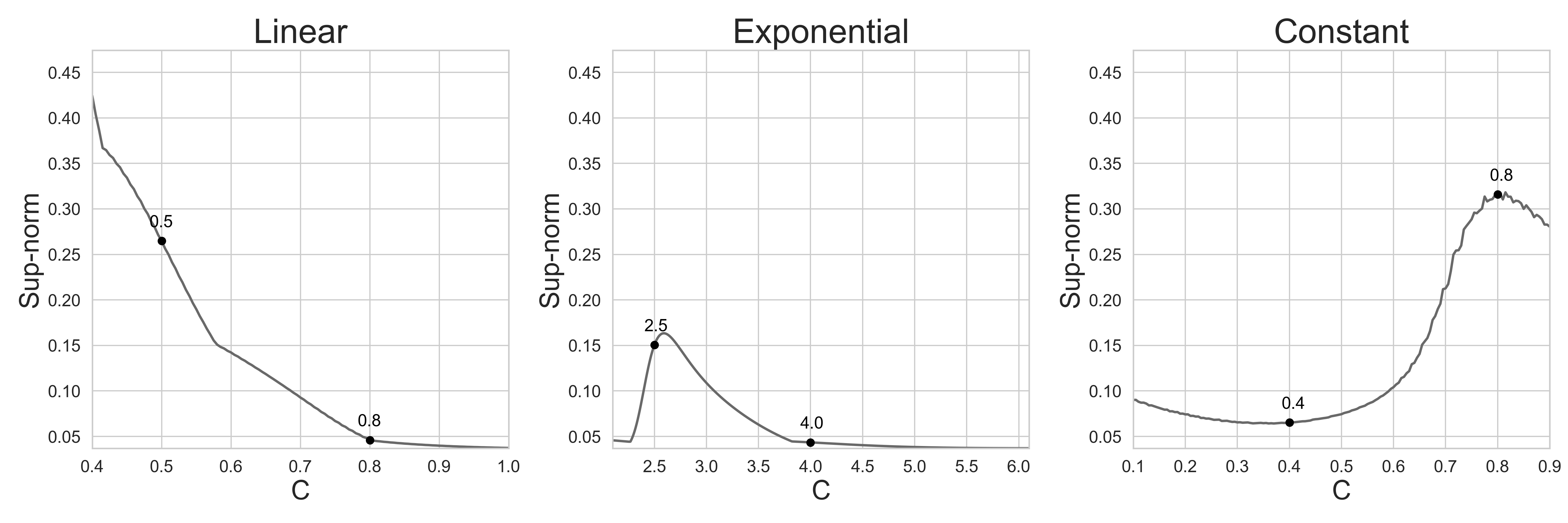}
    \caption{The sup-norm function $\mathcal{S}(c)$ for linear (left), exponential (middle), and constant (right) centipede games, based on predictions from the calibrated QDCH.}
    \label{fig:optimal_supnorm_Q}
\end{figure}

Following the same calibration procedure on the pilot data, we calibrate QDCH with $\hat{\tau} = 2.60$ and $\hat{\lambda} = 0.05$ 
(Obs. = 120, S.E. of $\hat{\tau} = 0.76$, S.E. of $\hat{\lambda} = 0.01$, log-likelihood $= -45.29$).  
Using these estimates, we compute the CDFs predicted by the calibrated QDCH for each class of games under the three elicitation methods, and then calculate the sup-norm distances between these distributions.
The sup-norm functions are shown in Figure~\ref{fig:optimal_supnorm_Q}, and the sup-norm 
distances for the selected games are reported in Table~\ref{tab:RN_E_prediction_QDCH}.

In general, the predictions of QDCH vary across the three elicitation
methods due to the effect of quantal responses. Accordingly, 
Table~\ref{tab:RN_E_prediction_QDCH} reports the pairwise sup-norm distances 
between the distributions under each method.
From the table, we can first see  that despite the presence of 
quantal response effects, the calibrated QDCH, like DCH, does not predict 
any significant difference between the direct response method and the full strategy method.

\begin{table}[htbp!]
\centering
\caption{The Predicted Sup-norm of the Selected Centipede Games by QDCH}
\label{tab:RN_E_prediction_QDCH}
\renewcommand{\arraystretch}{1.2}
\begin{threeparttable}
\begin{tabular}{ccccccccc}
\hline
 & \multicolumn{2}{c}{Linear} &  & \multicolumn{2}{c}{Exponential} &  & \multicolumn{2}{c}{Constant} \\ \cline{2-3} \cline{5-6} \cline{8-9} 
 & $c=0.5$ & $c=0.8$ &  & $c=2.5$ & $c=4$ &  & $c=0.4$ & $c=0.8$ \\
 & (small) & (large) &  & (small) & (large) &  & (small) & (large) \\ \hline
\multirow{2}{*}{Sup-norm (RS vs. DR)} & \multirow{2}{*}{\textbf{0.265$^\dagger$}} & \multirow{2}{*}{0.046} & \multirow{2}{*}{} & \multirow{2}{*}{0.151} & \multirow{2}{*}{0.043} & \multirow{2}{*}{} & \multirow{2}{*}{0.065} & \multirow{2}{*}{\textbf{0.316$^\dagger$}} \\
 &  &  &  &  &  &  &  &  \\
\multirow{2}{*}{Sup-norm (RS vs. FS)} & \multirow{2}{*}{\textbf{0.430$^\dagger$}} & \multirow{2}{*}{0.143} & \multirow{2}{*}{} & \multirow{2}{*}{0.118} & \multirow{2}{*}{0.166} & \multirow{2}{*}{} & \multirow{2}{*}{0.091} & \multirow{2}{*}{\textbf{0.367$^\dagger$}} \\
 &  &  &  &  &  &  &  &  \\ 
\multirow{2}{*}{Sup-norm (FS vs. DR)} & \multirow{2}{*}{0.198\phantom{*}} & \multirow{2}{*}{0.097} & \multirow{2}{*}{} & \multirow{2}{*}{0.192} & \multirow{2}{*}{0.203} & \multirow{2}{*}{} & \multirow{2}{*}{0.026} & \multirow{2}{*}{0.051\phantom{*}} \\
 &  &  &  &  &  &  &  &  \\ 
 \hline
\end{tabular}
\begin{tablenotes}
\footnotesize
\item $\dagger$ indicates that the game is expected to yield statistically significant 
strategy method effects under the Kolmogorov–Smirnov test 
at the 1\% significance level, given a sample of 192 participants.
\end{tablenotes}
\end{threeparttable}
\end{table}

However, QDCH makes strikingly different qualitative predictions compared to DCH when comparing the direct response and full strategy methods to the reduced strategy method.  
Rather than predicting stronger strategy method effects in all three Large games, as DCH does, the calibrated QDCH qualitatively predicts that the strategy method effects in the Small Linear game ($c = 0.5$) and the Large Constant game ($c = 0.8$) are stronger than those in the remaining four games.  

\bigskip
\noindent\textbf{QDCH Strategy Method Effect:} \emph{The calibrated QDCH predicts that
\begin{itemize}
    \item[1.] there is no significant difference between the direct response method and 
    the full strategy method across all six games;
    \item[2.] the differences between the direct response method/full strategy method
    and the reduced strategy method are larger in the Small Linear game and
    the Large Constant game than in the other four games.
\end{itemize}}
\bigskip

To evaluate this prediction, we reproduce the analysis of the relative magnitudes of the effects in Table~\ref{tab:diff_method_effect_HvsL} by grouping the Small Linear game ($c = 0.5$) and the Large Constant game ($c = 0.8$) 
as games with large effects, denoted $\mbox{Large}^Q$, and classifying the remaining four games as games with small effects, denoted $\mbox{Small}^Q$.  
We then compute the mean differences in terminal nodes between elicitation methods for both groups, denoted $\Delta_L^Q$ and $\Delta_S^Q$, and use $\Delta_L^Q - \Delta_S^Q$ as a measure of the relative magnitude of the strategy method effects.  
These quantities are reported in Table~\ref{tab:diff_method_effect_HvsL_QDCH}.

\begin{table}[htbp!]
\centering
\caption{Relative Magnitudes of Strategy Method Effects (Grouped Under QDCH)}
\label{tab:diff_method_effect_HvsL_QDCH}
\renewcommand{\arraystretch}{1.35}
\begin{threeparttable}
\begin{tabular}{cccccc}
\hline
 &  & \multicolumn{2}{c}{$\Delta$ in Terminal Nodes} &  &  \\ \cline{3-4}
 &  & $\Delta_L^{Q}$ & $\Delta_S^{Q}$ &  & $\Delta_L^{Q} - \Delta_S^{Q}$ \\ \hline
\multirow{2}{*}{$\mbox{RS} - \mbox{DR}$} & \multirow{2}{*}{} & 0.312 & 0.057 &  & 0.255 \\
 &  & $p=0.003$ & $p=0.355$ &  & $p=0.057$ \\
& & & & & \\
\multirow{2}{*}{$\mbox{RS} - \mbox{FS}$} & \multirow{2}{*}{} & 0.417 & 0.003 &  & 0.414 \\
 &  & $p=0.000$ & $p=0.681$ &  & $p=0.003$ \\
 & & & & & \\
\multirow{2}{*}{$\mbox{FS} - \mbox{DR}$} & \multirow{2}{*}{} & -0.104\phantom{-} & 0.055 &  & -0.159\phantom{-} \\
 &  & $p=0.440$ & $p=0.488$ &  & $p=0.255$ \\ \hline
\end{tabular}
\begin{tablenotes}
\footnotesize
\item Statistical inferences of $\Delta_L^{Q}$ and $\Delta_S^{Q}$ are based on the signed-rank test, while those of $\Delta_L^{Q} - \Delta_S^{Q}$ are based on the rank-sum test, with $p$-values reported below. 
\end{tablenotes}
\end{threeparttable}
\end{table}

From Table~\ref{tab:diff_method_effect_HvsL_QDCH}, we first observe that, when comparing the full strategy method to the direct response method, neither $\Delta_L^Q$, $\Delta_S^Q$, nor $\Delta_L^Q - \Delta_S^Q$ differs significantly from zero, consistent with the prediction of QDCH.  
Furthermore, when comparing the reduced strategy method with the direct response method/full strategy method, we find significant strategy method effects in the $\mbox{Large}^Q$ games, but not in the $\mbox{Small}^Q$ games.  
More importantly, the relative magnitudes of these effects differ significantly across the two groups of games: $\Delta_L^Q - \Delta_S^Q = 0.255$ for the RS vs. DR comparison (rank-sum test, $p$-value = 0.057) and $\Delta_L^Q - \Delta_S^Q = 0.414$ for the RS vs. FS comparison 
(rank-sum test, $p$-value = 0.003), consistent with the qualitative predictions of the calibrated QDCH.  
These findings indicate that the relative magnitudes of the observed strategy method effects align with the predictions of the calibrated QDCH.

\begin{result}
For the comparison between the direct response method/full strategy method 
and the reduced strategy method, the strategy method effects are significantly
stronger in the small linear game and the large constant game than in the other four games, 
consistent with the qualitative prediction of the calibrated QDCH.
\end{result}

\subsection{Additional Robustness Analyses}
\label{appendix:additional_results}

\subsubsection{Order Effect}

An important robustness check in our within-subject design is to assess whether the sequence in which the elicitation methods were implemented had any influence on participants’ behavior. In the experiment, we deliberately alternated the order of the two strategy-method treatments across sessions---full strategy (FS) and reduced strategy (RS)---while consistently implementing the direct response (DR) treatment last. This design choice was intended to avoid \textit{feedback contamination}: no feedback was provided between games or between the first two treatments, and the direct response was placed at the end because sequential play inherently reveals the outcome to participants. 

\begin{table}[htbp!]
\centering
\caption{Comparison Between the FR Order and the RF Order}
\label{tab:aggregate_level_tests_two_orders}
\renewcommand{\arraystretch}{1.5}
\begin{adjustbox}{width=\columnwidth,center}
\begin{threeparttable}
\begin{tabular}{lccccccccccc}
\hline
 & \multicolumn{3}{c}{Direct Response Method} &  & \multicolumn{3}{c}{Reduced Strategy Method} &  & \multicolumn{3}{c}{Full Strategy Method} \\ \cline{2-4} \cline{6-8} \cline{10-12} 
Mean Terminal Nodes & FR Order & RF Order & $p$-values &  & FR Order & RF Order & $p$-values &  & FR Order & RF Order & $p$-values \\ \hline
Pooled Data & 3.812 & 3.906 & 0.653 &  & 4.056 & 3.948 & 0.543 &  & 3.851 & 3.872 & 0.918 \\
\multicolumn{1}{r}{} & (2.110) & (2.084) &  &  & (2.011) & (2.085) &  &  & (2.026) & (2.070) &  \\ \hline
Small Linear & 4.604 & 4.604 & 0.940 &  & 4.917 & 4.896 & 0.946 &  & 4.167 & 4.458 & 0.412 \\
\multicolumn{1}{r}{} & (1.717) & (1.811) &  &  & (1.525) & (1.723) &  &  & (1.783) & (1.620) &  \\
Large Linear & 5.271 & 5.521 & 0.618 &  & 5.667 & 5.083 & 0.407 &  & 5.562 & 5.417 & 0.918 \\
\multicolumn{1}{r}{} & (1.655) & (1.414) &  &  & (1.296) & (2.009) &  &  & (1.353) & (1.706) &  \\
Small Exponential & 5.292 & 5.438 & 0.933 &  & 5.562 & 5.208 & 0.208 &  & 5.417 & 5.271 & 0.578 \\
\multicolumn{1}{r}{} & (1.485) & (1.039) &  &  & (1.059) & (1.443) &  &  & (1.115) & (1.220) &  \\
Large Exponential & 4.604 & 4.688 & 0.760 &  & 4.625 & 4.771 & 0.468 &  & 4.479 & 4.542 & 0.657 \\
\multicolumn{1}{r}{} & (1.350) & (1.277) &  &  & (1.130) & (1.212) &  &  & (1.354) & (1.485) &  \\
Small Constant & 1.167 & 1.292 & 0.282 &  & 1.396 & 1.417 & 0.852 &  & 1.521 & 1.500 & 0.655 \\
\multicolumn{1}{r}{} & (0.373) & (0.538) &  &  & (0.729) & (0.886) &  &  & (0.790) & (0.979) &  \\
Large Constant & 1.938 & 1.896 & 0.727 &  & 2.167 & 2.312 & 0.575 &  & 1.958 & 2.042 & 0.814 \\
\multicolumn{1}{r}{} & (0.876) & (0.918) &  &  & (0.874) & (1.102) &  &  & (1.020) & (1.190) &  \\ \hline
\end{tabular}
\begin{tablenotes}
\item[1.] The standard deviations are reported in parentheses.
\item[2.] The statistical inference is based on the rank-sum test 
between the two orders.
\end{tablenotes}
\end{threeparttable}
\end{adjustbox}
\end{table}

To evaluate whether there is any significant order effect,
we compare the distributions of terminal nodes between the FR and RF Orders within
each elicitation method. Table~\ref{tab:aggregate_level_tests_two_orders} reports the
corresponding results. The ``FR Order'' and ``RF Order'' columns display the mean terminal 
nodes and their standard deviations (in parentheses) for the FS-then-RS and RS-then-FS sessions, respectively.
We use the rank-sum test to assess whether the distributions of terminal nodes differ 
significantly across the two orders for each elicitation method; the resulting $p$-values are 
reported in the ``$p$-values'' columns.

As shown in the ``Pooled Data'' row of Table~\ref{tab:aggregate_level_tests_two_orders}, there is no 
compelling evidence that the terminal node distributions differ significantly between the two ordering 
conditions for any of the three elicitation methods.
When we further disaggregate the analysis to the game level, the conclusion remains unchanged.
The second part of Table~\ref{tab:aggregate_level_tests_two_orders} reports the corresponding $p$-values from 
the rank-sum tests, which remain statistically insignificant across all games and elicitation methods.
These results indicate that the terminal node distributions are statistically indistinguishable between the 
two orders.
Thus, we conclude that there is no significant order effect.

\begin{result}
There is no statistically significant difference between the FR and RF orders.
\end{result}

\subsubsection{Robustness Analysis for Structural Estimation}

As a robustness analysis, we disaggregate the structural estimation in Section~\ref{section:structural_estimation} by difference classes of games 
to examine whether our main findings persist across different payoff structures (Tables~\ref{tab:simplified_model_estimation_Linear_appendix}, \ref{tab:simplified_model_estimation_Exponential_appendix}, and \ref{tab:simplified_model_estimation_Constant_appendix}, respectively). 
Across all game classes, we find that incorporating quantal responses into DCH (i.e., QDCH) continues to substantially improve model fit. QDCH consistently yields higher likelihood scores than DCH across all elicitation methods and in the pooled data.
Likelihood ratio tests confirm that this improvement is statistically significant in all classes ($p$-values $<$ 0.001), indicating that the quantal response effect is robust across different game structures.

Next, we assess whether the observed deviations from strategic equivalence are primarily driven by the quantal response component or the DCH mechanism by comparing QDCH with AQRE using the Vuong test.
For linear games, QDCH significantly outperforms AQRE across all treatments as well as in the pooled data (pooled $p$-value $< 0.001$).
For exponential games, the pattern is similar: QDCH outperforms AQRE in both the reduced and direct response treatments and in the pooled data (pooled $p$-value $< 0.001$), although the difference is not statistically significant under the full strategy method ($p$-value $= 0.539$).
This suggests that the DCH mechanism may play only a limited role under the full strategy method for exponential games. For constant games, QDCH also outperforms AQRE in the pooled data (pooled $p$-value $< 0.001$), but the difference is less pronounced than in the other game classes. In fact, QDCH and AQRE differ significantly at the 1\% level only in the pooled data.

Overall, the analysis confirms the robustness of our main conclusions.
Both the quantal response and DCH mechanisms contribute to the strategy method effect, though their relative importance varies across game structures and elicitation methods.
Quantal responses consistently lead to substantial improvements in model fit, whereas the influence of the DCH mechanism appears more context-dependent—stronger in linear games and weaker in constant games.

\begin{table}[H]
\centering
\caption{Estimation Results and Model Comparisons for Linear Games}
\label{tab:simplified_model_estimation_Linear_appendix}
\renewcommand{\arraystretch}{1.4}
\begin{adjustbox}{width=1.2\columnwidth,center}
\begin{threeparttable}
\begin{tabular}{cccccccccccccccc}
\hline
 & \multicolumn{3}{c}{Reduced Strategy Method} &  & \multicolumn{3}{c}{Full Strategy Method} &  & \multicolumn{3}{c}{Direct Response Method} &  & \multicolumn{3}{c}{Pooled Data} \\ \cline{2-4} \cline{6-8} \cline{10-12} \cline{14-16} 
 & QDCH & DCH & AQRE &  & QDCH & DCH & AQRE &  & QDCH & DCH & AQRE &  & QDCH & DCH & AQRE \\ \hline
$\tau$ & 1.741 & 0.665 & --- &  & 3.100 & 0.875 & --- &  & 2.544 & 1.042 & --- &  & 2.363 & 0.875 & --- \\
S.E. & (0.222) & (0.084) & --- &  & (0.379) & (0.034) & --- &  & (0.297) & (0.079) & --- &  & (0.175) & (0.025) & --- \\
$\lambda$ & 0.026 & --- & 0.014 &  & 0.046 & --- & 0.027 &  & 0.032 & --- & 0.019 &  & 0.040 & --- & 0.019 \\
S.E. & (0.004) & --- & (0.001) &  & (0.005) & --- & (0.001) &  & (0.003) & --- & (0.001) &  & (0.003) & --- & (0.001) \\ \hline
LL & \textbf{-398} & -444 & -425 &  & \textbf{-496} & -618 & -518 &  & \textbf{-360} & -419 & -386 &  & \textbf{-1283} & -1488 & -1359 \\
$\Delta\%$ &  & 11.6\% & 6.8\% &  &  & 24.6\% & 4.4\% &  &  & 16.4\% & 7.2\% &  &  & 16.0\% & 5.9\% \\ \hline
LRT &  & 93.3 & --- &  &  & 243.7 & --- &  &  & 117.9 & --- &  &  & 411.1 & --- \\
Vuong &  & --- & 3.425 &  &  & --- & 2.389 &  &  & --- & 3.167 &  &  & --- & 4.406 \\
$p$-value &  & $<0.001$\phantom{$<$} & $<0.001$\phantom{$<$} &  &  & $<0.001$\phantom{$<$} & 0.017 &  &  & $<0.001$\phantom{$<$} & 0.002 &  &  & $<0.001$\phantom{$<$} & $<0.001$\phantom{$<$} \\ \hline
\end{tabular}
\begin{tablenotes}
\small
\item[1.] The estimations for the reduced and full strategy 
method treatments are based on 384 choices of reduced and 
complete strategies, respectively. The estimation for the direct response 
method treatment is based on 920 decision-node choices.
\item[2.] $\Delta\%$ indicates the percentage improvement in likelihood relative to QDCH.
\item[3.] For the Vuong test of the direct response method treatment,
we account for the interdependence of decision-node choices by 
conducting the test on the 192 observations of terminal nodes.
\end{tablenotes}
\end{threeparttable}
\end{adjustbox}
\end{table}

\begin{table}[H]
\centering
\caption{Estimation Results and Model Comparisons for Exponential Games}
\label{tab:simplified_model_estimation_Exponential_appendix}
\renewcommand{\arraystretch}{1.4}
\begin{adjustbox}{width=1.2\columnwidth,center}
\begin{threeparttable}
\begin{tabular}{cccccccccccccccc}
\hline
 & \multicolumn{3}{c}{Reduced Strategy Method} &  & \multicolumn{3}{c}{Full Strategy Method} &  & \multicolumn{3}{c}{Direct Response Method} &  & \multicolumn{3}{c}{Pooled Data} \\ \cline{2-4} \cline{6-8} \cline{10-12} \cline{14-16} 
 & QDCH & DCH & AQRE &  & QDCH & DCH & AQRE &  & QDCH & DCH & AQRE &  & QDCH & DCH & AQRE \\ \hline
$\tau$ & 4.754 & 1.828 & --- &  & 7.037 & 1.303 & --- &  & 4.561 & 1.485 & --- &  & 5.218 & 1.584 & --- \\
S.E. & (0.197) & (0.347) & --- &  & (2.819) & (0.087) & --- &  & (0.295) & (0.101) & --- &  & (0.214) & (0.098) & --- \\
$\lambda$ & 0.012 & --- & 0.011 &  & 0.019 & --- & 0.017 &  & 0.015 & --- & 0.012 &  & 0.017 & --- & 0.014 \\
S.E. & (0.002) & --- & (0.001) &  & (0.002) & --- & (0.001) &  & (0.001) & --- & (0.001) &  & (0.001) & --- & (0.001) \\ \hline
LL & \textbf{-393} & -453 & -400 &  & \textbf{-469} & -533 & -470 &  & \textbf{-341} & -380 & -371 &  & \textbf{-1229} & -1374 & -1257 \\
$\Delta\%$ &  & 15.3\% & 1.8\% &  &  & 13.6\% & 0.2\% &  &  & 11.4\% & 8.8\% &  &  & 11.8\% & 2.3\% \\ \hline
LRT &  & 118.9 & --- &  &  & 126.7 & --- &  &  & 76.6 & --- &  &  & 290.3 & --- \\
Vuong &  & --- & 3.670 &  &  & --- & 0.614 &  &  & --- & 5.218 &  &  & --- & 3.781 \\
$p$-value &  & $<0.001$\phantom{$<$} & $<0.001$\phantom{$<$} &  &  & $<0.001$\phantom{$<$} & 0.539 &  &  & $<0.001$\phantom{$<$} & $<0.001$\phantom{$<$} &  &  & $<0.001$\phantom{$<$} & $<0.001$\phantom{$<$} \\ \hline
\end{tabular}
\begin{tablenotes}
\small
\item[1.] The estimations for the reduced and full strategy 
method treatments are based on 384 choices of reduced and 
complete strategies, respectively. The estimation for the direct response 
method treatment is based on 938 decision-node choices.
\item[2.] $\Delta\%$ indicates the percentage improvement in likelihood relative to QDCH.
\item[3.] For the Vuong test of the direct response method treatment,
we account for the interdependence of decision-node choices by 
conducting the test on the 192 observations of terminal nodes.
\end{tablenotes}
\end{threeparttable}
\end{adjustbox}
\end{table}

\begin{table}[H]
\centering
\caption{Estimation Results and Model Comparisons for Constant Games}
\label{tab:simplified_model_estimation_Constant_appendix}
\renewcommand{\arraystretch}{1.4}
\begin{adjustbox}{width=1.2\columnwidth,center}
\begin{threeparttable}
\begin{tabular}{cccccccccccccccc}
\hline
 & \multicolumn{3}{c}{Reduced Strategy Method} &  & \multicolumn{3}{c}{Full Strategy Method} &  & \multicolumn{3}{c}{Direct Response Method} &  & \multicolumn{3}{c}{Pooled Data} \\ \cline{2-4} \cline{6-8} \cline{10-12} \cline{14-16} 
 & QDCH & DCH & AQRE &  & QDCH & DCH & AQRE &  & QDCH & DCH & AQRE &  & QDCH & DCH & AQRE \\ \hline
$\tau$ & 3.270 & 2.176 & --- &  & 1.043 & 0.666 & --- &  & 7.870 & 1.228 & --- &  & 2.731 & 1.235 & --- \\
S.E. & (0.342) & (0.333) & --- &  & (0.364) & (0.200) & --- &  & (5.187) & (0.504) & --- &  & (0.240) & (0.076) & --- \\
$\lambda$ & 0.020 & --- & 0.022 &  & 0.031 & --- & 0.010 &  & 0.008 & --- & 0.008 &  & 0.016 & --- & 0.012 \\
S.E. & (0.002) & --- & (0.001) &  & (0.012) & --- & (0.001) &  & (0.003) & --- & (0.001) &  & (0.001) & --- & (0.001) \\ \hline
LL & \textbf{-346} & -381 & -353 &  & \textbf{-761} & -771 & -767 &  & \textbf{-180} & -204 & -180 &  & \textbf{-1302} & -1374 & -1338 \\
$\Delta\%$ &  & 10.1\% & 2.0\% &  &  & 1.3\% & 0.8\% &  &  & 13.3\% & 0.0\% &  &  & 5.5\% & 2.8\% \\ \hline
LRT &  & 71.0 & --- &  &  & 20.3 & --- &  &  & 48.7 & --- &  &  & 142.3 & --- \\
Vuong &  & --- & 2.140 &  &  & --- & 1.823 &  &  & --- & 0.427 &  &  & --- & 4.829 \\
$p$-value &  & $<0.001$\phantom{$<$} & 0.032 &  &  & $<0.001$\phantom{$<$} & 0.068 &  &  & $<0.001$\phantom{$<$} & 0.670 &  &  & $<0.001$\phantom{$<$} & $<0.001$\phantom{$<$} \\ \hline
\end{tabular}
\begin{tablenotes}
\small
\item[1.] The estimations for the reduced and full strategy 
method treatments are based on 384 choices of reduced and 
complete strategies, respectively. The estimation for the direct response 
method treatment is based on 302 decision-node choices.
\item[2.] $\Delta\%$ indicates the percentage improvement in likelihood relative to QDCH.
\item[3.] For the Vuong test of the direct response method treatment,
we account for the interdependence of decision-node choices by 
conducting the test on the 192 observations of terminal nodes.
\end{tablenotes}
\end{threeparttable}
\end{adjustbox}
\end{table}

\end{document}